\documentclass[structabstract, a4paper]{aa}  
\usepackage{graphicx}
\usepackage{txfonts}
\usepackage[utf8]{inputenc}
\usepackage{dcolumn}
\usepackage{supertabular}
\usepackage{rotating}
\usepackage{longtable}
\usepackage{lscape}
\usepackage{float}
\usepackage{subfig}
\usepackage{url}
\usepackage{color}
\usepackage{natbib}
\usepackage[dvipsnames]{xcolor}
\usepackage{ulem}

\usepackage[colorinlistoftodos]{todonotes}

\bibpunct{(}{)}{;}{a}{}{,} 

\newcommand{\kms}{\mbox{km\,s$^{-1}$}}

\newcommand{\dd}{$^{\circ}$}

\newcolumntype{d}[1]{D{.}{\cdot}{#1}}
\newcolumntype{.}{D{.}{.}{-1}}

\newcommand{\hii}{H{\sc ii}}

\newcommand{\uchii}{UC\,H{\sc ii}}

\newcommand{\mum}{$\mu$m}

\newcommand{\lsun}{L$_\odot$}
\newcommand{\msun}{M$_\odot$}

\begin{document}

\title{New detections of (sub)millimeter hydrogen radio recombination lines towards high-mass star-forming clumps}
\authorrunning{Kim et al.}
\titlerunning{High-frequency recombination line survey}

\author{W.-J.\,Kim\inst{1}\thanks{Member of the International Max Planck Research School (IMPRS) for Astronomy and Astrophysics at the Universities of Bonn and Cologne.}
        \and 
                J.\,S.\,Urquhart\inst{2}
                \and
                F.\,Wyrowski\inst{1} 
                \and
        K.\,M.\,Menten\inst{1}
                \and 
        T.\,Csengeri\inst{1}
}

\institute{ Max-Planck-Institut f\"ur Radioastronomie, Auf dem H\"ugel 69, 53121 Bonn, Germany\\ \email{wjkim@mpifr-bonn.mpg.de, kim@iram.es} \and  School of Physical Sciences, University of Kent, Ingram Building, Canterbury, Kent CT2\,7NH, UK}

\date{ drafted / Received 2017 / Accepted 2018}

\abstract
{}{Previous radio recombination line (RRL) observations of dust clumps identified in the APEX Telescope Large Area Survey of the Galaxy (ATLASGAL) have led to the detection of a large number of RRLs in the 3mm range. Here, we aim to study their excitation with shorter wavelength (sub)millimeter radio recombination line (submm-RRL) observations.}
{We made observations of submm-RRLs with low principal quantum numbers ($n$ $\leq$ 30)  using the APEX 12\,m telescope, toward 104 \hii\ regions associated with massive dust clumps from ATLASGAL. The observations covered the H25$\alpha$, H28$\alpha$, and H35$\beta$ transitions. Toward a small subsample the H26$\alpha$, H27$\alpha$, H29$\alpha$, and H30$\alpha$ lines were observed to avoid contamination by molecular lines at adjacent frequencies.}
{We have detected submm-RRLs (signal-to-noise $\geq$ 3\,$\sigma$) from compact \hii\ regions embedded within 93 clumps. The submm-RRLs are approximately a factor of two brighter than the mm-RRLs and consistent with optically thin emission in local thermodynamic equilibrium (LTE). The average ratio (0.31) of the measured H35$\beta$/H28$\alpha$ fluxes is close to the LTE value of 0.28. No indication of RRL maser emission has been found.
The Lyman photon flux, bolometric, and submm-RRL luminosities toward the submm-RRL detected sources present significant correlations. The trends of dust temperature and the ratio of bolometric luminosity to clump mass, $L_{\rm bol}/M_{\rm clump}$, indicate that the \hii\ regions are related to the most massive and luminous clumps. By estimating the production rate of ionizing photons, $Q$, from the submm-RRL flux, we find that the $Q$(H28$\alpha$) measurements provide estimates of the Lyman continuum photon flux consistent with those determined from 5\,GHz radio continuum emission. Six RRL sources show line profiles that are a combination of a narrow and a broad Gaussian feature. The broad features are likely associated with high-velocity ionized flows. } 
{We have detected submm-RRLs toward 93 ATLASGAL clumps. Six RRL sources have high-velocity RRL components likely driven by high-velocity ionized flows. 
Their observed properties are consistent with thermal emission that correlates well with the Lyman continuum flux of the \hii\ regions.
The sample of \hii\ regions with mm/submm-RRL detections probes, in our Galaxy, luminous clumps ($L_{\rm bol} > 10^4$\,\lsun) with high $L_{\rm bol}/M_{\rm clump}$. We also provide suitable candidates for further studies of the morphology and kinematics of embedded, compact \hii\ regions with the Atacama Large Millimeter/submillimeter Array (ALMA). } 

\maketitle



\section{Introduction}\label{sec:intro}

Massive stars affect their environment in various ways, thereby shaping the evolution of galaxies and leading to chemical enrichment of the interstellar medium (\citealt{zinnecker2007}). Due to fast timescales of high-mass star formation, their whole evolution takes place whilst they are still deeply embedded in their natal clump. Indeed, the high-mass star or cluster is hidden at optical and even near-infrared wavelengths. For these reasons, it is more challenging to make observations of high-mass star formation than those of low-mass star formation, and also they are rare and located at much larger distances \citep{schuller2009,urquhart2018_agal_full}. For these reasons, the feedback between young OB stars and their molecular clouds is still not well understood. 

Young massive stars are hot and thus create small regions of hot ionized gas around them that expands into their local environment and interacts with the surrounding molecular gas; these are known as compact \hii\ regions. The \hii\ region phase is the last stage of the massive star formation process. Studying this phase is not only important due to the exciting stars' influence on their environment, but it is also key to understand how high-mass stars obtain masses above 10\,\msun\ \citep{churchwell2010}. Indeed, the youngest and most compact \hii\ regions (hyper- and ultra-compact) are often associated with various star formation signposts such as molecular outflows, infall motions, and methanol and water masers, which themselves are pumped by radiation and shocks (\citealt{urquhart2013_cornish,urquhart2015_mathanol}). 
These signposts suggest that the mass assembly process is still active.

The gas in \hii\ regions surrounding newly born OB stars has velocity dispersions of 25$-$35\,\kms\ determined by a mix of thermal  ($\sim$\,20\,\kms) and turbulent motions (\citealt{wilson2009}). The kinematics in the ionized gas can be investigated using radio recombination lines (RRLs), particularly toward sources that are deeply embedded in dense molecular clumps. In these dense clumps, observations of Lyman $\alpha$, H$\alpha,$ and other ultraviolet, optical, or near-infrared recombination lines are strongly attenuated due to high levels of visual extinction (up to hundreds of mag.) while RRL emission can still escape out of the clumps. Therefore, RRLs are an excellent tool for analyzing the distribution and kinematics of the ionized gas associated with \hii\ regions. 

Interestingly, strong maser emission has been observed from submillimeter RRLs toward only a few sources \citep{thum1995,martin-pintado2002,contreras2017}. Therefore, it is interesting to investigate how widespread RRL maser emission is towards \hii\ regions. 

Submillimeter-RRL observations provide kinematics of \hii\ regions and probe the Lyman continuum photon production rate, which can be used as a measure of star formation rates (SFRs) in external galaxies \citep{scoville2013,bendo2017}. These rates have been compared with those measured by other SFR tracers such as ultraviolet continuum emission, optical/near-infrared recombination lines, mid-/far-infrared continuum emission, and radio continuum emission and have been found to give consistent results (e.g., NGC5253, \citealt{bendo2017}).

In our previous study of hydrogen RRLs at millimeter wavelengths (\citealt{kim2017}), we identified 178 mm-RRL sources toward 976 compact dust clumps selected from the Atacama Pathfinder EXperiment (APEX) Telescope Large Area Survey of the Galaxy (ATLASGAL) Compact Source Catalogues and GaussClump Source Catalogue (CSCs; \citealt{contreras2013,urquhart2014_csc}, and GCSC; \citealt{csengeri2014}). 
These mm-RRLs were found to be associated with embedded \hii\ regions identified from cm-wavelength radio continuum surveys (e.g., \citealt{urquhart_radio_south,urquhart2013_cornish}) and were used to derive the properties of the ionized gas, which in turn were compared with the properties of their natal molecular clumps. The properties of the mm-RRLs were found to be consistent with properties derived from the 6\,cm radio continuum.

To further characterize the properties of these \hii\ regions, we have re-observed many of these sources targeting RRLs at submillimeter wavelengths, which not only provide brighter flux than mm-RRLs but also allow us to investigate the RRL excitation and whether maser emission is influencing the intensity of RRLs.
We have, therefore, carried out submillimeter RRL (submm-RRL) observations using the APEX 12m telescope to investigate further the \hii\ regions detected in the course of our mm-RRL survey.

The observations and data reduction are explained in Sect.\,\ref{sec:obs}. 
The general properties of the detected submm-RRLs, including the detection rates, and a comparison between submm-RRL and mm-RRL are presented in Sect.\,\ref{sec:results}. The associations with \hii\ regions and molecular clumps are described in Sect.\,\ref{sec:hii_mol}. The derivation of the ionizing Lyman photon production rate from the submm-RRL data is described in Sect.\,\ref{sec:q_subrrl}. Several sources with peculiar line profiles are discussed in Sect.\,\ref{sec:interseting}. Lastly, we present a summary of our main results in Sect.\,\ref{sec:sum_con}. 

\section{Observation and data reduction}\label{sec:obs}
\subsection{Source selection and observational setup}\label{sec:source}

The sources for this targeted survey with the APEX 12\,m diameter submillimeter telescope \citep{gusten2006} have been selected based only on their peak mm-RRL intensity as measured from our surveys with the Institut de Radioastronomie Millim\'etrique (IRAM) 30\,m and the Mopra 22\,m telescopes (\citealt{kim2017}). For the selection, the intensity threshold was set to 0.2\,K of the peak mm-RRL intensity. This threshold was chosen by comparing the intensity of stacked mm-RRL profiles\footnote{Stacked transitions from the IRAM 30\,m data are H39$\alpha$, H40$\alpha$, H41$\alpha,$ and H42$\alpha$, and in the case of the Mopra 22\,m data transitions are H41$\alpha$ and H42$\alpha$ (see \citealt{kim2017} for details).}  to H26$\alpha$ RRL data that were observed by APEX as part of a different molecular line survey toward the ATLASGAL ``Top 100'' sample (see \citealt{giannetti2014,konig2017} for details). In total, 104 ATLASGAL sources were selected from the mm-RRL catalog \citep{kim2017}, and the submm-RRL observations with the APEX 12\,m telescope were carried out primarily in 2015 (Project IDs: M0025-95 and M0018-96) with some additional observations being made in 2016.

The APEX observations used the dual frequency First Light APEX Submillimeter Heterodyne (FLASH) receiver \citep{klein2014} for the H25$\alpha$, H27$\alpha$, H28$\alpha,$ and H35$\beta$ transitions and the APEX-1 (HET230, \citealt{vassilev2008}) receiver for the H29$\alpha$ and H30$\alpha$ transitions. The observations mainly covered the H25$\alpha$, H28$\alpha,$ and H35$\beta$ transitions. However, the H27$\alpha$, H29$\alpha,$ and H30$\alpha$ transitions were also observed to obtain better line profiles toward some sources for which the other RRL profiles were found to be contaminated by emission from molecular lines. The spectra of the observed submm-RRL transitions, therefore, vary slightly from source to source. In Table\, \ref{tb:obs_source_list} we provide details of the transitions observed toward each source and indicate where they are detected. In Table\,\ref{tb:obs_log} we give the observed frequency, the oscillator strength of each transition, and the Full width at half maximum (FWHM) beam widths of the APEX beam at the frequency of the transition, the conversion factor between K and Jy, and the telescope's main beam efficiency, $\eta_{\rm MB}$\footnote{http://www.apex-telescope.org/instruments.}. The beam sizes given in  Table\,\ref{tb:obs_log} are calculated using $\Theta_{\rm FWHM} = 7.8''\, \times\,\left( 800/\nu \right)$\footnote{http://www.apex-telescope.org/telescope.}, where $\nu$ is a rest frequency of the transition in GHz.

\begin{table*}
\centering
\caption{\label{tb:obs_source_list} List of observed sources. }
\begin{tabular}{c c c c c c c }
\hline\hline
ID & ATLASGAL && RA & Dec. && Observed  \\
No. & clump name && $\alpha$(J2000) & $\delta$(J2000) &&  transition \\
\hline
1&AGAL010.151$-$00.344&&18:09:21.2&$-$20:19:28&&H25$\alpha$, H28$\alpha$, (H35$\beta$?)\\
2&AGAL010.168$-$00.362&&18:09:26.7&$-$20:19:03&&(H25$\alpha$?), H28$\alpha$, (H35$\beta$)\\
3&AGAL010.472$+$00.027&&18:08:37.9&$-$19:51:48&&(H26$\alpha^{c}$?)\\
4&AGAL010.624$-$00.384&&18:10:28.6&$-$19:55:46&&H25$\alpha$, H26$\alpha$, H28$\alpha$, H35$\beta$\\
5&AGAL011.936$-$00.616&&18:14:00.8&$-$18:53:24&&H25$\alpha$, H28$\alpha$, (H35$\beta$?)\\
 6&AGAL012.804$-$00.199&&18:14:13.5&$-$17:55:32&&H25$\alpha$, H26$\alpha$, H28$\alpha$, H35$\beta$\\
7&AGAL013.209$-$00.144&&18:14:49.3&$-$17:32:46&&H25$\alpha$, H28$\alpha$, (H35$\beta$?)\\
8&AGAL013.872$+$00.281&&18:14:35.6&$-$16:45:39&&H25$\alpha$, H28$\alpha$, H35$\beta$\\
9&AGAL015.024$-$00.654&&18:20:17.9&$-$16:11:30&&H25$\alpha$, H28$\alpha$, (H35$\beta$)\\
10&AGAL015.029$-$00.669&&18:20:22.4&$-$16:11:44&&H25$\alpha$, H26$\alpha$, H28$\alpha$, (H35$\beta$)\\
11&AGAL015.051$-$00.642&&18:20:18.7&$-$16:09:43&&(H25$\alpha$?), H28$\alpha$, (H35$\beta$)\\
12&AGAL018.301$-$00.389&&18:25:41.8&$-$13:10:21&&H25$\alpha$, H28$\alpha$, (H35$\beta$)\\
13&AGAL028.199$-$00.049&&18:42:58.1&$-$04:13:58&&H25$\alpha$, H28$\alpha$, (H35$\beta$)\\
14&AGAL029.954$-$00.016&&18:46:03.5&$-$02:39:24&&H25$\alpha$, H27$\alpha$, H28$\alpha$, H35$\beta$\\
15&AGAL030.753$-$00.051&&18:47:38.2&$-$01:57:51&&(H25$\alpha$), H28$\alpha$, (H35$\beta$)\\
\hline
\end{tabular}
\tablefoot{Non-detections are given in parentheses. Uncertain detections are marked with a question mark, ``?'', and are considered as non-detections for all data analysis. A superscripted ``$c$'' indicates severe contamination by complex molecules or unidentified lines.  The full table is available in Appendix\,\ref{appendix:full_tb}.}
\end{table*}

\begin{figure}
\centering
\includegraphics[width = 0.50\textwidth]{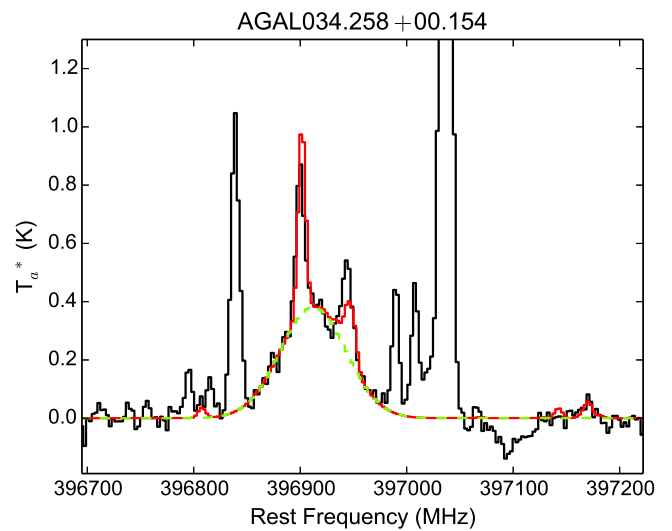}
\caption{\label{fig:identify_mol} Example of the identification of molecular lines blended with submm-RRL emission (black line) in AGAL034.258+00.154. The red profile shows a combined profile of the modeled profiles of CH$_{3}$OCHO lines using WEEDS and a Gaussian profile of the H25$\alpha$ RRL. The latter is also shown as the green dashed line. Other emission lines are unidentified molecular lines.}
\end{figure}

\begin{table*}
\centering
\caption{\label{tb:obs_log} Observed transitions and observational parameters.} 
\begin{tabular}{c c c c c c c c c }
\hline\hline
RRL &Frequency & Oscillator & Observed & rms  & Beam size & K to Jy & $\eta_{\rm MB}$  & Receiver\\
transition  & (MHz) & strength$^{*}$ & sources & (Jy) & ($''$) & (Jy\,K$^{-1}$) &  &\\
\hline
H25$\alpha \cdots \cdots$ & 396900.866 & 5.0541050 & 85 & 1.24  & 15.7 & 41 & 0.73 &FLASH\\
H26$\alpha \cdots \cdots$ & 353622.776 & 5.2449384 & 27 & 1.64  & 17.6 & 41 & 0.73 &FLASH\\
H27$\alpha \cdots \cdots$ & 316415.451 & 5.4357673 & 13 & 0.43  & 19.7 & 41 & 0.73 &FLASH\\
H28$\alpha \cdots \cdots$ & 284250.594 & 5.6265921 & 85 & 0.50  & 22.0 & 39 & 0.75 &FLASH\\ 
H29$\alpha \cdots \cdots$ & 256302.056 & 5.8174132 & 7  & 0.44  & 24.3 & 39 & 0.75 &APEX-1 (HET230)\\ 
H30$\alpha \cdots \cdots$ & 231900.947 & 6.0082311 & 30 & 0.40  & 26.9 & 39 & 0.75 &APEX-1 (HET230)\\
\hline
H35$\beta \cdots \cdots$  & 282332.932 & 1.0002938 & 85 & 0.45 & 22.1 & 39 & 0.75 &FLASH\\
\hline
\end{tabular}
\tablebib{(*)~\citet{goldwire1968}.}
\end{table*}

\subsection{Data reduction and identification of blended molecular lines}\label{sec:reducation}

In the submillimeter wavelength regime, the spectra of hot molecular cores, that is, warm, dense regions around embedded massive young stellar objects, are dominated by many lines from complex molecules. As a result, molecular lines can overlap with the submm-RRLs, making the detection and line fitting more complicated, particularly when trying to distinguish between potential RRL maser emission, which can have spectral features as narrow as typical molecular lines \citep{thum1995,martin-pintado2002}. Therefore, it is necessary to identify any molecular line contribution to observed narrow features on a broad profile.

To identify narrow line features, we used WEEDS within the Continuum and Line Analysis Single-dish Software (CLASS) of the GILDAS package\footnote{https://www.iram.fr/IRAMFR/GILDAS/doc/html/class-html/class.html.}; CLASS is also used for the RRL data reduction. Figure\,\ref{fig:identify_mol} shows an example of identified emission lines of the CH$_{3}$OCHO molecule and a submm-RRL (H25$\alpha$) toward AGAL034.258$+$00.154. We fit the CH$_{3}$OCHO molecular emission assuming local thermal equilibrium (LTE) and that the lines are exposed to the same physical conditions (single value of excitation temperature, column density, linewidth, and peak velocity to all transitions of this molecule within observed frequency bands). The fits to the CH$_{3}$OCHO line data obtained from WEEDS and the H25$\alpha$ line are shown in red and green, respectively. The broad profile of the H25$\alpha$ line  is consistent with that of other submm-RRL transitions observed towards this source with regard to peak velocity, intensity, and linewidth. In addition, we find that, while the H26$\alpha$ transition is not blended with molecular lines nearby in frequency, it is contaminated in several cases by attenuated strong emission from a CS (7$-$6) transition ($\nu$ = 342883\,MHz) originating from the other side-band of the receiver. 

All narrow features could be identified as blended complex lines from molecules. The observed broad features originate from hydrogen submm-RRLs. We, therefore, only use the line parameters determined from fits to the broad velocity features for the analysis in this paper. A polynomial baseline of order one to three was fitted to the line-free channels in a 300\,\kms\ wide velocity range and subtracted from the spectrum of each transition. The average root mean square (rms) values for each transition are given in Table\,\ref{tb:obs_log}; these were determined from the line-free channels, which have a velocity resolution of 2.1\,\kms. In some cases the submm-RRLs have been smoothed to a velocity resolution of 4.1\,\kms\ to increase the signal-to-noise ratio. We find that the majority of the submm-RRLs can be fitted with a single Gaussian component. However, for a  handful of sources, the profiles have significant red or blue shifted wings and these were fitted with two components, a narrower component that fits the central portion of the line and a much broader component that fits the high-velocity wings. We will discuss those sources with two Gaussian RRL components in Sect.\,\ref{sec:interseting}. Furthermore, we re-reduced all data of individual and stacked mm-RRLs presented in \cite{kim2017} using the same velocity resolution as for the submm-RRLs to facilitate comparisons of the mm- and submm-RRLs (see Fig.\,2 for examples).

\begin{table}
\centering
\caption{\label{tb:detection} List of detected source numbers and detection rates for each transition.}
\begin{tabular}{cccccc}
\hline\hline
RRL & Observed  & Detected& Detection \\
transition & sources &  sources & Rate (\%)\\
\hline
H$n\alpha \, \, \cdots \cdots$ & 104 & 93& 89\\ 
\hline
H25$\alpha \cdots \cdots$ & 85 & 43 & 51\\
H26$\alpha \cdots \cdots$ & 27 & 16 & 59\\
H27$\alpha \cdots \cdots$ & 13 & 12 & 92\\
H28$\alpha \cdots \cdots$ & 85 & 79 & 93\\ 
H29$\alpha \cdots \cdots$ &  7 &  5 & 71\\ 
H30$\alpha \cdots \cdots$ & 30 & 26 & 86\\
\hline
H35$\beta \cdots \cdots$  & 85 & 35 & 41\\
\hline
\end{tabular}
\tablefoot{The H$n\alpha$ means the staked RRL of the detected submm-RRL transitions for an individual source.}
\end{table}
\setlength{\tabcolsep}{2pt}
\begin{table}
\centering
\caption{\label{tb:line_para} Gaussian line parameters of the individual submm-RRL transitions and stacked submm-RRLs. }
\begin{tabular}{c c c c c . . }
\hline\hline
ID & RRL & $\varv_{\rm Peak}$ & $\Delta \varv$ & Area & \multicolumn{1}{c}{Peak} & \multicolumn{1}{c}{rms}  \\
No. & transition & (\kms) & (\kms)    & (Jy \kms) & \multicolumn{1}{c}{(Jy)} & \multicolumn{1}{c}{(Jy)} \\
\hline
1&H$n\alpha$ & $+$21 & 32$\pm$1 & 156.5$\pm$5.7 & 4.64 &0.46 \\
 &H25$\alpha$ & $+$22 & 24$\pm$9 & 80.7$\pm$20.3 & 3.21 & 1.10 \\
 &H28$\alpha$ & $+$20 & 31$\pm$2 &  165.6$\pm$7.7 & 4.99 & 0.63 \\
 &H35$\beta$ & $-$ & $-$ &  $-$ & $-$ & 0.56 \\ 
\hline
2&H$n\alpha$ & $+$11 & 17$\pm$1 & 71.7$\pm$4.2 &  3.88 & 0.47 \\
 &H25$\alpha$ & $-$ & $-$ & $-$ & $-$ & 1.15 \\
 &H28$\alpha$ & $+$11 & 21$\pm$3 & 77.7$\pm$8.0 & 3.43 & 0.66 \\
 &H35$\beta$ & $-$ & $-$ & $-$ & $-$ & 0.78 \\
\hline
3&H26$\alpha$ & $-$ & $-$ & $-$ & $-$ & 2.01 \\
\hline
4&H$n\alpha$ & $+$0 & 29$\pm$1 & 310.6$\pm$7.7 & 10.00 & 0.49\\
 &H25$\alpha$ & $+$3 & 33$\pm$4 & 254.2$\pm$77.5 & 7.18 & 1.23 \\
 &H26$\alpha$ & $+$6 & 38$\pm$3 & 336.5$\pm$28.6 & 8.36 & 1.47 \\
 &H28$\alpha$ & $+$0 & 29$\pm$1 & 314.7$\pm$7.0 & 10.20 & 0.48 \\
 &H35$\beta$ &  $+$1 & 30$\pm$2 & 104.0$\pm$7.2 & 3.23 & 0.60 \\
\hline
\end{tabular}
\tablefoot{All line parameters are available at the CDS via anonymous ftp. The ID number corresponds to the ID number in Table.\,\ref{tb:obs_source_list}.}
\end{table}
\setlength{\tabcolsep}{6pt}

\section{Results}\label{sec:results}

\subsection{Detection rate}

Since the intensity of RRLs increases with the frequency of the RRL \citep{gordon2002}, high detection rates of submm-RRLs were expected from the observed mm-RRL sources. Table\,\ref{tb:detection} gives the detection rate for each submm-RRL transition. The detection rates are lower than 100\% because our observations are less sensitive than the mm-observations; the non-detections toward six sources might be due to poorer flux sensitivity (for H28$\alpha$, rms = 0.50\,Jy compared to the rms values of the Mopra and IRAM observations, which were 0.20 and 0.05\,Jy, respectively.). Furthermore, the detections of submm-RRLs towards 11 clumps are ambiguous due to blending from molecular lines that make it difficult to separate the molecular and submm-RRLs profiles; we considered these sources as non-detections, which are indicated in Table\,\ref{tb:obs_source_list}.

\begin{figure}
\centering
\includegraphics[width = 0.50\textwidth, trim= 0.5cm 1.0cm 0.5cm 2.0cm, clip]{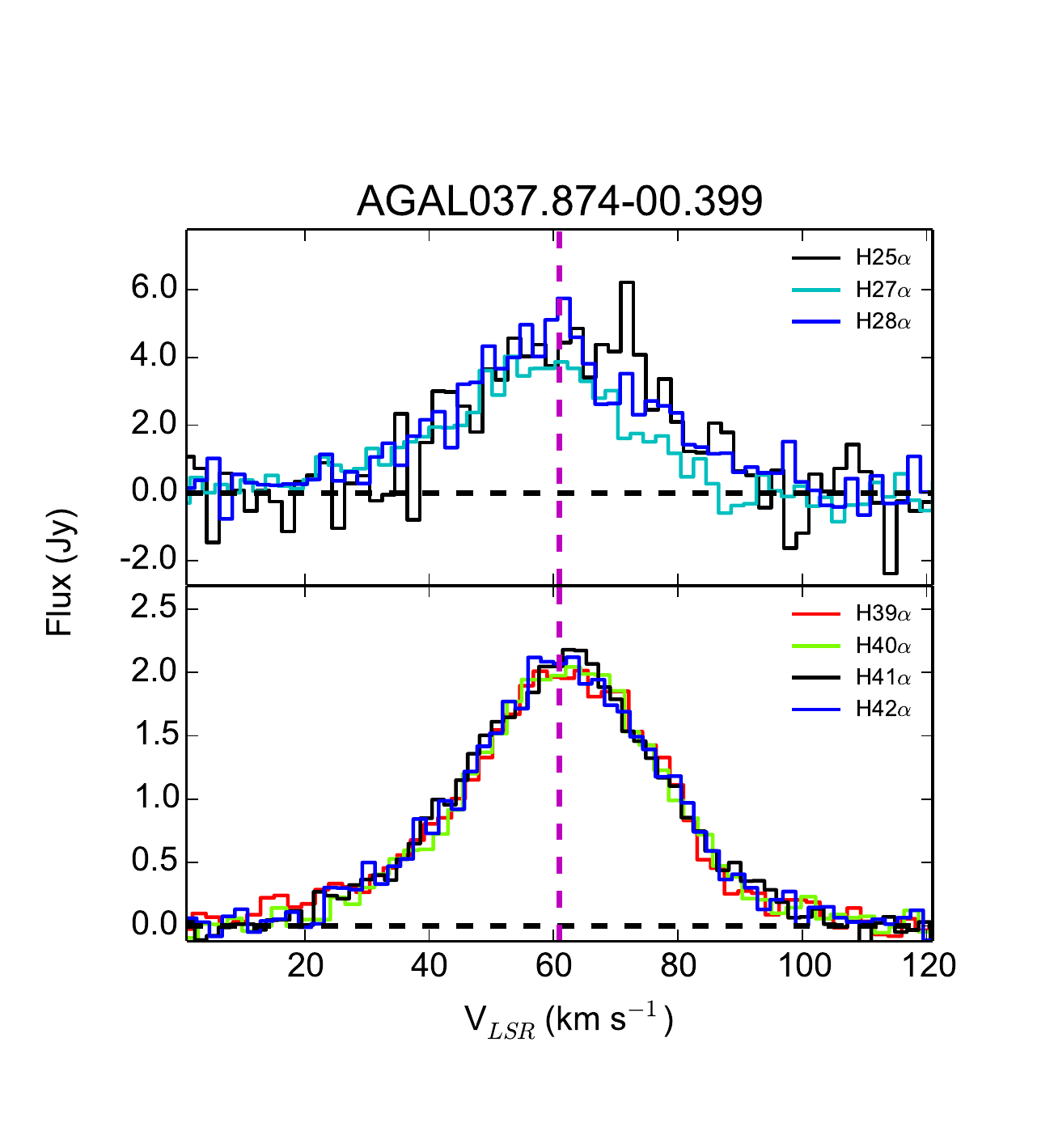}
\includegraphics[width = 0.50\textwidth, trim= 0.5cm 1.0cm 0.5cm 2.0cm, clip]{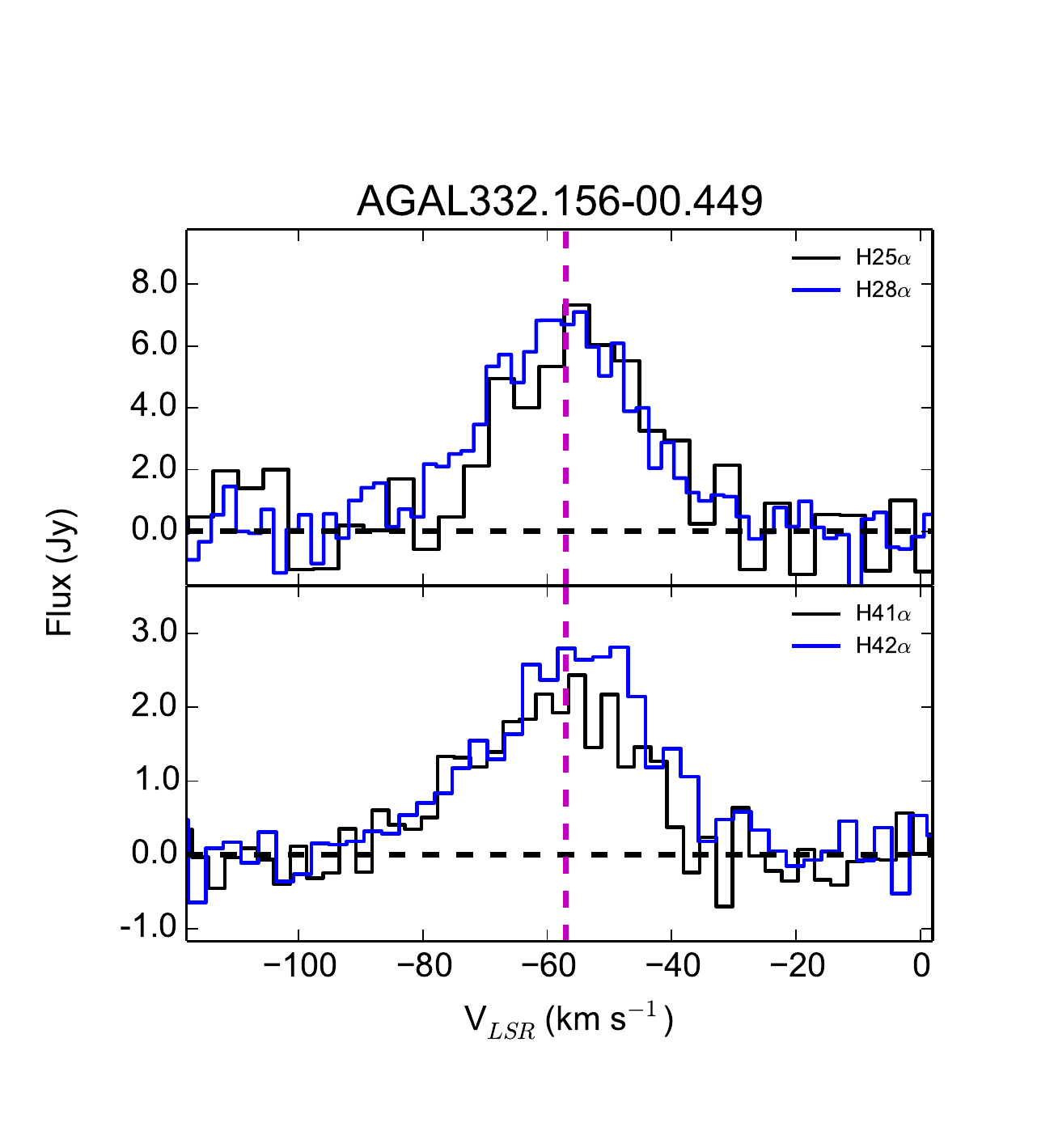}
\caption{\label{fig:spec} Examples of detected submm-RRLs (upper panels) of AGAL037.874$-$00.399 and AGAL332.156$-$00.449 and their corresponding mm-RRLs (lower panels) from \cite{kim2017}. The vertical purple dashed lines indicate the systemic velocity measured with H$^{13}$CO$^{+}$ (1$-$0) lines. The different colors present different submm/mm-RRL transitions detected in this paper and \cite{kim2017}. In flux unit, 1\,Jy corresponds to $\sim$\,0.025\,K for the APEX 12\,m, 0.169\,K for the IRAM 30m, and 0.045\,K for the Mopra 22\,m data.}
\end{figure}

Furthermore, if the blending is not too severe as in the case of the H25$\alpha$ of AGAL034.258$+$00.154 (Fig.\,\ref{fig:identify_mol}), the molecular emission and RRL can be reliably separated, and the detection of the submm-RRL is considered valid. The H$n\alpha$ and H35$\beta$ submm-RRL transitions (signal-to-noise (S/N) ratio $\geq$ 3$\sigma$) have been detected towards a large fraction of the observed sources; these are detected towards 93 (89\,\%) and 34 (40\,\%) clumps, respectively. The detection rate of the H35$\beta$ compared with the H$n\alpha$ transitions is consistent with the results of the mm-RRL survey at a similar S/N level.

\subsection{Properties of submm-RRLs}

\begin{figure}
\centering
\includegraphics[width = 0.50\textwidth, trim= 0.5cm 0.5cm 0.2cm 0.2cm, clip]{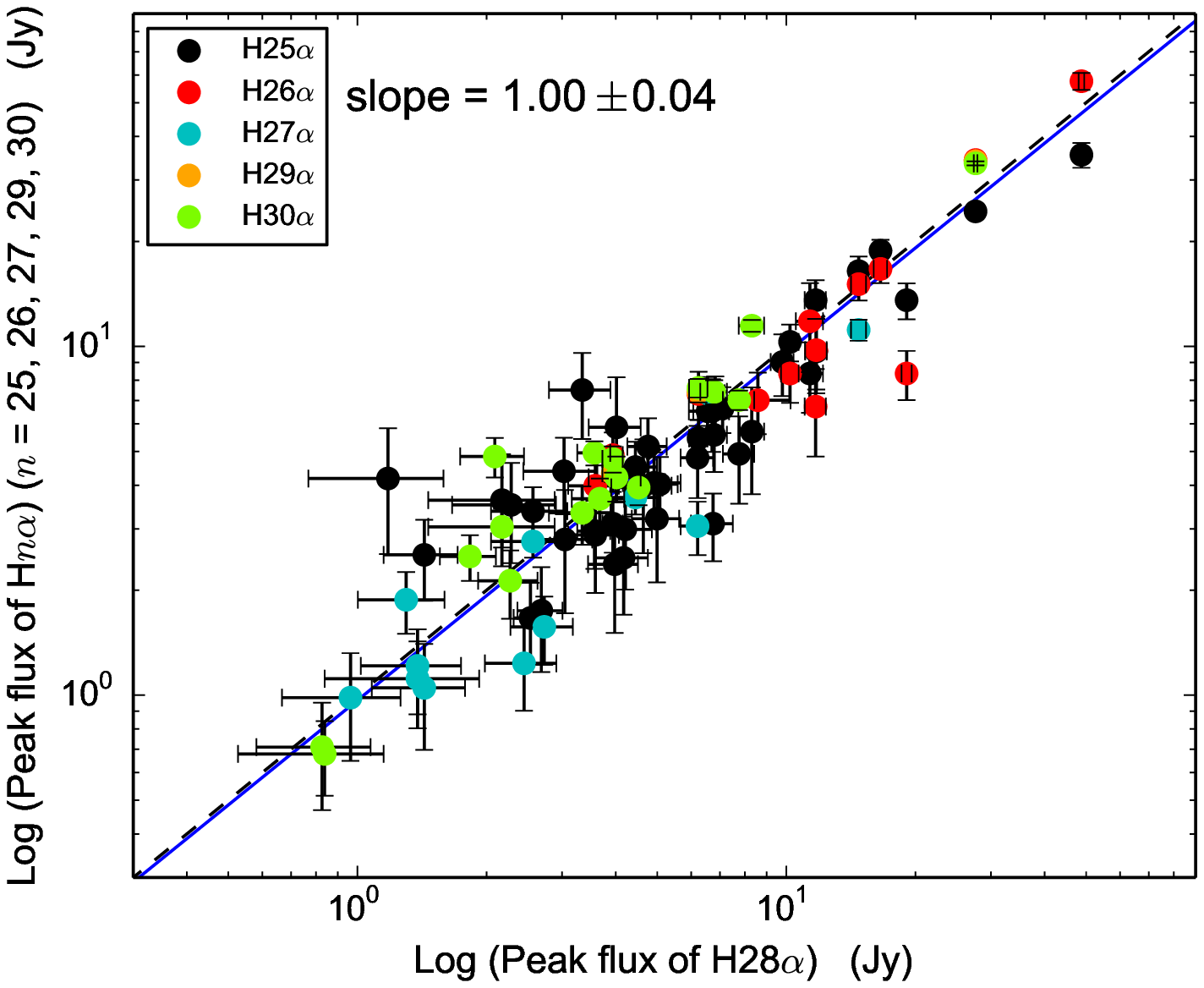}
\includegraphics[width = 0.50\textwidth, trim= 0.5cm 0.5cm 0.2cm 0.2cm, clip]{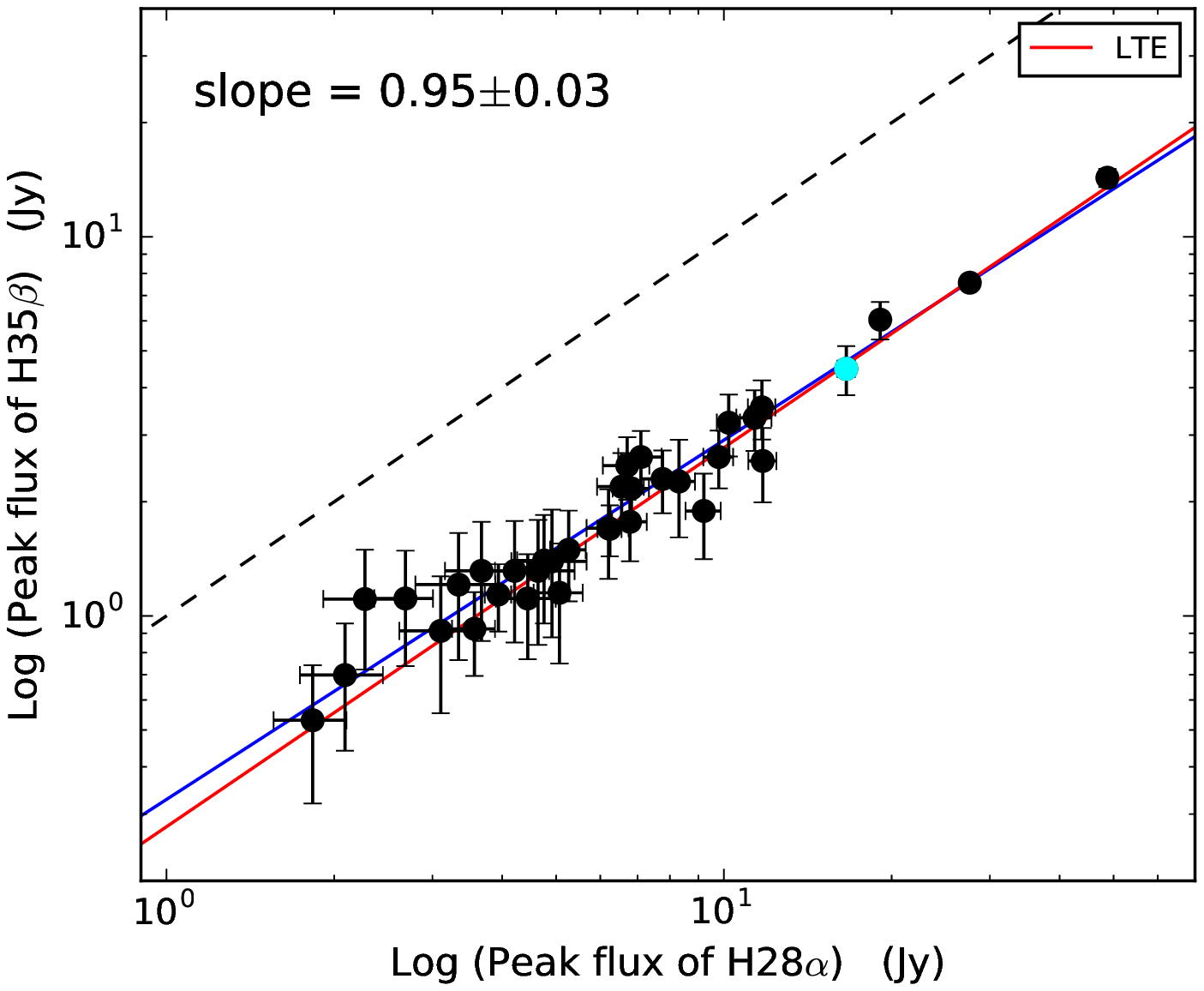}
\caption{\label{fig:submm_flux} Upper: Peak flux of the H28$\alpha$ transition versus peak flux of other transitions; H25$\alpha$ (black), H26$\alpha$ (red), H27$\alpha$ (cyan), H29$\alpha$ (orange), and H30$\alpha$ (bright green). All the plotted sources are detected in the H28$\alpha$ transition, and thus there are a few sources missing, which are only detected in other lines. 
Lower: Peak flux of the H28$\alpha$ transition versus peak flux of H35$\beta$. In both plots, equal fluxes are indicated by the black dashed line, and blue lines indicate the best-fit determined by BCES bisector fits to all data points. The red line in the right panel presents the predicted LTE value.}
\end{figure}

Figure\,\ref{fig:spec} displays examples of detected submm-RRL spectra (upper panels) and the corresponding mm-RRL spectra (lower panels; \citealt{kim2017}) toward AGAL037.874$-$00.399 and AGAL332.156$-$0.449. The profiles of the submm- and mm-RRLs show good agreement in peak velocity, linewidth, and profile shape, however, we note a significant difference in the intensity, with the submm-RRLs being approximately a factor of two brighter. Figure\,\ref{fig:submm_flux} shows a comparison of the peak flux of the H28$\alpha$ line to the peak fluxes measured for the other submm-RRL transitions (i.e., $n$ = 25, 26, 27, 29, and 30) (upper panel), and also for the H35$\beta$ transition (lower panel). We chose the H$28\alpha$ transition as the main transition for these comparisons because this transition is mainly covered by all the observations and is detected with a reasonable S/N level toward the observed sources.

Since we observed multiple submm-RRL transitions toward the majority of our sources, we can investigate whether there is a significant flux difference for different quantum numbers. Overall the peak fluxes determined from the different submm-RRLs agree with each other within the uncertainties (as shown in the upper panel of Fig.\,\ref{fig:submm_flux}): the slope determined from a linear bivariate correlated errors and intrinsic scatter bisector (BCES bisector){\footnote{Bivariate correlated errors and intrinsic scatter (BCES) fit is a direct extension of ordinary least-squares regression with measurement errors of two variables (e.g., X and Y) \citep{akritas1996}. Here, BCES bisector fit is one of the methods when it is not apparent which variable is considered as the dependent one and which as independent. It gets the line that bisects the BCES(X|Y) and BCES(Y|X) lines, which produce a different slope \citep{akritas1996}. The code is offered from \url{http://home.strw.leidenuniv.nl/~sifon/pycorner/bces/}. }} fit to the data is 1.00$\pm$0.04. In addition to the best-fit, the Pearson correlation coefficient ($r$) for the whole data set is 0.95 with $p-$value $\ll$ 0.0001, revealing that the RRL fluxes are very highly correlated. Although some sources show deviations from the primary trend, their S/N ratio tends to be poor. The consistency of the submm-RRL fluxes indicates that there is no extreme density gradient toward the \hii\ regions on angular scales of a few tens of arc-seconds, which would lead to differences in the measured fluxes \citep{dupree1970}. By comparing the linewidths of the various submm-RRLs similarly as we have just done for the fluxes, we also find a moderate correlation ($r = 0.53$ with $p-$value $\ll$ 0.0001). The correlation coefficient is considerable but the uncertainty in the linewidths tends to be significantly higher than for the peak fluxes.

The lower panel of Fig.\,\ref{fig:submm_flux} shows the comparison of peak fluxes of the H28$\alpha$ and H35$\beta$ transitions. As we have seen, the detection rate of the H35$\beta$ line is lower than that of the H28$\alpha$ line (Table\,\ref{tb:detection}) since the H35$\beta$ lines are about six times fainter than the H28$\alpha$ lines. The slope from BCES bisector fitting is 0.95 and the correlation coefficient ($r$) of  0.99 with a small $p-$value $\ll$ 0.0001 reveals a significant correlation between both transitions. If maser excitation contributed to the observed fluxes, a low ratio of H35$\beta$/H28$\alpha$ fluxes is expected to be quite similar to the ratio measured in the submillimeter wavelength regime toward MWC349 (H35$\beta$/H28$\alpha$ $\sim$ 0.078). The peak flux ratios of H28$\alpha$ and H35$\beta$ in the lower panel of Fig.\,\ref{fig:submm_flux} are, however, consistent with the predicted LTE value. The average value of H35$\beta$/H28$\alpha$ toward each source, 0.31, is only slightly above the LTE prediction value of 0.28 \citep{dupree1970,thum1995} but close to the value of 0.32 measured from typical \hii\ regions in the Galaxy \citep{thum1995}.

In our previous analysis of the mm-RRLs (\citealt{kim2017}), we identified two sources in which the observed fluxes could have been the result of maser emission; these were AGAL034.258+00.154 and AGAL043.166+00.011. They were identified by their low H$\beta$/H$\alpha$ ratios, which is suggestive of non-LTE conditions due to an increase in the H$\alpha$ intensity from stimulated maser amplification (as seen in MWC349A; \citealt{thum1995}).  The submm-RRL profiles for these two sources reveal no features that would indicate the presence of  maser emission similar to those seen towards MWC349A at submillimeter wavelengths and their profiles are consistent with pure thermal emission. As discussed in the previous paragraph, the ratio of H35$\beta$/H28$\alpha$ of AGAL043.166+00.011 (cyan circle) shown in the lower panel of Fig.\,\ref{fig:submm_flux} does not show a significant offset from the LTE value. In the case of AGAL034.258+00.154, it was impossible to determine this ratio due to contamination of the H35$\beta$ line. The H$n\alpha$ line profiles of the source, however, already showed no maser emission. We can, therefore, rule out, based on the new submillimeter data, that AGAL034.258+00.154 and AGAL043.166+00.011 are associated with RRL maser emission.

\subsection{Properties of submm-RRLs in comparison to mm-RRLs}

\begin{table}
\centering
\caption{\label{tb:linewidth_ratio} Ratios of the H$n\alpha$ linewidth to the H42$\alpha$ linewidth.}
\begin{tabular}{c c c c }
\hline \hline
   & \multicolumn{3}{c}{Linewidth ratio} \\\cline{2-4}
Submm-RRL  & Median ($\sigma$) & Maximum & Minimum \\
transition & (\kms) & (\kms)  & (\kms) \\
 \hline
H25$\alpha$/H42$\alpha$ & 1.0 (0.08) & 1.2 & 1.0\\
H26$\alpha$/H42$\alpha$ & 1.0 (0.09) & 1.1 & 0.8 \\
H27$\alpha$/H42$\alpha$ & 0.9 (0.28) & 1.5 & 0.8\\
H28$\alpha$/H42$\alpha$ & 1.0 (0.13) & 1.5 & 0.8\\
H30$\alpha$/H42$\alpha$ & 1.0 (0.05) & 1.1 & 0.9\\
H39$\alpha$/H42$\alpha$ & 1.0 (0.07) & 1.2 & 0.9\\
H40$\alpha$/H42$\alpha$ & 1.0 (0.09) & 1.3 & 0.9\\
H41$\alpha$/H42$\alpha$ & 1.0 (0.07) & 1.2 & 0.8\\
\hline
\end{tabular}
\tablefoot{The H$n\alpha$ means the staked RRL of the detected submm-RRL transitions for an individual source. It only contains sources with  peak fluxes greater than 8\,$\sigma$. In the case of the H29$\alpha$, there are no sources that have both H29$\alpha$ and H42$\alpha$ detections $\geq$\,8\,$\sigma$. }
\end{table}

Table\,\ref{tb:linewidth_ratio} compares the linewidth ratios of detected H$n\alpha$ lines (i.e., n = 25, 26, 27, 28, 30, 39, 40, 41) to the H42$\alpha$ line (in this table we only consider detections with a S/N ratio $>$\,8\,$\sigma$). The low quantum numbers correspond to higher frequencies. Here we chose the H$42\alpha$ transition as the main transition to investigate intrinsic linewidths caused by non-thermal motions.

The median ratios are very close to unity for the whole range of transitions (H41$\alpha$ to H25$\alpha$) and are all within one standard deviation. The similarity in linewidth implies that the different transitions are probing the same volume of gas under the same physical conditions. We note that there are some relative deviations from unity (see maximum and minimum ratios in Table\,\ref{tb:linewidth_ratio}). However, only a few particular sources seem to be affected, which might be related to micro-turbulence in the inner \hii\ regions \citep{jaffe1999}.

Since the linewidths and intensities are comparable in the whole range of submm-RRLs, it is feasible to stack the transitions of adjacent quantum numbers to improve their S/N ratio towards each source. However, even by stacking the RRLs we still found some sources with poor S/N levels. We therefore plotted the uncertainties in the linewidth and peak intensity to determine reliable thresholds that can be used to select high S/N sources. From the distribution of the uncertainties, we chose the S/N ratio threshold to be larger than 8\,$\sigma$ for both the linewidths and peak fluxes of mm/submm-RRLs to allow for a reliable comparison with the stacked 3mm data.

\begin{figure}
\centering
\includegraphics[width = 0.49\textwidth]{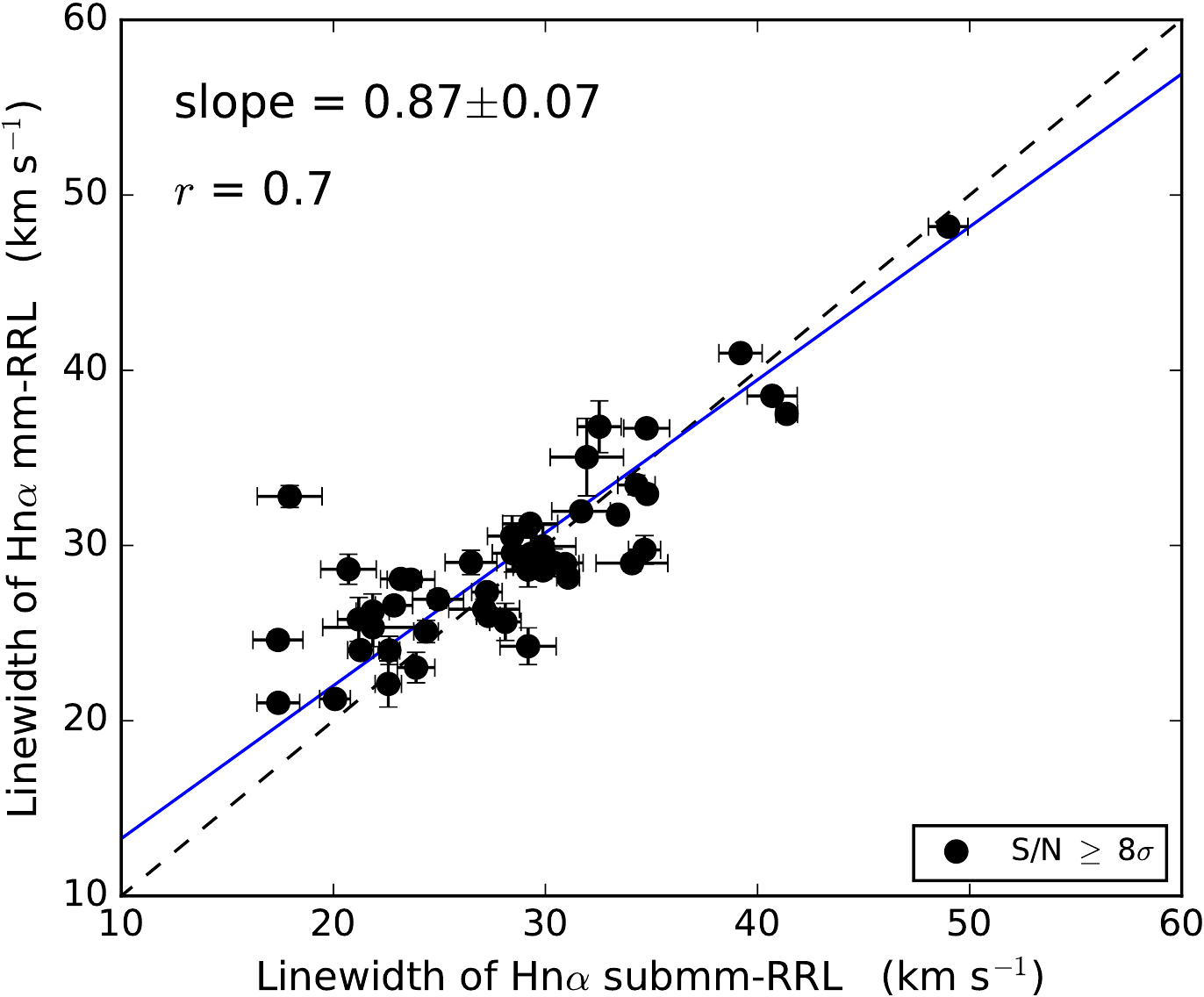}
\includegraphics[width = 0.49\textwidth]{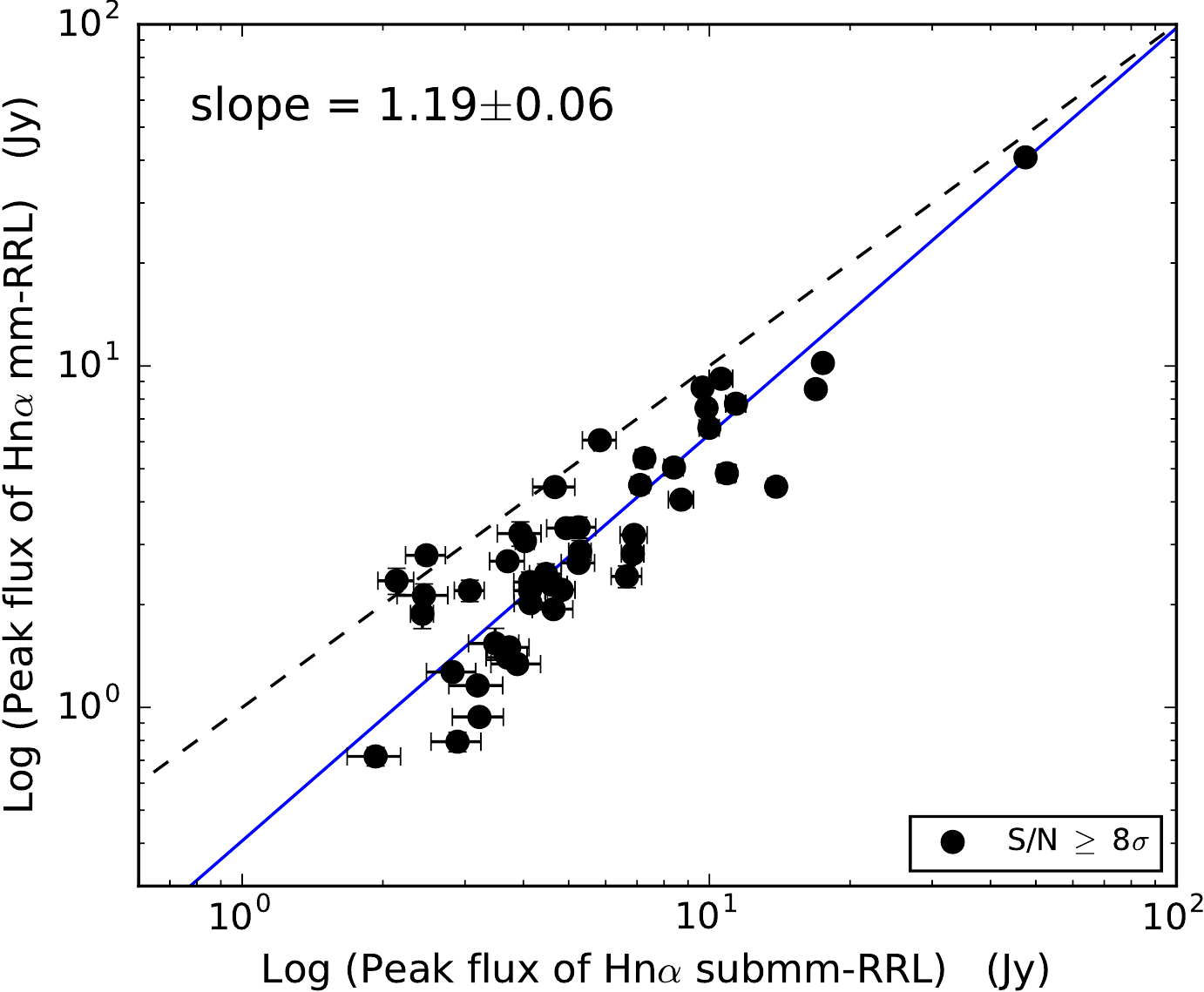}
\caption{\label{fig:submm_mm} Upper: Linewidth comparison of stacked mm- and submm-RRLs. The black filled circles indicate sources with a peak flux with S/N\,$\geq$\,8\,$\sigma$ of both RRLs. Lower: Peak flux comparison of stacked mm- and submm-RRL. The black symbols are the same as in the upper panel. The black dashed lines indicate the locus of equal linewidths and fluxes. The blue lines show the best fit from BCES bisector fitting. }
\end{figure}

Figure\,\ref{fig:submm_mm} compares the linewidth and peak flux for the stacked mm-RRLs and submm-RRLs. In the upper panel of Fig.\,\ref{fig:submm_mm}, the distribution of high-reliability linewidths of mm-RRLs and submm-RRLs reveals a good correlation (Pearson correlation coefficient $r$ = 0.7 with $p-$value $\ll$ 0.0001). A linear least square BCES bisector fit to all of the data points is close to the line of equality. The median value and standard deviation of all the linewidths of mm-RRLs are 29.6\,\kms\ and 5.3\,\kms, and those values for submm-RRLs are 28.7\,\kms\ and 6.2\,\kms. Those standard deviations of mm-RRLs and submm-RRLs are just a factor of two-three more than their velocity resolution ($\sim$2\,\kms). In addition, the sources with a good S/N ratio (black circles) are even more tightly clustered around the line of equality. This linewidth comparison and the result in Table\,\ref{tb:linewidth_ratio} are consistent with the hypothesis that both the mm-RRLs and submm-RRLs are probing the same ionized gas and that we can discard any significant influence of pressure broadening on the fitted parameters, which would have affected the mm-RRLs stronger than the submm-RRLs. In general, the pressure broadening is negligible at (sub)millimeter wavelengths except that toward some HII regions with high electron density the broadening effect still can affect broadening on lines even at millimeter wavelengths.

\begin{table*}
\tiny
\centering
\caption{\label{tb:clump_properties} Summary of properties of molecular clumps and embedded sources associated with RRLs discussed in the text.}
\begin{tabular}{c c c c c c c c c c c c}
\hline \hline
ATLASGAL & $\varv_{\rm sys}$ &1st $\varv$ & 1st $\Delta\varv$ & 2nd $\varv$ & 2nd $\Delta\varv$ & Dist. & $T_{\rm dust}$ & Log $L_{\rm bol}$ & Log $M_{\rm clump}$ & Log $L/M$ & Log Ly \\
name &(\kms) &(\kms) & (\kms) & (\kms) & (\kms) & (kpc) & (K) & ($L_{\odot}$) & ($M_{\odot}$) &  & (photon s$^{-1}$) \\
\hline
AGAL012.804$-$00.199&$+$35.5 &$+$33.4 &28.7 &$+$46.1& 39.4& 2.6&$-$&$-$&$-$&$-$& 48.9 \\
AGAL029.954$-$00.016&$+$97.5 &$+$98.7 &23.7 &$+$85.8&33.1& 5.2&35.5&5.7&3.6&2.1&48.9 \\
AGAL034.258$+$00.154&$+$57.5 &$+$54.8 &30.6 &$+$40.9 &57.6& 1.6&29.2&4.8&3.2&1.6&47.6 \\
AGAL043.164$-$00.029&$+$16.0 &$+$9.1 &33.5 &$+$28.2&46.0&11.1&31.2&6.2&4.5&1.7&49.5 \\
AGAL043.166$+$00.011&$+$12.7 &$+$6.7 &34.8 &$+$37.9&45.8&11.1&33.3&6.9&5.0&1.9&49.3 \\
AGAL045.121$+$00.131&$+$59.0 &$+$62.2 &31.0&$+$58.6 &57.3& 8.0&34.5&6.0&3.9&2.2&49.2 \\
\hline
\end{tabular}
\tablefoot{The full table for all sources is only available at the CDS via anonymous ftp. The line parameters have been obtained with Gaussian fits to the stacked 3mm RRLs. Columns from the third to sixth are only for these selected sources since the linewidths are obtained from two Gaussian components fit in Section\,\ref{sec:interseting}. 
The first $\varv$ and second $\varv$ are the local standard of rest velocities of the peaks of narrow and broad RRL components of these sources. The first $\Delta\varv$ and second $\Delta\varv$ are linewidths of the narrow and broad RRL components. From left to right, other columns are the clump name, the systemic velocity ($\varv_{\rm sys}$) of the dense clump, heliocentric distance (Dist.), dust temperature ($T_{\rm dust}$), bolometric luminosity ($L_{\rm bol}$) of all RMS sources embedded in the ATLASGAL clump, clump mass ($M_{\rm clump}$) , ratio of bolometric luminosity over clump mass ($L_{\rm bol}/M_{\rm clump}$), and Lyman photon flux (Ly) from radio continuum emission at 6\,cm wavelength.}
\tablebib{Properties of the ATLASGAL clumps: \cite{konig2017,urquhart2018_agal_full}. $\varv_{\rm sys}$: \cite{kim2017}. 6\,cm radio continuum emission: \cite{becker1994,white2005,urquhart_radio_south,urquhart_radio_north,purcell2013}. }
\end{table*}

In the lower panel of Fig.\,\ref{fig:submm_mm}, we compare the peak fluxes of mm- and submm-RRLs. The linear least-squares BCES bisector fit to all of the data reveals the presence of an offset from the line of equality, with the submm-RRL fluxes being significantly higher. The shift in the distribution from the line of equality is approximately a factor of two from mm (86$-$100\,GHz) to submm (231$-$397\,GHz) wavelengths. 
This shift agrees with the estimated difference in line intensities of Fig.\,3 of \cite{peters2012}, assuming a \hii\ region with an electron density of $n_{e} = 5\times10^{5}$\,cm$^{-3}$ and a temperature of $T_{e} = 10^{4}$\,K. This difference is due to the increase in the line to continuum ratio with increasing frequency expected for RRLs. Furthermore, the RRL intensity computation by \cite{peters2012} only considers the line emissivity and does not include stimulated emissions. The good agreement we have found with the theoretical predictions is consistent with the detected mm-RRLs and submm-RRLs being emitted from optically thin \hii\ regions. Nevertheless, there are some sources where the fluxes deviate significantly from the observed trend possibly indicating the presence of some \hii\ regions with different physical conditions.

\section{Association with molecular clumps}\label{sec:hii_mol}

Comparing the peak velocities of the ionized gas determined from the submm- and mm-RRLs and those of the molecular gas determined from the H$^{13}$CO$^{+}$ (1$-$0) transition \citep{kim2017}, we find peak velocity differences ($\varv_{\rm H^{13}CO^{+}} - \varv_{\rm RRL}$) of them to be in good agreement with a standard deviation of 4.8\,\kms. This is consistent with the hypothesis that the RRLs are associated with young \hii\ regions that are either embedded or still associated with their natal clouds. Thus, the properties of these \hii\ regions may be related to the properties of their molecular clumps. We have, therefore, tried to identify correlations between the ionized and molecular gas using properties of radio emission from the Co-Ordinated Radio 'N' Infrared Survey for High-mass star formation (CORNISH) \citep{hoare2012,purcell2013} and Red Midcourse Space Experiment (MSX) (\citealt{lumsden2013}) 5\,GHz radio continuum surveys \citep{urquhart_radio_south,urquhart_radio_north} and the properties of dust clumps determined from submm-dust emission of the ATLASGAL survey \citep{schuller2009_full,konig2017,urquhart2018_agal_full}.

\begin{figure}
\centering
\includegraphics[width = 0.50\textwidth]{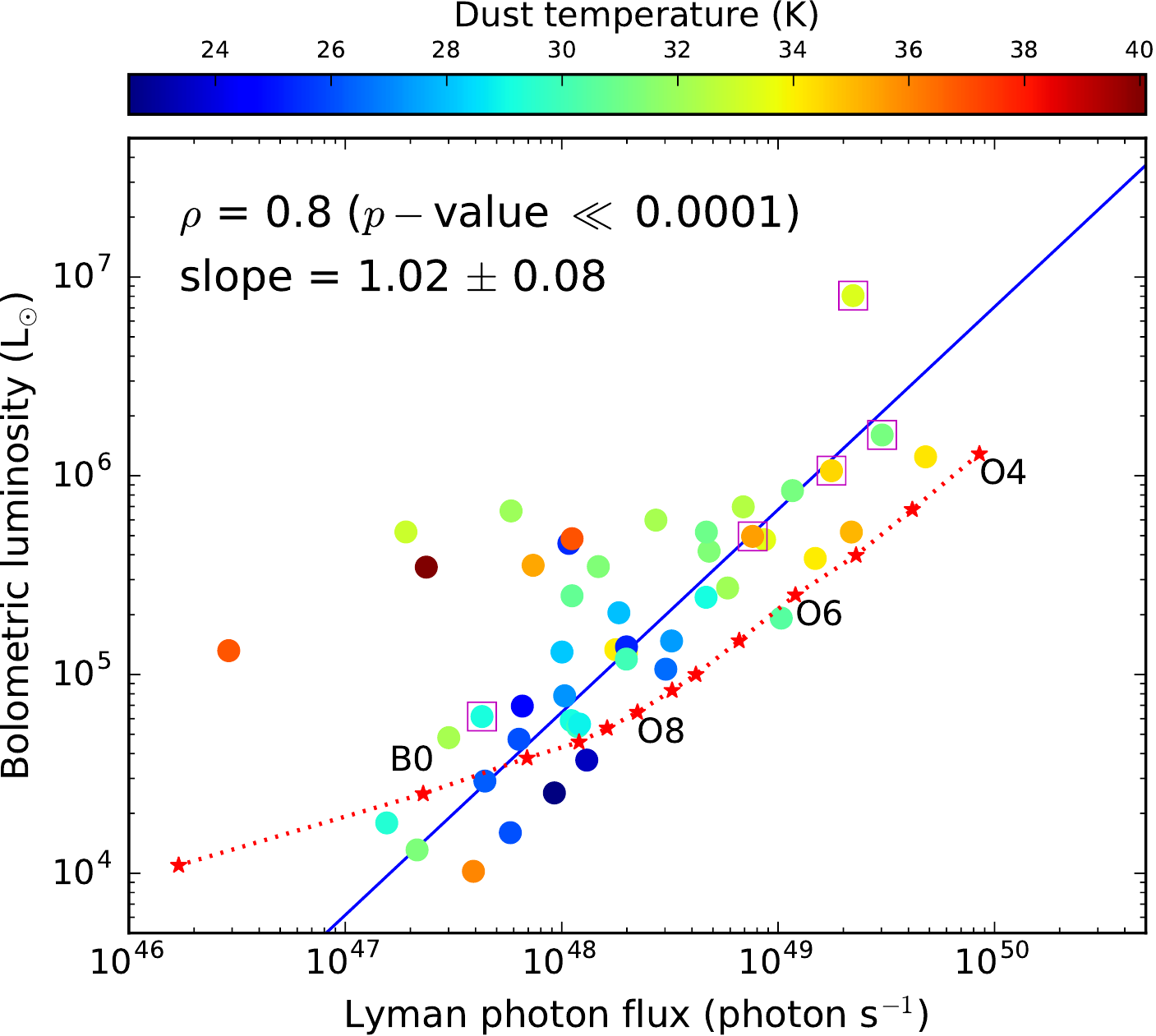}
\caption{\label{fig:lyman} Bolometric luminosity ($L_{\rm bol}$) as a function of Lyman continuum photon flux of 6\,cm radio continuum emission integrated over the beam of the 3\,mm observations. The colors indicate dust temperatures in the molecular clumps. The blue line shows the result of a fit to the data points using BCES bisector fitting. The purple squares superposed on the circles are particular sources with two Gaussian component RRLs (see Table\,\ref{tb:clump_properties} and Sect.\,\ref{sec:interseting}). The red dotted line and red stars indicate bolometric luminosity and Lyman continuum photon flux corresponding to each spectral type (O4\,$-$\,B0.5) for ZAMS stars from \cite{panagia1973}. }
\end{figure}

In Table.\,\ref{tb:clump_properties}, we provide a summary of the properties of molecular clumps and \hii\ regions toward a small sample of selected sources (the full table is available via Centre de Donn\'ees astronomiques de Strasbourg (CDS). The sources selected for this table are those that are associated with non-Gaussian RRL profiles, which are discussed in Sect.\,\ref{sec:interseting}. The peak velocities and linewidths for the first and second Gaussian components were obtained from fits to the stacked mm-RRLs. The distance, dust temperature ($T_{\rm dust}$), bolometric luminosities ($L_{\rm bol}$) of embedded objects, clump mass ($M_{\rm clump}$), and $L_{\rm bol}/M_{\rm clump}$ ratio are taken from \cite{urquhart2018_agal_full} and the Lyman photon flux is determined from the 6\,cm radio continuum sources  correlated with the mm-RRLs \citep{kim2017}.

Figure\,\ref{fig:lyman} shows the relation between the Lyman continuum photon flux emitted by the embedded massive star(s) and the bolometric luminosity generated by the whole cluster of protostellar objects embedded in the molecular clumps. 
The distribution of bolometric luminosities is shifted above the track of the zero-age main sequence (ZAMS) luminosities for the spectral type of a single star \citep{panagia1973}, probably because the bolometric luminosity measured is the contribution of a whole cluster of embedded objects \citep{urquhart2013_cornish}. Only some sources with Lyman photo fluxes equivalent to B-type stars lie below the ZAMS values (red dotted line). Such excess photon fluxes of the B-type stars agree well with the results of several previous studies (e.g., \citealt{sanchez_monge2013,urquhart2013_cornish}). The authors showed that the distributions of young B-type \hii\ regions are consistent with the predicted blackbody with the same radius and temperature as a ZAMS star rather than the prediction of Lyman continuum emission (red dotted line).

\begin{figure}
\centering
\includegraphics[width = 0.50\textwidth]{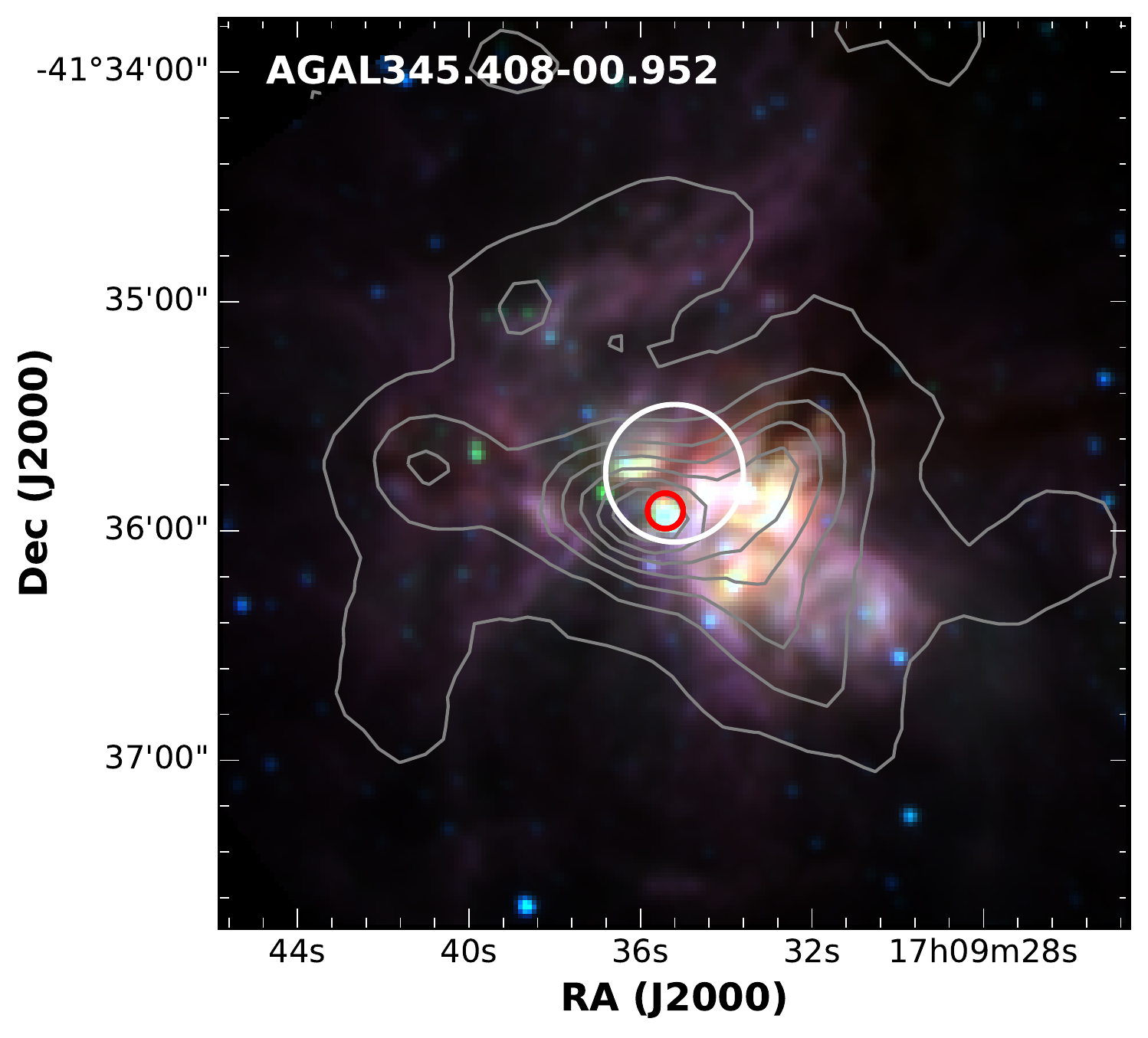}
\caption{\label{fig:outlier} GLIMPSE IRAC three-color composite image (blue; 3.6\,\mum, green; 4.6\,\mum, and red; 8\,\mum) of one (AGAL345.408$-$00.952) of the three outliers in Fig.\,\ref{fig:lyman}. The gray contours represent the 870\,\mum\ dust continuum emission from the ATLASGAL survey. The white circle indicates the size of Mopra beam (FWHM, 36$''$) and the red circle presents a RMS radio continuum source \citep{urquhart_radio_south}.}
\end{figure}

Except for a few outliers, we find a significant correlation between the two parameters (Spearman correlation coefficient excluding the three outliers, $\rho$ = 0.8 with $p-$value $\ll$ 0.0001). The correlation coefficient and the distribution (the BCES bisector fit with a blue line) show a clear correlation, but both bolometric luminosity and Lyman photon flux have a significant distance dependence. We have estimated a partial Spearman correlation ($r_{\rm AB, C}$) test of bolometric luminosity and Lyman photon flux to eliminate their distance dependence using Eq.\,4. of \cite{urquhart2013_methanol}. We obtained a partial Spearman correlation coefficient of 0.5 and Student’s $t-$value of 3.8. These results of the partial Spearman correlation test also show there is a reasonable correlation between the two parameters, rejecting the null hypothesis and being independent of distance. The good correlation between the Lyman flux and the bolometric luminosity would suggest that the luminosity of the clumps is dominated by the most massive stars in the forming protocluster (cf. \citealt{urquhart2013_cornish}). The sources that are best fitted with two Gaussian components, indicated by the purple squares, tend to be the most luminous in the sample except for one (AGAL034.258$+$00.154). By looking at the dust temperature, we find they tend to increase with the bolometric luminosity and the Lyman photon flux increases, although there are some deviations from this trend. This is consistent with the feedback (radiation, outflows, and strong winds) from the embedded \hii\ regions increasing the temperature in surrounding molecular material. The increase in the dust temperature is likely to be linked to the evolutionary status of the central objects and as their luminosities increase so do the temperatures of the natal clumps.

\begin{figure}
\centering
\includegraphics[width = 0.50\textwidth]{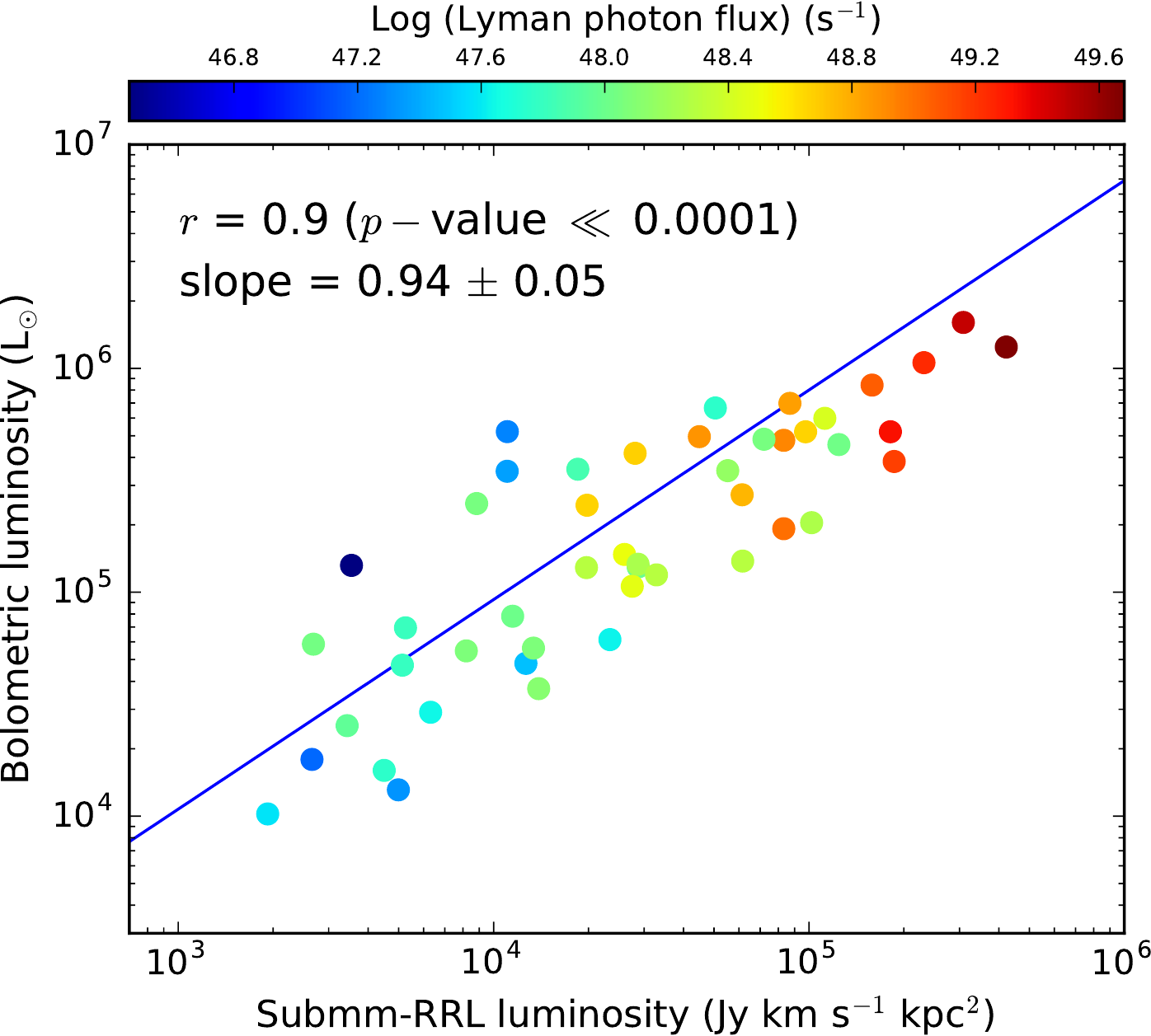}
\caption{\label{fig:submmrrl_bollum_ly} Luminosity of the submm-RRLs versus $L_{\rm bol}$ of embedded central objects. The color bar indicates the Lyman photon flux of ionizing stars associated with the dust clumps and the submm-RRLs. The best linear fit is presented with the blue line. }
\end{figure}

\begin{figure}
\centering
\includegraphics[width = 0.50\textwidth]{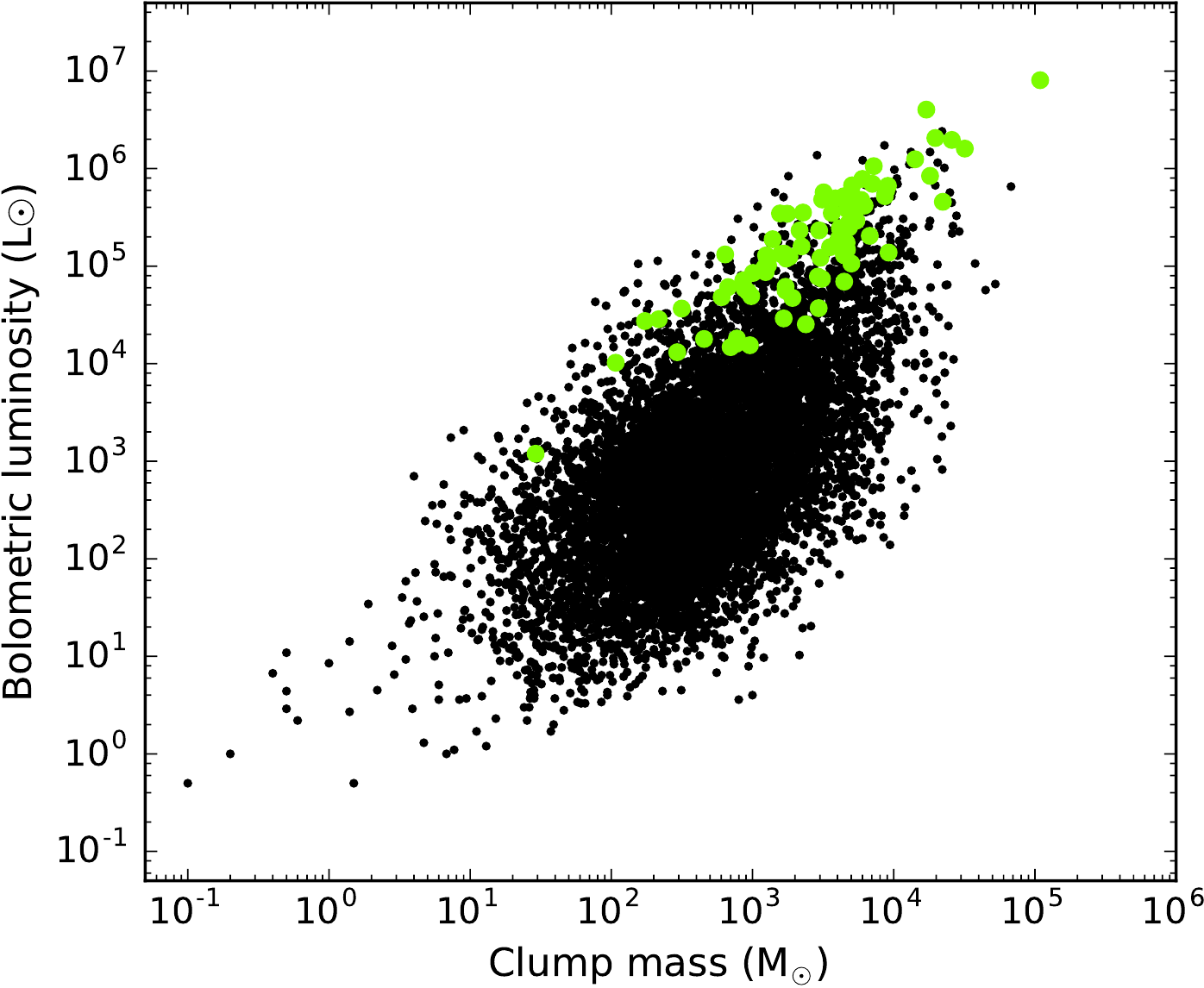}
\caption{\label{fig:mass_bollum} Bolometric luminosity ($L_{\rm bol}$) as a function of $M_{\rm clump}$ toward the full ATLASGAL sample \citep{urquhart2018_agal_full} that indicates both non-detected and unobserved sources by this submm-RRL survey}. The bright green circles indicate sources with a submm-RRL detection.
\end{figure}

In Fig.\,\ref{fig:lyman}, three outliers (AGAL010.168$-$00.362, AGAL331.546$-$00.067, and AGAL345.408$-$00.952) significantly deviate to the left from the linear fit. These all have higher bolometric luminosities compared to the Lyman photon flux than the other sources, and we also note that two have a significantly higher dust temperatures ($T_{\rm dust} \geq$ 36.5\,K; AGAL010.168$-$00.362 and AGAL345.408$-$00.952). The cause of the high dust temperatures toward the outliers is likely to be that sizes of the radio continuum sources are much smaller than the aperture used for the infrared and submm photometry and subsequent spectral energy distribution (SED) analysis (see \citealt{konig2017} for details of the photometry and fitting). Indeed, inspection of the Infrared Array Camera (IRAC) three-color composite images  (3.6\,\mum, 4.6\,\mum, and 8\,\mum) of the Galactic Legacy Infrared Mid-Plane Survey Extraordinaire (GLIMPSE) reveals that all three are associated with extended mid-infrared emission (see Fig.\,\ref{fig:outlier} for example). It is also clear from these images that the bolometric luminosity is arising from an extended star formation complex while the radio emission is associated with only a small part of the complex.

Figure\,\ref{fig:submmrrl_bollum_ly} presents comparisons of three parameters, the luminosity of the submm-RRLs, bolometric luminosity, and Lyman photon flux toward the submm-RRL detected clumps. In addition to the strong correlation seen in Fig.\,\ref{fig:lyman}, the luminosity of the submm-RRLs shows pronounced correlations with the bolometric luminosity and Lyman photon flux. The Pearson ($r$) correlation coefficient gives a high correlation (0.9) with a small $p-$value $\ll$ 0.0001. The three parameters increase altogether. It means that bright and hot stars emit stronger UV radiation and thus increase brightness of submm-RRLs and furthermore bolometric luminosity.

\begin{figure}
\centering
\includegraphics[width = 0.50\textwidth]{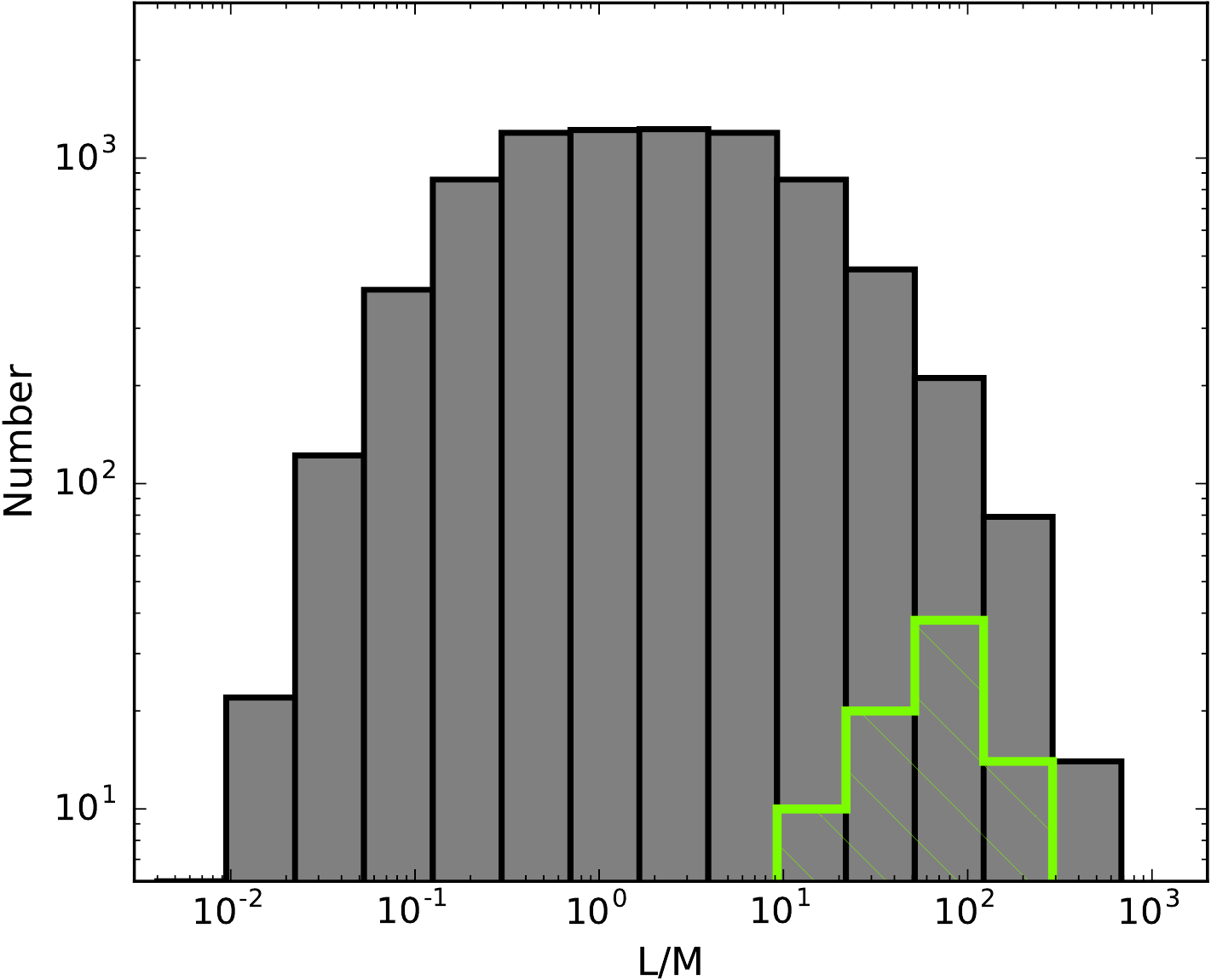}
\caption{\label{fig:l_m_hist} Histogram of $L_{\rm bol}/M_{\rm clump}$ for the full ATLASGAL sources (gray) and the submm-RRL sources (bright green). }
\end{figure}

The good correlations between Lyman photon flux, submm-RRL, and bolometric luminosities and the high dust temperature ($T_{\rm dust}\,>\,$24\,K) suggest that these submm-RRL sources are already quite evolved. We have plotted the clump mass as a function of the bolometric luminosity for all the ATLASGAL clumps \citep{urquhart2018_agal_full} which contain non-detected and unobserved sources (black dots), and submm-RRL sources (green circles) in Fig.\,\ref{fig:mass_bollum}. 
It is evident from this plot that the submm-RRL sources in this survey are a part of the most luminous sources in the whole inner-Galactic plane mapped by ATLASGAL. Figure\,\ref{fig:l_m_hist} shows a histogram of the bolometric luminosity to clump mass ratio ($L_{\rm bol}/M_{\rm clump}$) for the full ATLASGAL sample with the distribution of the submm-RRL associated clumps over-plotted in green. The $L_{\rm bol}/M_{\rm clump}$ ratio is often used as a diagnostic for evolution (e.g., \citealt{eden2013}) and shows that the submm-RRL sources are some of the most evolved young stellar objects in the Galaxy. The mm- and submm-RRLs analysis presented here and in \citet{kim2017} have, therefore, parameterized a sample of some of the most luminous and evolved compact objects in the Galaxy.

\section{Photoionizing photon production rate, $Q$}\label{sec:q_subrrl}

If we assume the detected submm-RRL lines probe most of the photoionizing stars in clumps within the beam of APEX, it is possible to measure the photoionizing photon production rate, $Q$. In principle, Lyman photon flux and $Q$ measure the same quantity: the number of emitted photons per second. 
The photoionizing photon production rate, $Q,$ can be obtained from the submm-RRL flux using Eq.\,(\ref{eq:q_value}) given by \cite{scoville2013} and \cite{bendo2017},
\begin{equation}\label{eq:q_value}
\begin{aligned}
& \frac{Q({\rm H}n\alpha)}{\rm s^{-1}} = 3.99 \times 10^{24} \left(\frac{\alpha_{B}}{\rm cm^{3}\,s^{-1}}\right) \\
&~~~~~~~~~~~~~~~~\times \left( \frac{\epsilon_{\nu}}{\rm erg\,s^{-1}\,cm^{-3}} \right)^{-1}\left( \frac{\nu}{\rm GHz}\right)\left(\frac{D}{\rm kpc}\right)^{2}\left(\frac{\int F_{\nu}\,d\varv}{\rm Jy\,\kms} \right),
\end{aligned}
\end{equation}

\noindent where $\alpha_{B}$ is the total recombination coefficient of the sum over captures to all levels above the ground level in Case B \citep{osterbrock1989}.  The effective emissivity, $\epsilon_{\nu}$, is at a frequency $\nu$ of observed RRL transition. These $\alpha_{B}$ and $\epsilon_{\nu}$ parameters vary with the electron density and temperature.   

\begin{figure}
\centering
\includegraphics[width= 0.5\textwidth]{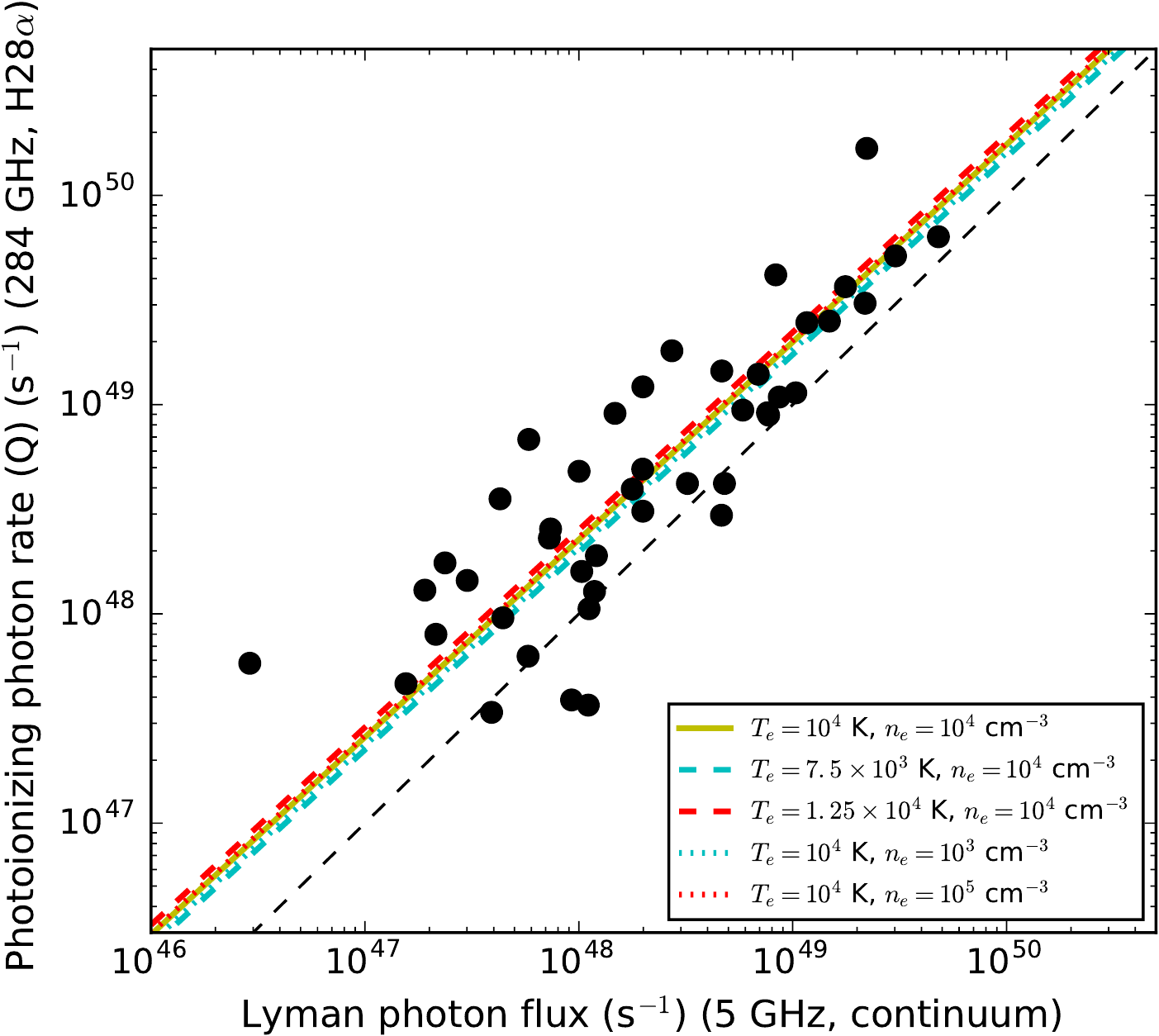}
\caption{\label{fig:lyman_q_subrrl} Photoionzing photon rate ($Q$) measured by the submm-RRLs (this paper) as a function of Lyman photon flux estimated by 5\,GHz radio continuum emission from CORNISH and RMS surveys \citep{purcell2013,urquhart_radio_south}. Those two parameters have the same unit (s$^{-1}$), the number of photons per time. The equality of both axes is indicated with a black dashed line. The color dotted and dashed lines present the best linear fits to the data points calculated for different electron temperature and density in Eq.\,(\ref{eq:q_value}). The black data points are estimated with $T_{e}$ = $10^{4}$\,K and $n_{e}$ = 10$^{4}$\,cm$^{-3}$.}
\end{figure}

We estimated the $Q$ value for the H28$\alpha$ data since this transition is detected toward the majority of the observed sources. The $\alpha_{B}$ and $\epsilon_{\nu}$ for our calculations are taken from the published values by \cite{storey1995}. Figure\,\ref{fig:lyman_q_subrrl} shows the relationship (black circles) between the Lyman photon flux measured by 5\,GHz radio continuum emission and the estimated $Q({{\rm H28}\alpha})$ using the $\alpha_{B}$ and $\epsilon_{\nu}$ based on the chosen $T_{e}$ = $10^{4}$\,K and $n_{e}$ = 10$^{4}$\,cm$^{-3}$. The cyan and red dashed and dotted lines present variations of the fits resulting from $Q$ values computed using different electron densities and temperatures ($T_{e}$ = $10^{4}$\,K and $n_{e}$ = 10$^{4}$\,cm$^{-3}$). The variations are not significant in ranges of electron density ($10^{3} - 10^{5}$\,cm$^{-3}$) and temperature ($7.5\times10^{3} - 1.25\times10^{4}$\,K). There is a slight shift between the Lyman photon flux and  the photoionizing rate $Q({\rm H}28\alpha)$ indicating that the single dish recombination line measurement picks up more photoionizing photons than the interferometric continuum measurements, which might resolve out emission. Overall,  these two measurements show good agreement, and this confirms that the H28$\alpha$ lines are thermally excited. Furthermore, it confirms that the $Q$ values measured by mm/submm RRLs are useful to measure star formation rates on extragalactic scales.

\section{Sources of interest}\label{sec:interseting}

\begin{figure*}
\includegraphics[width= 0.33\textwidth, trim= 0.5cm 0.5cm 0.2cm 0.2cm, clip]{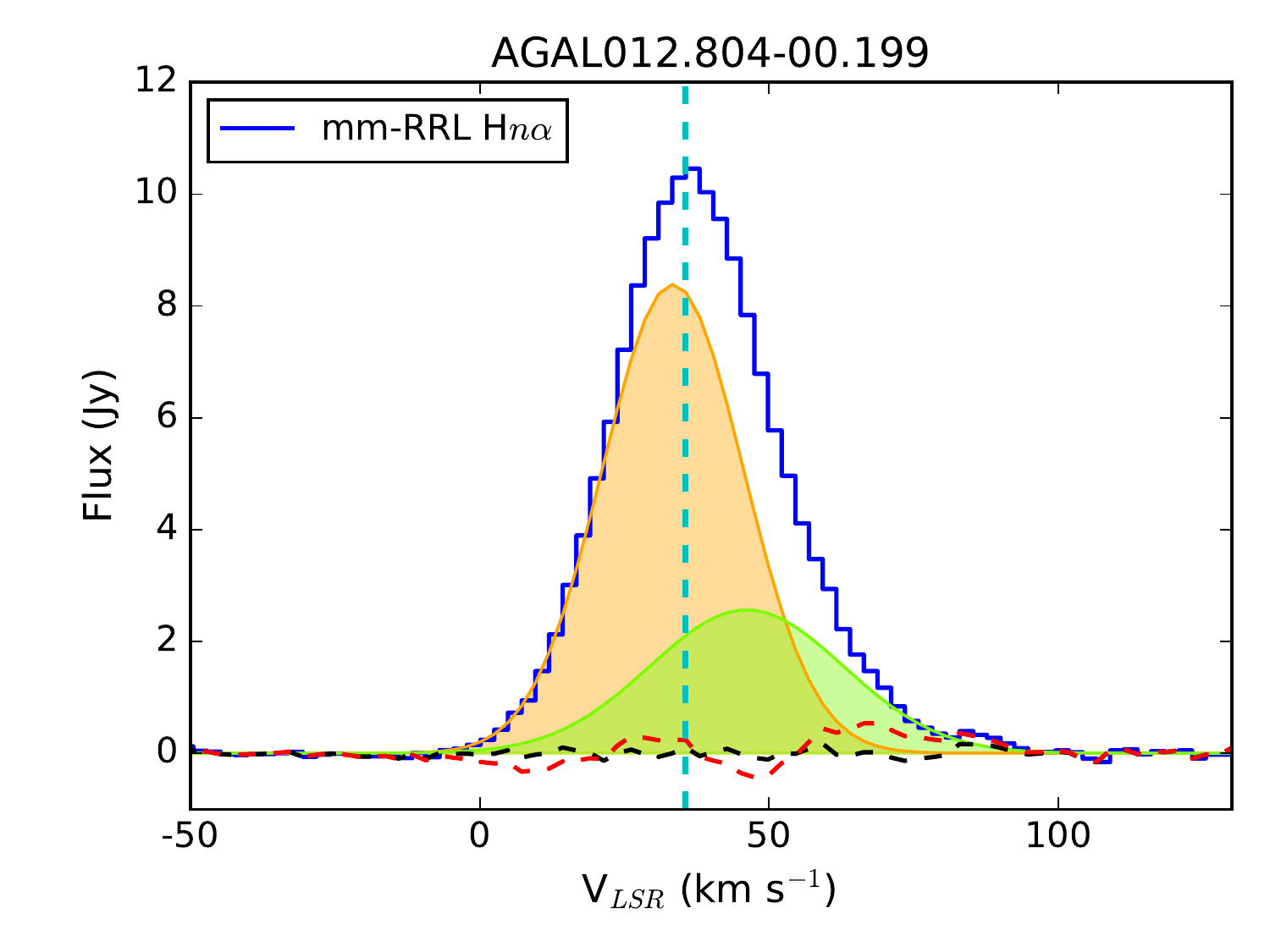}
\includegraphics[width= 0.33\textwidth, trim= 0.5cm 0.5cm 0.2cm 0.2cm, clip]{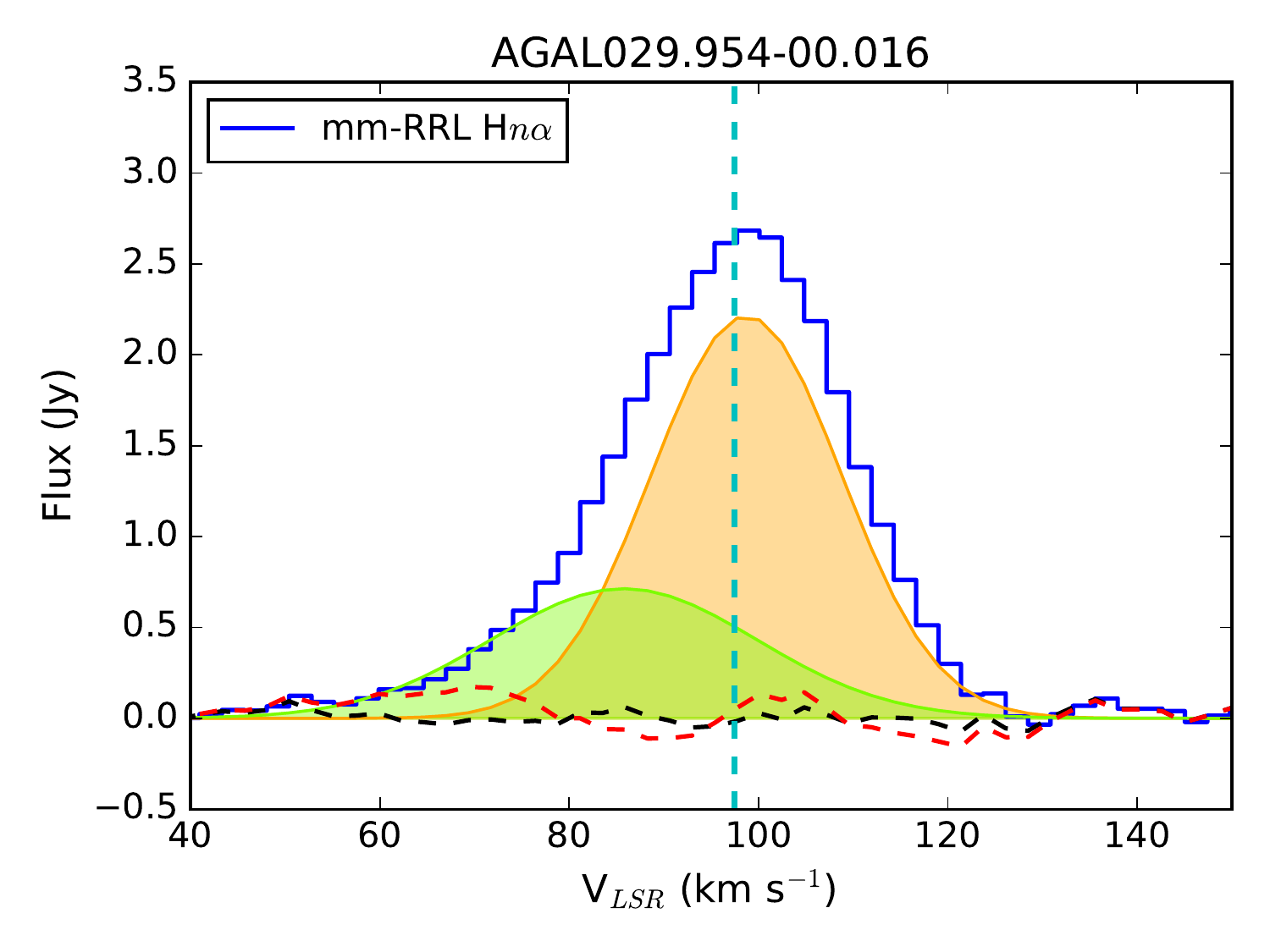}
\includegraphics[width= 0.33\textwidth, trim= 0.5cm 0.5cm 0.2cm 0.2cm, clip]{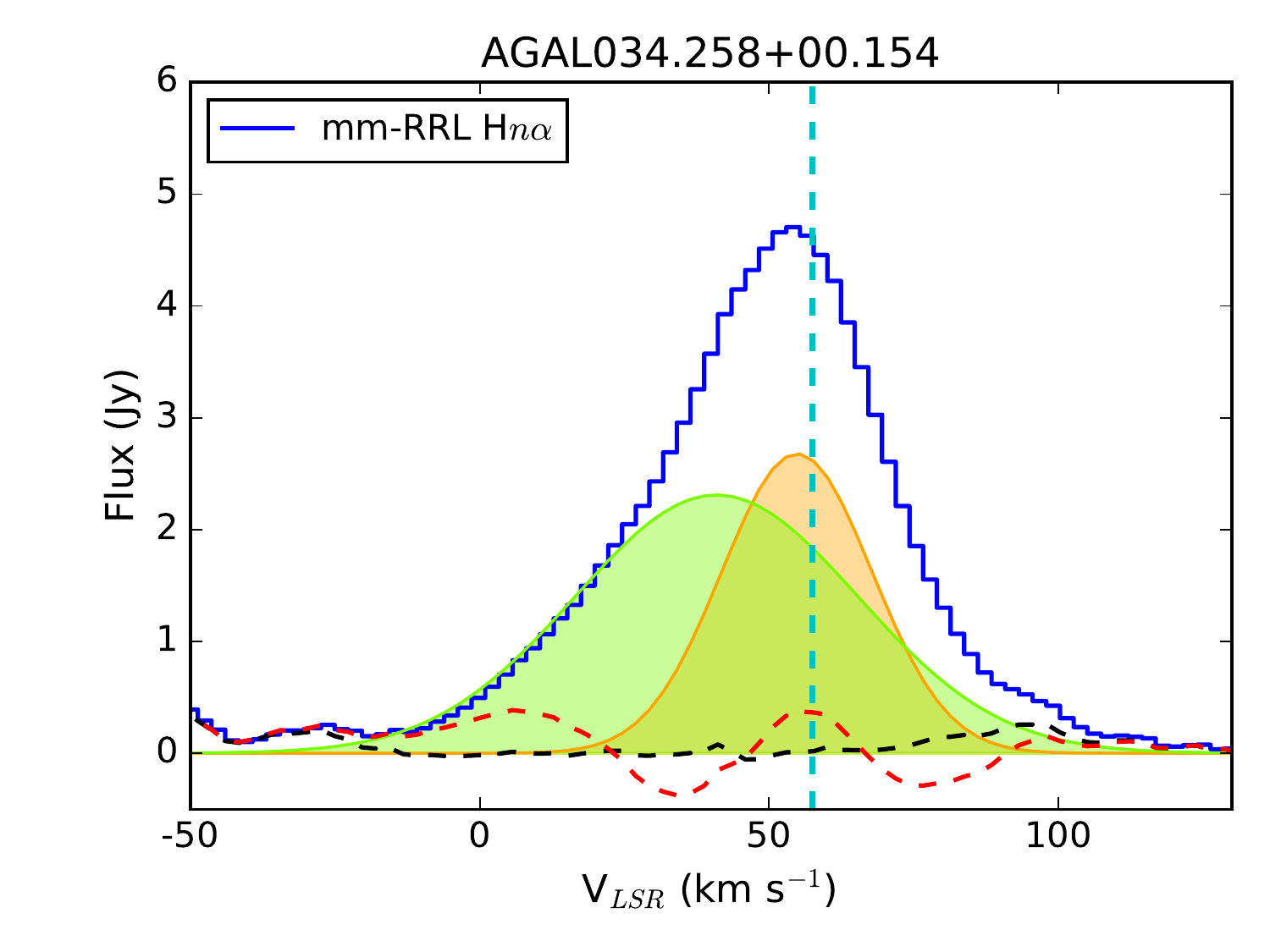}
\includegraphics[width= 0.33\textwidth, trim= 0.5cm 0.5cm 0.2cm 0.2cm, clip]{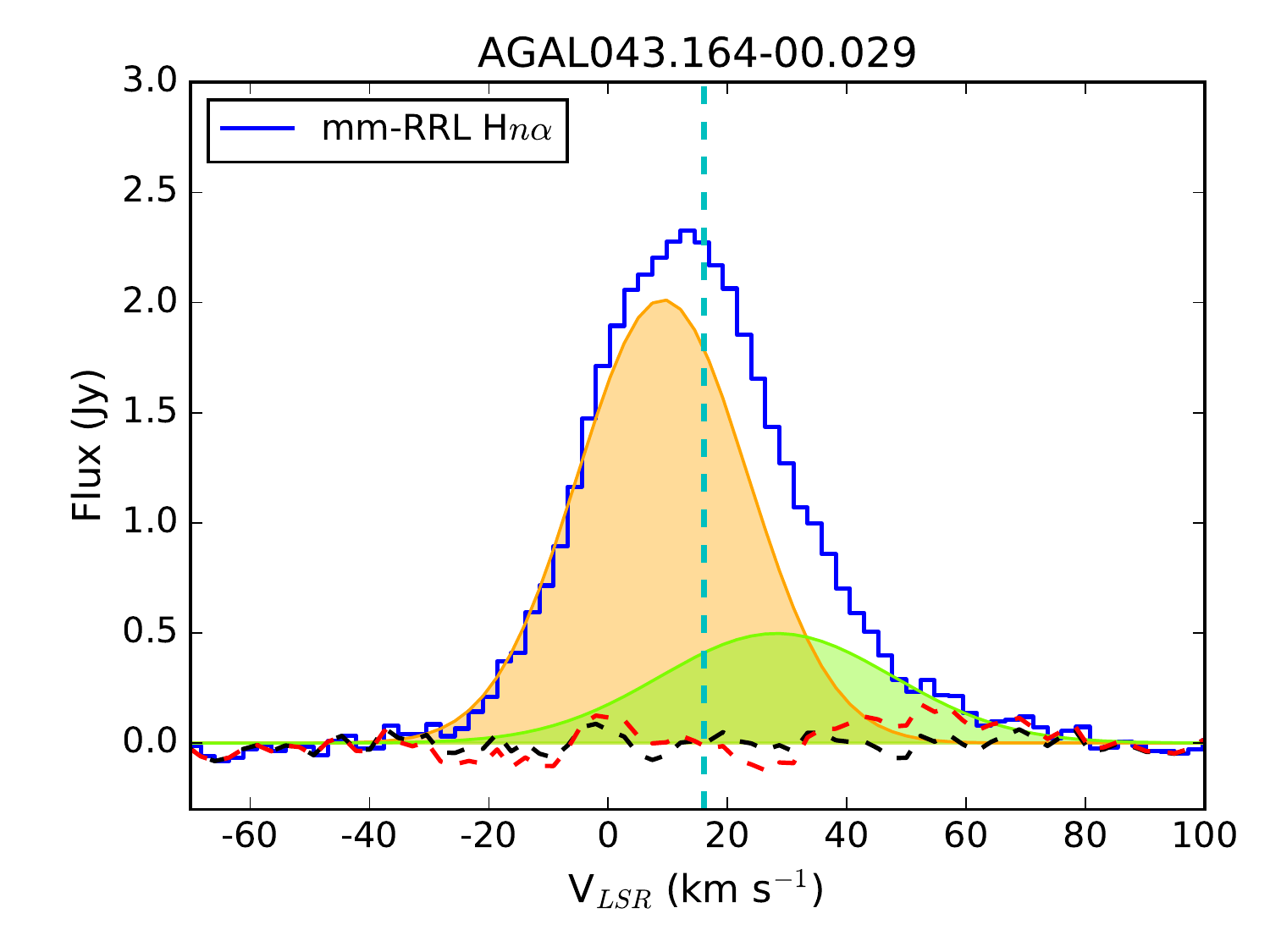}
\includegraphics[width= 0.33\textwidth, trim= 0.5cm 0.5cm 0.2cm 0.2cm, clip]{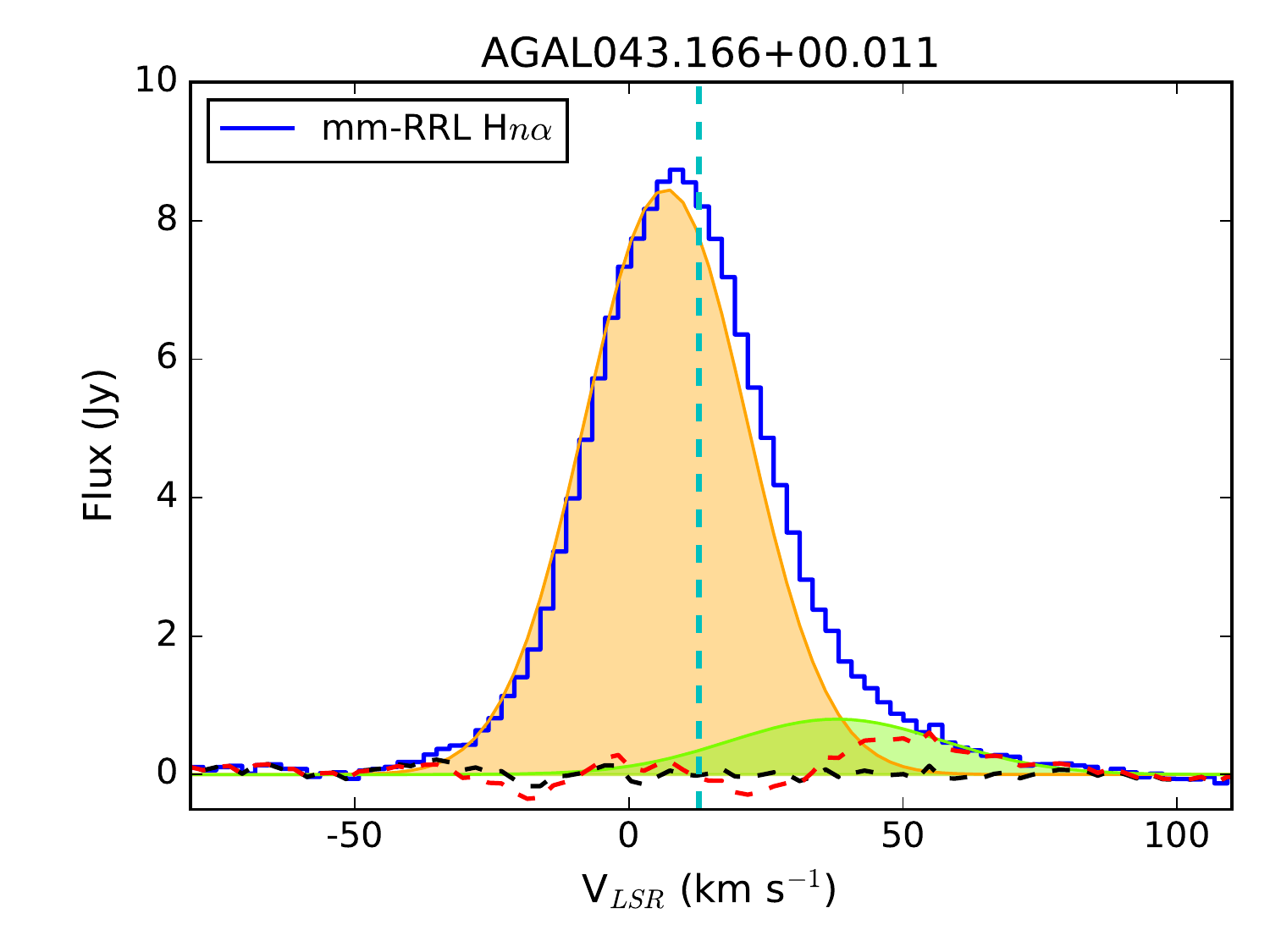}
\includegraphics[width= 0.33\textwidth, trim= 0.5cm 0.5cm 0.2cm 0.2cm, clip]{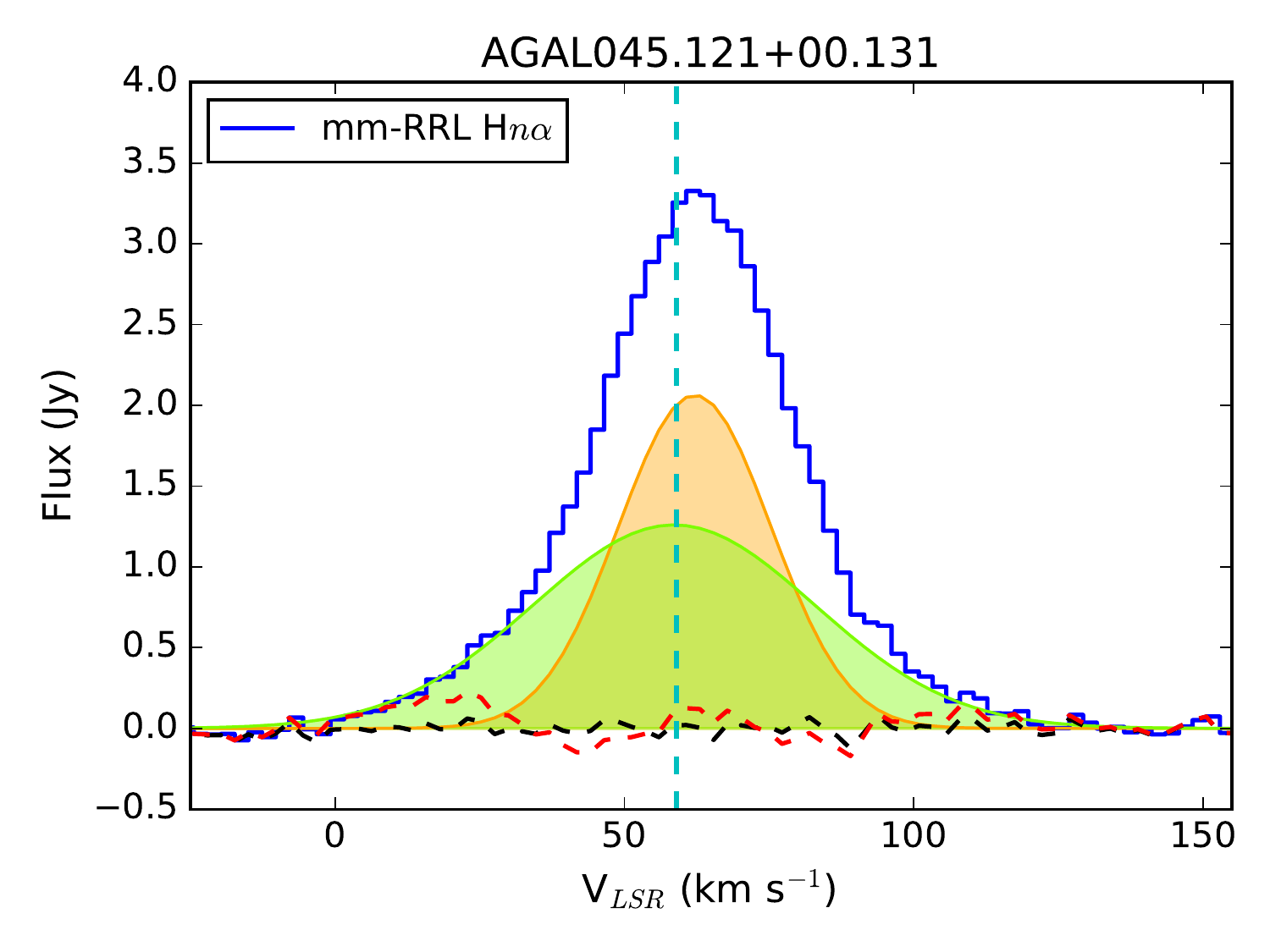}
\caption{\label{fig:multi_gausw} Green and yellow filled areas show two Gaussian fitted profiles of the H$n\alpha$ transition.
The back dashed line is a residual noise signal of the two Gaussian profiles. On the other hand, the red dashed line indicates a residual noise from a one-component Gaussian fit to the data. The vertical cyan dashed line shows the systemic velocity of the clump measured by H$^{13}$CO$^{+}$ (1$-$0).}
\end{figure*}

In Sect.\,2.2 we mentioned six sources where the RRL spectral profile deviates significantly from the expected Gaussian line shape due to the presence of excess emission seen at higher velocities. We show the stacked mm-RRL spectral profiles for these sources in Fig.\,\ref{fig:multi_gausw}. In all of these cases, fitting a single Gaussian profile to the RRL results in large residuals. However, the large residuals can be significantly reduced using a two-component fit to these data. One component is centered at the systemic velocity of the sources and a second, broader component is fitted to the emission from the higher velocity gas. The parameters of the Gaussian fits of the two components are given in Table\,\ref{tb:clump_properties}. In this section, we have a more detailed look at these sources to understand the nature of these additional high-velocity components. Although the non-Gaussian profiles are observed in both the submm- and mm-RRLs, we have chosen to fit the mm-RRLs as they tend to have a higher S/N ratio and provide therefore more reliable results.

\begin{figure*}
\includegraphics[height= 0.3\textwidth]{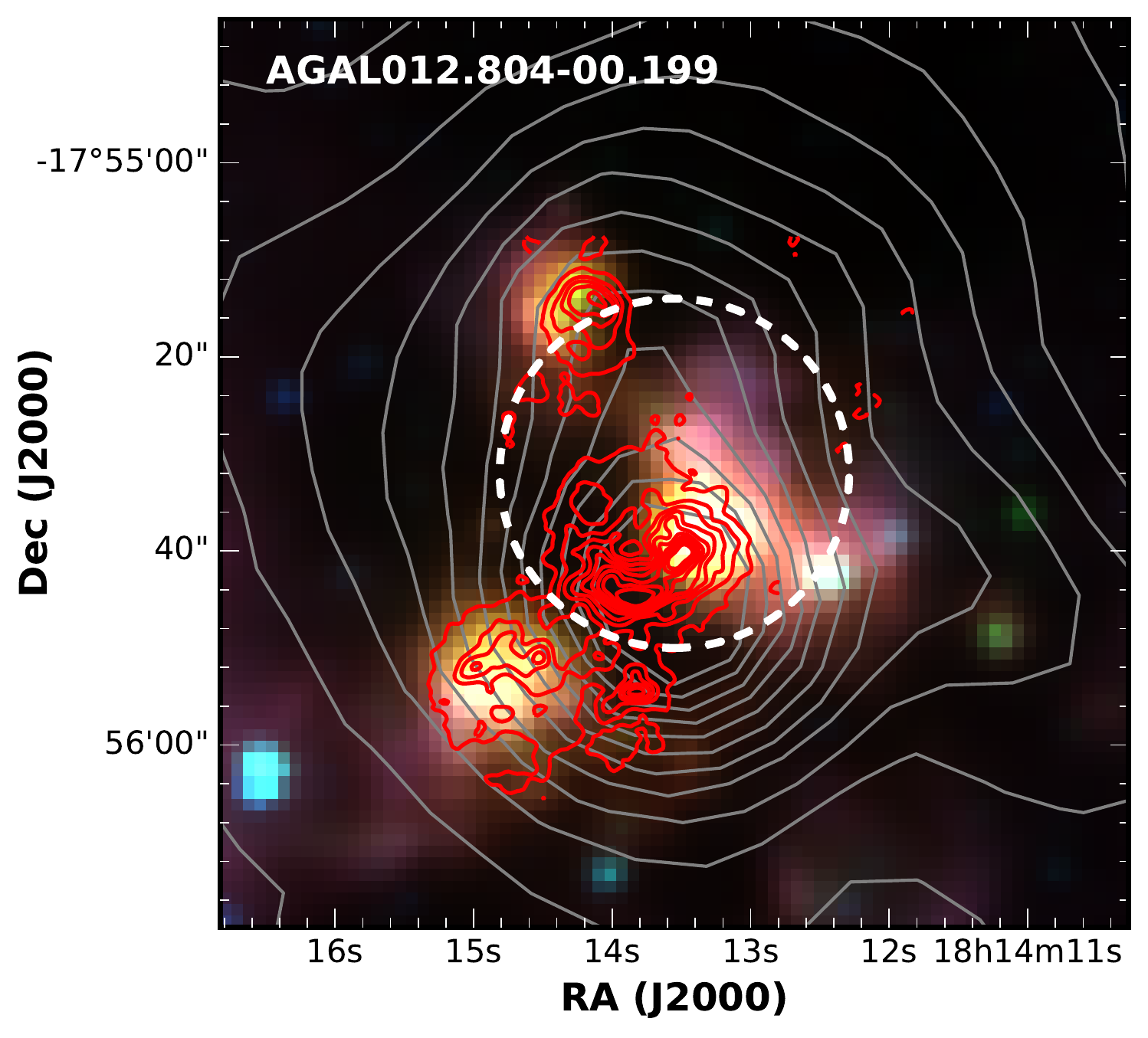}
\includegraphics[height= 0.3\textwidth]{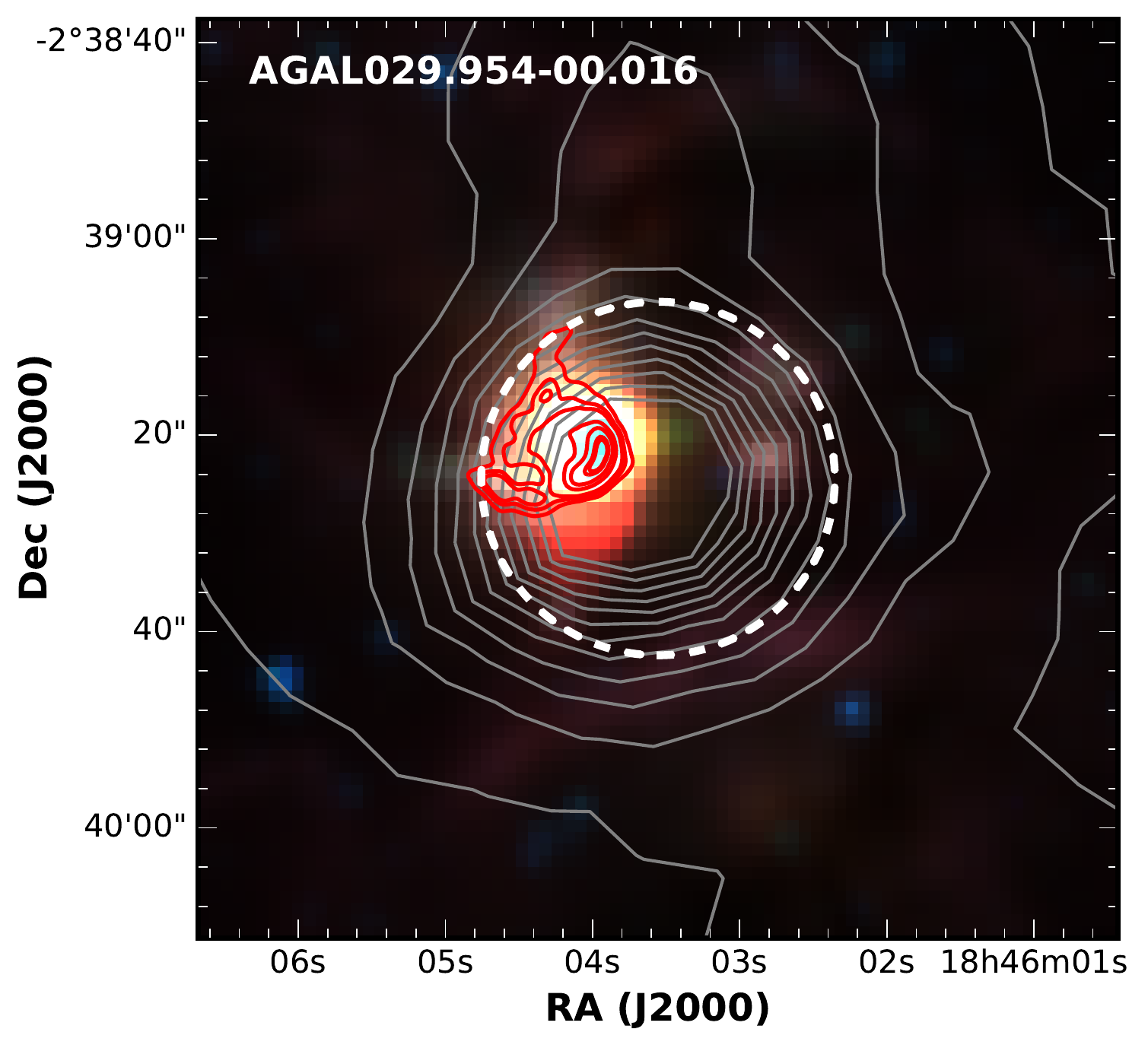}
\includegraphics[height= 0.3\textwidth]{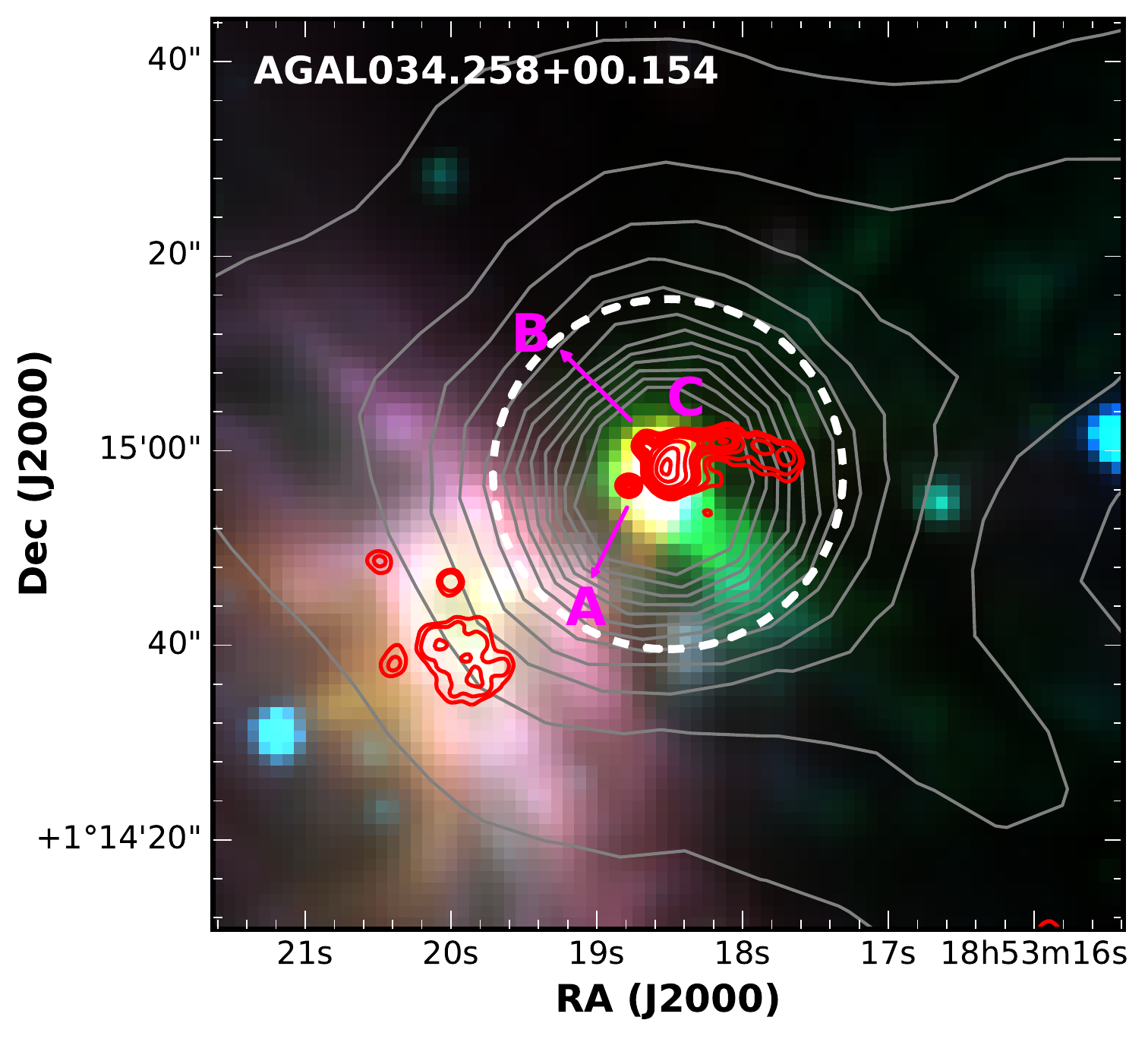}
\includegraphics[height= 0.3\textwidth]{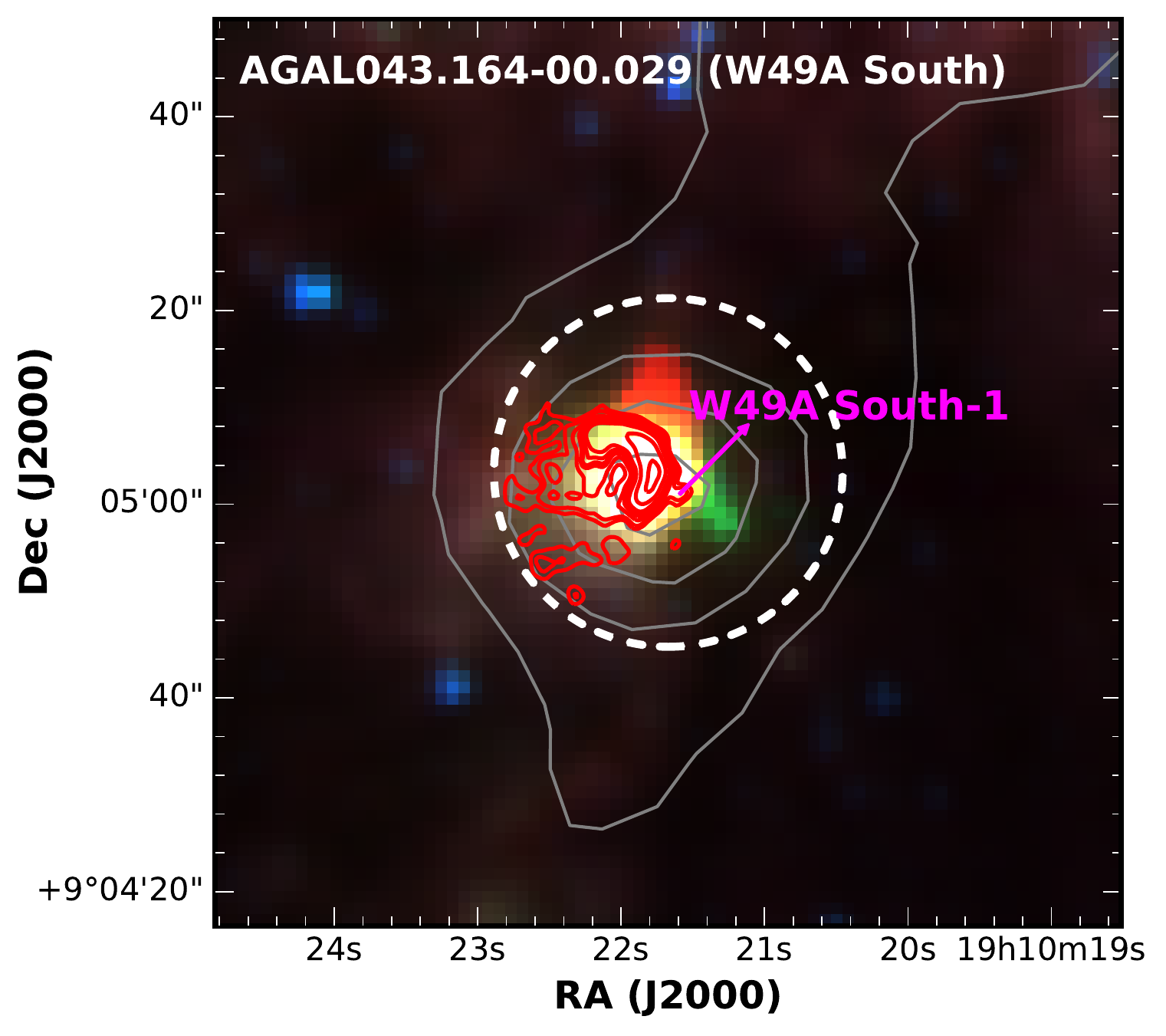}
\includegraphics[height= 0.3\textwidth]{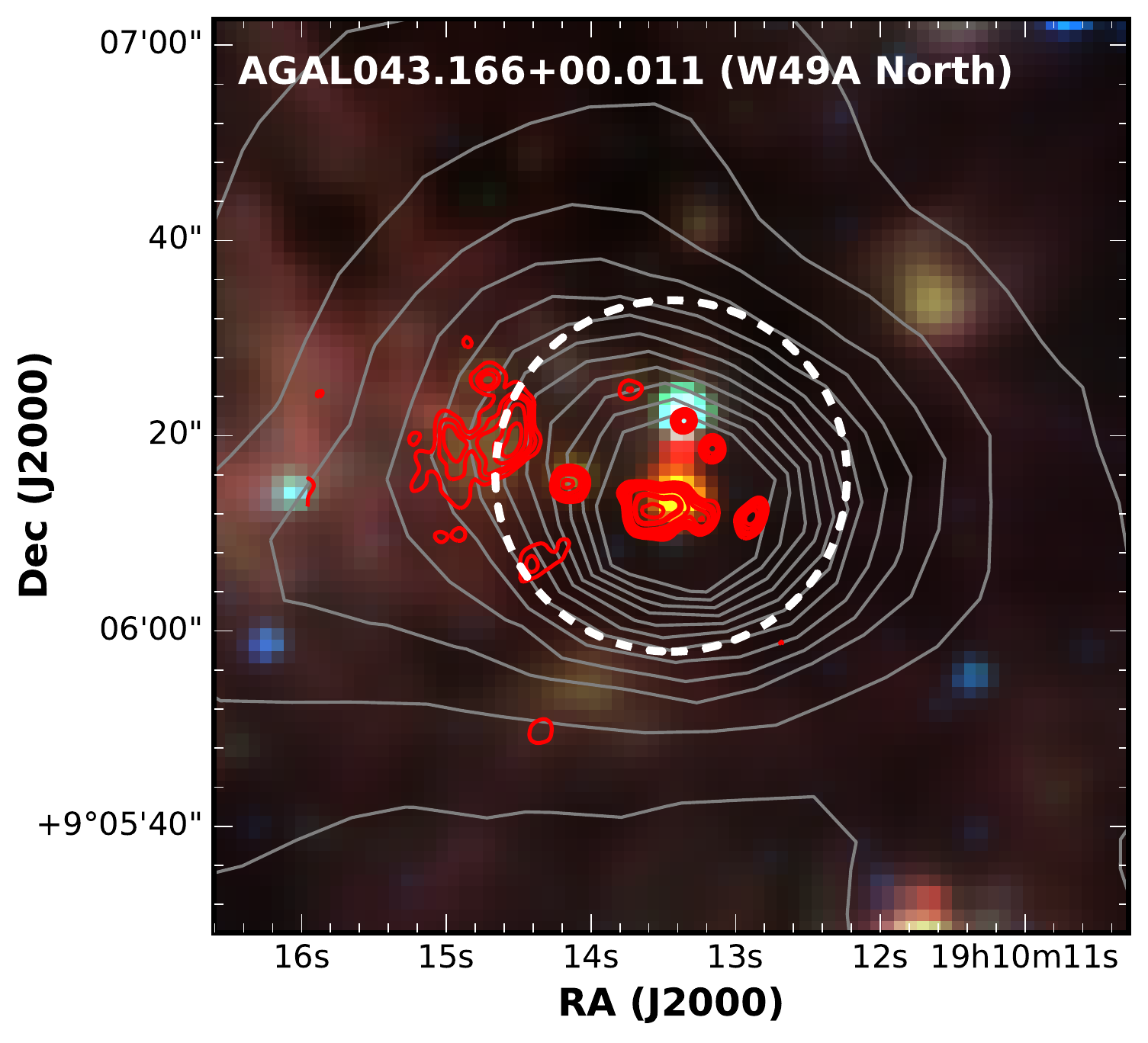}
\includegraphics[height= 0.3\textwidth]{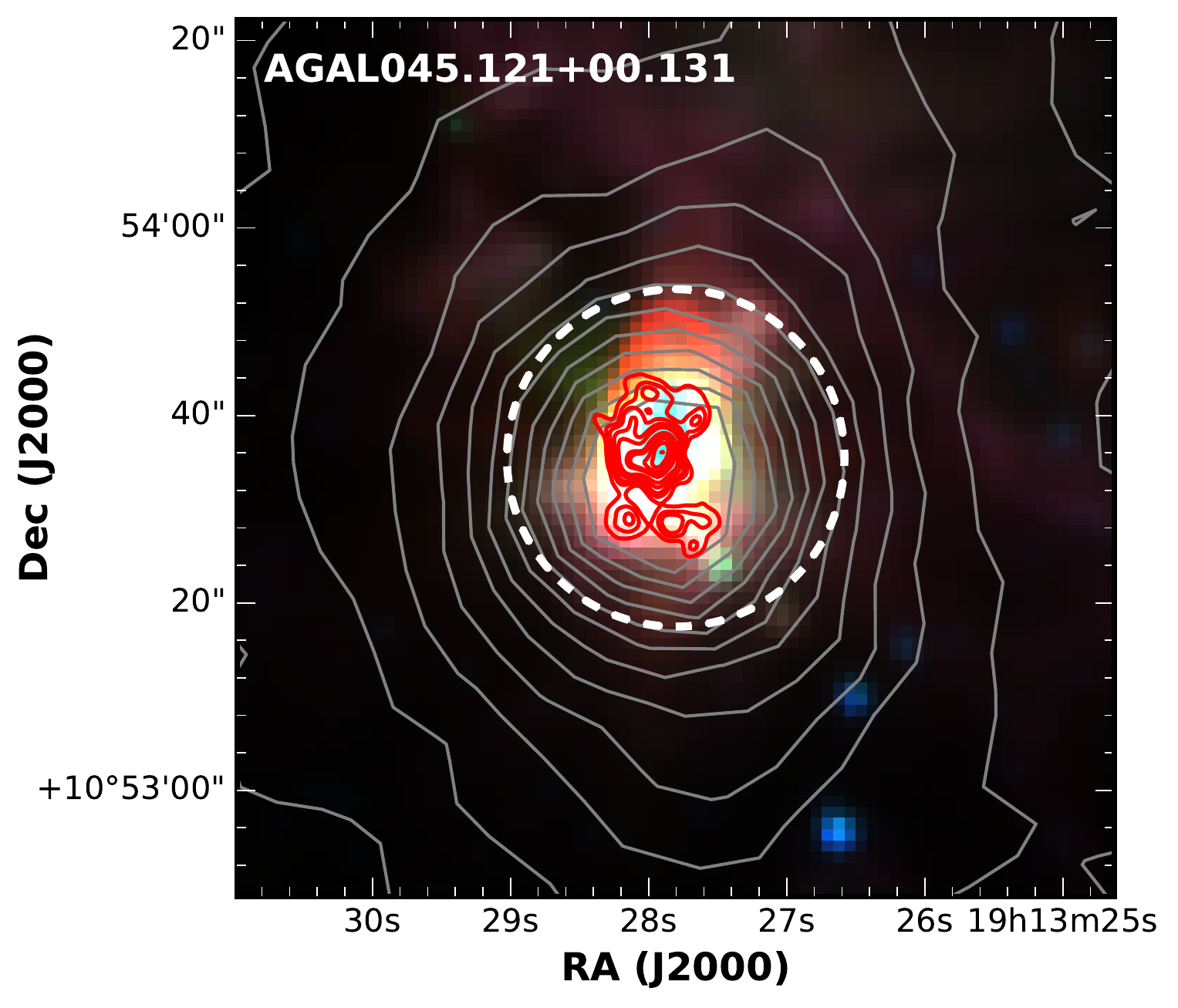}
\caption{\label{fig:multi_mid} GLIMPSE IRAC three-color composite images (blue; 3.6\,\mum, green; 4.6\,\mum, and red; 8\,\mum). The gray contours trace the 870\,\mum\ dust continuum emission from the ATLASGAL survey. The red contours show the  5\,GHz radio continuum emission from the CORNISH survey. The cyan dashed circles are centered on the pointing positions of the IRAM observation with FWHM of 29$''$.}
\end{figure*}

To facilitate the discussion of these sources we complement the spectra presented in Fig.\,\ref{fig:multi_gausw} with three-color mid-infrared images that show the local environment of these regions in Fig.\,\ref{fig:multi_mid}. It is interesting to note that all of the sources where a high velocity component has been identified are located in the first quadrant. However, these sources were all targeted in the observing campaign with the IRAM 30\,m telescope, resulting in higher sensitivity observations than the observations of southern targets with the Mopra 22\,m telescope, which therefore might have missed the weaker, broader components in many other sources. We also note that three of the sources show red-shifted emission, two show blue-shifted emission, and one appears to have both blue and red-shifted components. The higher-velocity gas is likely due to either an expanding flow away from the \hii\ regions \citep{tenorio-tagle1979a,tenorio-tagle1979b} or bow-shock generated by wind-blowing high-mass stars moving at a supersonic velocity \citep{van_buren1990,van_buren1992} or possibly the combination of both phenomena.

\subsection{AGAL012.804$-$00.199}

This \hii\ region is part of the W33 star-forming complex, and so we have adopted a distance of 2.6\,kpc to this clump (\citealt{immer2012a}). The radio emission consists of two distinct sources, a bipolar structure located towards the center of the clump and a more compact, roughly circular source located to the north. All of the radio emission is correlated with bright mid-infrared emission seen in the GLIMPSE image  (see the upper left panel of Fig.\,\ref{fig:multi_mid}). The position of the IRAM beam is slightly offset from the peak of the bipolar radio emission, but the brightest lobe of bipolar structure falls within the FWHM beam size. However, emission from the weaker bipolar lobe and the compact source to the north are also likely to be picked up in the outer part of the beam. 

The mm-RRL spectrum is shown in the upper left panel of Fig.\,\ref{fig:multi_gausw}; this reveals the presence of a red wing, which might be associated with an ionized flow perhaps arising from one of the bipolar lobes. The systemic velocity of the clump is 35.4\,\kms, which is broadly consistent with the velocity of the strongest RRL component and so it is the weaker and broader component that is associated with the faster-moving gas. \cite{bieging1978} also reported different velocity RRL components at low-frequency (1.7, 5, and 8.6\,GHz) using the Effelsberg 100\,m telescope. They suggested that the components with different velocities are associated with flows of ionized gas, which is consistent with our interpretation.

\subsection{AGAL029.954$-$00.016}

This source is associated with W43 South at a distance of 5.2\,kpc (\citealt{reid2014}) and has a systemic velocity of 97.2\,\kms. The IRAM beam is centered on the peak of the dust emission, and the radio emission is slightly offset to the west; however, it is fully covered by the beam. The radio emission has a cometary morphology with a bright head pointing towards the center of the clump and a tail-like structure that extends away from the center of the clump into the lower-column density region. The radio emission is correlated both spatially and morphologically with the strong mid-infrared emission (see the upper middle panel of Fig.\,\ref{fig:multi_mid}). The clump has a mass of $\sim 4 \times 10^3$\,\msun\ and a dust temperature of 35.5\,K and so feedback from the \hii\ region is having a significant impact on the physical properties of its host clump.

The RRL profile is shown in the upper middle panel of Fig.\,\ref{fig:multi_gausw} and reveals the presence of a high-velocity blue-shifted component; the peak of this component is offset by $\sim$15-20\,\kms\ from the systemic velocity of the \hii\ region and the host clump. The high-velocity blue-shifted component corresponds to the velocity range of a Br$\gamma$ line (around 70 $\sim$ 90\,\kms) at the position of the head \citep{martin-hernandez2003}. The linewidth of a Br$\gamma$ toward the head of the \hii\ region is also found to be $\sim$20\,\kms\ broader than that from the tail, but their Br$\gamma$ observations could not explain it with either bow-shock or champagne flow models. In addition, they also showed the existence of a velocity gradient toward the tail \citep{martin-pintado2002}.

\cite{zhu2008}, however, showed that  [Ne\,{\sc ii}] line emission and its velocity maps toward this source well agreed with a model of stellar-wind pressure-driven \hii\ regions with a viewing angle of 135\dd. In this model, the cometary morphology of the \hii\ region is caused by either  relatively slow motion of the star through the molecular cloud or a density gradient in the cloud \citep{zhu2008,arthur2006}. According to the model, emission lines from the ionized gas of a cometary head show a broad linewidth with a negative velocity offset concerning the velocity of molecular gas \citep{zhu2008}. In fact, our broad RRL also has a negative velocity relative to the systemic velocity although it is not clear whether the broad RRL comes from the head of this source. Besides, there is a hot core in front of the \uchii\ region, and HCO$^{+}$ and SiO molecular lines \citep{maxia2001} show the line densities are higher toward the head than the tail. Therefore, our broader RRL component is likely to be associated with accelerated ionized gas toward the tail along the shell created by the relatively slow motion of the star through the the interstellar medium.

\subsection{AGAL034.258$+$00.154}

This ATLASGAL clump is associated with the G34.26+0.15 Complex. The radio emission consists of three radio continuum components; an extended source that has a cometary morphology and two compact regions to the west and north-west; these are referred to as C, A, and B, respectively (following \citealt{sewilo2011}). The clump has a mass of $\sim$ 1500\,\msun\ and dust temperature of $\sim$30\,K. The mid-infrared image (upper right panel of Fig.\,\ref{fig:multi_mid}) reveals that this region is located at the edge of a larger \hii\ region located to the south-west (region of extended mid-infrared emission). Besides, the morphologies of radio and mid-IR emissions are not so well correlated with each other toward this source. The reason is likely to be extinction \citep{hoare2007}.

The RRL profile shows a strong blue-shifted wing and is fitted with two components with similar intensities but different linewidths. The velocity of the narrowest component corresponds to the systemic velocity of the clump, but it is offset by $\sim$15\,\kms\ from the velocity of the broader component. There is a possibility that some of the RRL emission is coming from the larger  \hii\ region located to the south-west given that it falls within the outer part of the IRAM beam. This complex has been observed by \citet{sewilo2011} at higher resolution in continuum and the H53$\alpha$ RRL. Their study separated the emission from the three radio sources and determined the velocity of A and C to be $\sim$50\,\kms\ and that of B to be $\sim$65\,\kms. All of these are broadly consistent with the systemic velocity of the clump and the narrower RRL component. However, we note that source C has a very broad linewidth (49\,\kms) and so may be responsible for the  excess seen in the IRAM RRL spectra. \cite{jaffe1999} also observed 3-mm hydrogen RRLs towards  this source with the IRAM 30m telescope and also reported the presence of a similar board linewidth high-velocity component, which is associated with source C.
   
It is currently unclear from the present data whether the blue-shifted wing seen in the IRAM spectra is the result of contamination from the nearby extended \hii\ region or a combination of the three radio sources identified, however, the latter seems more likely. Furthermore, the cometary shape of source C has been well explained by the bow-shock model created by wind-blowing massive stars moving at a supersonic velocity with respect to the molecular gas \citep{van_buren1990}.

\subsection{W49A: AGAL043.164$-$00.029 and AGAL043.166$+$00.011}

These two \hii\ regions are located in the W49A star-forming complex. It is located at a distance of $\sim$11\,kpc \citep{zhang2013} and is one of the two most active star formation regions in the Galaxy (\citealt{urquhart2014_rms}). The region AGAL043.164$-$00.029 has a clump mass of $3\times 10^4$\,\msun, luminosity of $1.6\times 10^6$\,\lsun\ and systemic velocity of 14.3\,\kms\ while AGAL043.166$+$00.011 has a clump mass of $11\times 10^4$\,\msun, luminosity of $8\times 10^6$\,\lsun\ and systemic velocity of 2.9\,\kms.

\subsubsection{AGAL043.164$-$00.029}

The radio emission associated with AGAL043.164$-$00.029 (lower left panel of Fig.\,\ref{fig:multi_mid}) reveals a cometary morphology that is coincident with  strong mid-infrared emission; this is known as W49A South. The tail of the cometary structure extends to the west away from the peak submillimeter emission. In front of the head of the cometary \hii\ region, there is another very weak and compact \hii\ region; this is referred to as W49A South-1 \citep{de_pree1997}. The clump looks relatively isolated, so all of the RRLs likely arise from the \hii\ regions. The velocity of the strongest RRL component is correlated with the systemic velocity of the clump. The second component is red-shifted with a velocity 12\,\kms\ larger than the systemic velocity. 

This red-shifted component is likely to be associated with high-velocity ionized gas that forms the tail, and that is expanding into the lower-density region away from the head of the cometary structure; the head itself is coincident with the highest column density part of the clump. According to the RRL results of \cite{de_pree1997}, however, the velocity gradient over the cometary source and a broad linewidth toward the head supports the bow-shock model and a mass-loaded stellar wind in a molecular cloud with a density gradient, instead of the champagne flow model \citep{de_pree1997, zhu2008}. Nevertheless, their RRL velocity resolution was too poor to allow for a detailed investigation of the velocity structure of this source.

\subsubsection{AGAL043.166$+$00.011}

The region AGAL043.166$+$00.011 is the most massive and active clump in the Galaxy; it is known as W49 North. It is perhaps the only example of a young massive protocluster outside of the Galactic center region \citep{urquhart2018_agal_full}. The radio emission reveals a cluster of compact sources and this has previously been identified as being associated with the highest density of embedded \hii\ regions in the Galaxy (\citealt{urquhart2013_cornish,de_pree1997,de_pree2004}).  There is a good correlation between the compact sources seen in the mid-infrared image (lower middle panel of Fig.\,\ref{fig:multi_mid}), however, there are few compact radio sources that do not have an infrared counterpart, perhaps due to high extinction. 

The RRL spectrum reveals the presence of a relatively modest red-shifted wing that is offset by $\sim$30\,\kms\ from the systemic velocity of the clump. \cite{jaffe1999} also reported a broad mm-RRL component with an offset velocity from the systemic velocity towards this source. Given the large number of compact radio sources and the poor sensitivity to extended radio emission (due to the large heliocentric distance) it is not possible to investigate the nature of the high-velocity ionized gas associated with this region. Nevertheless, the 7-mm observations by \cite{de_pree2004} showed that the W49 North region has several broad RRL sources and these may be associated with signatures of ionized disk winds. However, it is not clear whether the broader component of our RRL is related to the broad RRLs of the 7-mm observations due to different spatial and flux sensitivities between both observations.

\begin{table*}
\small
\centering
\caption{\label{tb:alma_candidates} A list of potential ALMA candidates.}
\begin{tabular}{c c c . . . c . . c }
\hline\hline
&  &  &   & \multicolumn{2}{c}{RRL peak flux} & & \multicolumn{3}{c}{Radio continuum source} \\ \cline{5-6} \cline{8-10}
ATLASGAL & RA & Dec & \multicolumn{1}{c}{Dist.} & \multicolumn{1}{c}{mm} & \multicolumn{1}{c}{submm} & &\multicolumn{1}{c}{Size}  & \multicolumn{1}{c}{Peak flux} & Morphology \\
name & $\alpha$(J2000) & $\delta$(J2000) & \multicolumn{1}{c}{(kpc)} & \multicolumn{1}{c}{(Jy)} & \multicolumn{1}{c}{(Jy)} & & \multicolumn{1}{c}{($''$)} & \multicolumn{1}{c}{(mJy/beam)} & \\
\hline
AGAL010.624$-$00.384 &18:10:28.70&$-$19:55:49.1& 5.0& 6.6&10.0&& 4.6& 305.6&core-halo\\
AGAL012.804$-$00.199 &18:14:13.96&$-$17:55:44.9& 2.6&10.2&17.5&&16.2& 287.9&bipolar?\\
AGAL013.872$+$00.281 &18:14:35.81&$-$16:45:37.2& 3.9& 2.2& 4.8&&15.4&  24.7&cometary\\
AGAL032.797$+$00.191 &18:50:30.95&$-$00:01:56.5&13.0& 2.8& 6.9&&10.0&  279.1&cometary or multi-peaks\\
AGAL037.874$-$00.399 &19:01:53.59&  +04:12:51.7& 9.7& 2.0& 4.1&& 8.9&  255.7&cometary?\\
AGAL043.164$-$00.029 &19:10:22.01&  +09:05:03.2&11.3& 2.3& 4.6&& 9.6&  242.9&cometary\\
AGAL045.121$+$00.131 &19:13:27.96&  +10:53:35.7& 8.0& 3.2& 6.9&& 7.5&  299.7&cometary? or bipolar?\\
AGAL045.454$+$00.061 &19:14:21.31&  +11:09:11.7& 8.4& 1.4& 3.7&& 7.6&   61.8&unresolved\\
AGAL298.224$-$00.339 &12:10:01.18&$-$62:49:54.0&11.4& ...& 6.6&& 4.1&  369.9&cometary\\
AGAL298.859$-$00.437 &12:15:25.27&$-$63:01:17.0&10.1& ...& 4.0&& 2.6&   79.5&unresolved \\
AGAL324.201$+$00.121 &15:32:53.21&$-$55:56:11.8& 6.8& 2.2& 4.1&& 6.1&  286.6&cometary (\& unresolved) \\
AGAL330.954$-$00.182 &16:09:52.54&$-$51:54:54.9& 5.3& 4.1& 8.7&& 3.9&  444.4&unresolved \\
AGAL332.156$-$00.449 &16:16:40.53&$-$51:17:08.9& 3.6& 2.4& 6.7&&14.0&  101.2&irregular\\
AGAL332.826$-$00.549 &16:20:11.10&$-$50:53:15.4& 3.6& 4.9&10.9&& 3.4&  461.6&unresolved(bipolar?) \\
AGAL333.284$-$00.387 &16:21:31.70&$-$50:27:00.5& 3.6& 7.8&11.4&& 9.4&  421.4&irregular (cometary?) \\
AGAL333.604$-$00.212 &16:22:09.58&$-$50:05:59.6& 3.6&40.8&47.5&&10.6&  393.6&irregular \\
AGAL345.649$+$00.009 &17:06:15.99&$-$40:49:45.4& 1.4& 2.1& 2.5&& 5.1&  627.4&core-halo\\
\hline
\end{tabular}
\tablefoot{The coordinates provided here are coordinates of radio continuum sources in the ATLASGAL clumps instead of coordinates of the ATLASGAL clumps.}
\tablebib{Distance:~\citet{urquhart2013_methanol,lumsden2013,urquhart2018_agal_full}. Coordinate \& radio continuum source: \citet{purcell2013,urquhart_radio_south}.}
\end{table*}

\subsubsection{AGAL045.121$+$00.131}

This clump is located at a distance of 8\,kpc (\citealt{reid2014}) and has a systemic velocity of 58.7\,\kms. The clump has a mass of $\sim$7000\,\msun\ and a luminosity of $1\times 10^6$\,\lsun. The radio emission appears to have a bipolar structure extending to the north and south. The radio morphology is mirrored in the mid-infrared image (lower right panel of Fig.\,\ref{fig:multi_mid}). We note that discrete compact radio sources do appear in the radio emission instead of one bipolar structure and it is unclear if these indicate the presence of individual \hii\ regions or localized enhancements in the radio emission where the ionization front is impacting on a dense structure in the molecular gas. 

The RRL profile shows evidence for both red- and blue-shifted wings, although these are rather modest. The two components fitted to the RRL have similar velocities, both of which are correlated with the systemic velocity of the clump. The broader component is almost twice the linewidth of the narrow component and traces the high-velocity gas. The morphological distribution of the CORNISH continuum emission is correlated with the integrated [Ne {\sc ii}] line emission map presented by \cite{zhu2008}, but the [Ne {\sc ii}] observations did not fully resolve them. The 15.0\,\kms\ to 92.9\,\kms\ velocity channel maps of the [Ne {\sc ii}] emission to the north of this emission correspond to the most extended radio continuum and the mid-infrared source (see the lower right panel of Fig.\,\ref{fig:multi_mid}). On the other hand, the southern radio continuum emission structures correspond to the [Ne {\sc ii}] emission seen in the 36.2\,\kms\ to 71.6\,\kms\ velocity channels. Comparing the radio continuum emission with the [Ne {\sc ii}] velocity channel maps may indicate that the two Gaussian components do not arise from a bipolar flow, but may instead be the result of discrete radio sources sharing a similar peak velocity. High-resolution observations are required to obtain better understanding of the structure and kinematics of this region.


\subsection{Identification of an ALMA sample}

Despite many previous radio continuum observations, the morphology of many \hii\ regions still cannot be fully explained. Studying both the kinematics of ionized gas and the geometry of \hii\ regions provides some insight into their nature. However, the morphology and RRL profiles change with the orientation of our viewing angle and therefore detailed comparison of the observations with current theoretical models \citep{zhu2015,steggles2017} are also required to help us understand the different morphologies observed.

Given that most compact \hii\ regions detected have sizes of a few arc-second, high-resolution observations are required. Detailed studies have been previously conducted at cm-wavelengths  using the Very Large Array (VLA) and Australia Telescope Compact Array (ATCA) \citep{urquhart_radio_south,urquhart_radio_north,purcell2013}, however, cm-RRLs can be significantly affected by pressure broadening and the \hii\ regions can often be optically thick resulting in poorly constrained physical parameters. The ALMA telescope not only opens up the possibility to observe mm- and submm-RRLs at a similar resolution to the cm-RRLs observed by the VLA, but has several additional advantages: 1) the line to continuum ratio increases with frequency and so mm- and submm-RRLs are significantly brighter than at cm-wavelengths; 2) pressure broadening is reduced and is in most cases negligible; 3) the continuum emission of the \hii\ regions is optically thin; 4) there are many thermal molecular transitions that allow the ionized and molecular gas to be traced simultaneously and at a similar resolution. 

As previously shown, our sample consists of sources of the most massive clumps and most luminous \hii\ regions in the Galaxy, and with the radio continuum and mm- and submm-RRL data we can identify a well-characterized sample for follow-up with ALMA.  
Our selection criteria are:

\begin{enumerate}

\item The \hii\ regions should be resolved at high-resolution cm-continuum wavelengths ($\sim$2-3\arcsec) but at resolutions smaller than the ALMA primary beam at 850\,\mum\ (i.e., $<$18\arcsec). This ensures that we can investigate the kinematic structure of ionized gas over many beams and the source can be observed in a single pointed observation at the highest frequency.\\

\item The mm- and submm-RRLs should be bright and relatively isolated from contamination from thermal lines, which can be a significant issue at submillimeter wavelengths.\\

\item The \hii\ regions should be relatively isolated from nearby evolved \hii\ regions, which might contribute emission through the side lobes and make imaging difficult. \\

\end{enumerate}

Applying these criteria to the sample discussed in this paper we have identified 17 candidates for further detailed studies with ALMA (see Table\,\ref{tb:alma_candidates} and Fig.\,\ref{fig:mid-ir_alma}). We excluded several sources that satisfied criteria 2 and 3 above since good quality radio continuum data are not available for them. In Table\,\ref{tb:alma_candidates} we provide the coordinates of 6\,cm radio sources instead of the ATLASGAL dust clumps. Our list of potential ALMA candidates covers radio continuum sources with various radio properties and morphologies. The cometary morphology is the most common in the list, but there is some uncertainty in their morphological classification due to limited sensitivity to angular scales.

These selected sources have clean RRLs profiles at both submm/mm-wavelengths, and thus both wavelengths can be used for observations. Given that the \hii\ regions are optically thin, this means that both mm- and submm-observations are equally able to probe the properties of the ionized gas. As seen in Table\,\ref{tb:alma_candidates}, their submm-RRLs show brighter peak fluxes than their mm-RRLs. In fact, the line to continuum ratio at submm-wavelengths is a factor of two better at mm-wavelengths (see the lower panel of Fig.\,\ref{fig:submm_mm}, and Figs.\,3 and 4 in \citealt{peters2012}). Observations at submm-wavelengths will also have a factor of three higher resolution compared to mm-observations for the same configuration, however, the primary beam is also three times smaller, and so mm-observations may be more suitable for more extended \hii\ regions. The choice of what frequency at which to observe a particular source can, therefore, be tailored to ensure the region can be covered in a single pointing and that the same physical scales are probed (i.e., mm-wavelengths for near by and extended \hii\ regions and submm-wavelengths for more distant and compact \hii\ regions).

\section{Summary and conclusions}\label{sec:sum_con}

We present the results of a set of targeted submillimeter RRL observations of 104 compact \hii\ regions previously identified by \citet{kim2017}. We have observed the H25$\alpha$, H27$\alpha$, H28$\alpha$, H29$\alpha,$  H30$\alpha,$ and H35$\beta$ RRL transitions. The detection rates for the H$n\alpha$ and H$n\beta$ transitions are 89\% (93) and 40\% (34), respectively. These have been compared with the mm-recombination lines reported by \citet{kim2017} to investigate the physical properties of the ionized gas and its local environment.

We find that the submm-RRLs are approximately a factor of two brighter than the mm-RRLs, which is consistent with LTE, and that the RRL line emission from the \hii\ regions is optically thin. We also find a strong correlation between the velocities of the molecular gas and the ionized gas of the \hii\ region still being embedded in their host cloud. 

We find a significant correlation between the bolometric luminosity of the embedded clusters in their dust clumps and the ionizing photon flux produced by the stars. This indicates that the stars of the \hii\ regions provide the majority of the clumps' luminosity. We also find a trend for increasing dust temperature with the bolometric luminosity consistent with the feedback from the embedded \hii\ regions on its natal environment. Besides, the luminosity of the submm-RRLs presents robust correlations with both bolometric luminosity and Lyman photon flux. Comparing the physical properties of the \hii\ regions and their host clumps to the general population of dust clumps located in the inner-Galaxy, we find  the clumps with the submm-RRL detections to be some of the most massive and luminous in the Galaxy.  

We used the submm-RRL flux to measure the photoionizing production rate, $Q$, which corresponds to the Lyman photon flux.  The measured $Q$(H28$\alpha$) values for H28$\alpha$ lines are consistent with the Lyman photon flux measured from 5\,GHz radio continuum emission. This confirms the $Q$ as a tool for measuring star formation rates (SFRs) using submm-RRL observations in external galaxies, without the problems of dust obscuration and influences of evolved star and synchrotron emission on continuum fluxes, encountered by other methods to measure SFRs.

A single Gaussian component well describes the profiles of the majority of RRLs. However, we have identified six \hii\ regions where the spectral profiles show the presence of red and/or blue-shifted wings. These wings reveal the presence of a high-velocity component in the ionized gas. We investigate these sources in detail and find that the data towards four \hii\ regions are consistent with the presence of high-velocity flows associated with the head of cometary or bipolar \hii\ regions. The other two clumps are associated with clusters of compact \hii\ regions, and so the observed RRL profiles may result from the blending of the emission from several \hii\ regions. Higher-resolution observations are required to study the kinematics and physical properties of all of the regions in detail and to disentangle their morphology and long lifetime (about $10^{5}$ years).

Finally, we identify a sample of submm-RRL sources that would be suitable for high-resolution follow-up observations with ALMA. For the potential ALMA candidates,  17 \hii\ regions with mm/submm-RRL detections are selected with the following criteria: appropriate sizes for the ALMA beam; bright mm/submm-RRLs with no contamination by molecular lines; relatively isolated location of the \hii\ regions from nearby evolved \hii\ regions.

\section*{Acknowledgements}

\addcontentsline{toc}{section}{Acknowledgements} 
We would like to thank the referee for their constructive comments and suggestions that have helped to improve this paper. This paper is based on data acquired with the Atacama Pathfinder EXperiment (APEX). APEX is a collaboration between the Max Planck Institute for Radioastronomy, the European Southern Observatory, and the Onsala Space Observatory. This work was partly carried out within the Collaborative Research Council 956, sub-project A6, funded by the Deutsche Forschungsgemeinschaft (DFG). This document was produced using the Overleaf web application, which can be found at www.overleaf.com. Won-Ju\, Kim was supported for this research through a stipend from the International Max Planck Research School (IMPRS) for Astronomy and Astrophysics at the Universities of Bonn and Cologne. The ATLASGAL project is a collaboration between the Max-Planck-Gesellschaft, the European Southern Observatory (ESO) and the Universidad de Chile. It includes projects E-181.C-0885, E-078.F-9040(A), M-079.C-9501(A), M-081.C-9501(A) plus Chilean data.

\bibliographystyle{aa}
\bibliography{submmrrl}

\begin{appendix}

\onecolumn
\section{Continued table of list of observed sources}\label{appendix:full_tb}
\begin{longtable}{c c c c c }
\caption{\label{appetb:obs_source_list} List of Observed sources.}\\
\hline\hline
ID & ATLASGAL & RA & Dec. & Observed   \\
No. & clump name & $\alpha$(J2000) & $\delta$(J2000) &  transition \\
\hline
\endfirsthead
\caption{continued.}\\
\hline\hline
ID & ATLASGAL & RA & Dec. &  Observed   \\
No. & clump name &  $\alpha$(J2000) & $\delta$(J2000) &  transition \\
\hline
\endhead
\hline
\endfoot
16&AGAL030.818$-$00.056&18:47:46.4&$-$01:54:31&(H26$\alpha$) \\
17&AGAL030.866$+$00.114&18:47:15.5&$-$01:47:14&(H25$\alpha$), H28$\alpha$, (H35$\beta$) \\
18&AGAL031.243$-$00.111&18:48:45.1&$-$01:33:13&(H25$\alpha$), H28$\alpha$, (H35$\beta$) \\
19&AGAL031.281$+$00.062&18:48:12.1&$-$01:26:31&H30$\alpha$\\
20&AGAL031.412$+$00.307&18:47:34.2&$-$01:12:45&(H25$\alpha^{c}$?), (H26$\alpha^{c}$?), (H28$\alpha^{c}$?), (H35$\beta^{c}$?) \\
21&AGAL032.149$+$00.134&18:49:31.8&$-$00:38:08&H30$\alpha$\\
22&AGAL032.797$+$00.191&18:50:30.7&$-$00:02:01&H25$\alpha$, H28$\alpha$, H35$\beta$\\
23&AGAL033.133$-$00.092&18:52:08.3&$+$00:08:12&H30$\alpha$ \\
24&AGAL033.914$+$00.109&18:52:50.6&$+$00:55:29&(H25$\alpha$), H28$\alpha$, (H35$\beta$)\\
25&AGAL034.258$+$00.154&18:53:18.5&$+$01:14:58&H25$\alpha$, H26$\alpha$, H27$\alpha$,H28$\alpha$,(H35$\beta^{c}$?)\\
26&AGAL035.466$+$00.141&18:55:33.7&$+$02:19:12&H30$\alpha$ \\
27&AGAL035.579$-$00.031&18:56:22.7&$+$02:20:27&H30$\alpha$ \\
28&AGAL037.874$-$00.399&19:01:53.6&$+$04:12:52&H25$\alpha$,H27$\alpha$,H28$\alpha$,H35$\beta$ \\
29&AGAL043.148$+$00.014&19:10:11.0&$+$09:05:25&(H25$\alpha$?),H28$\alpha$,H35$\beta$\\
30&AGAL043.164$-$00.029&19:10:21.6&$+$09:05:03&H25$\alpha$,H28$\alpha$,H35$\beta$\\
31&AGAL043.166$+$00.011&19:10:13.4&$+$09:06:16&H25$\alpha$,H26$\alpha$,H28$\alpha$,H35$\beta$\\
32&AGAL043.178$-$00.011&19:10:18.7&$+$09:06:06&(H25$\alpha$),H28$\alpha$,(H35$\beta$) \\
33&AGAL043.236$-$00.047&19:10:33.7&$+$09:08:21&H25$\alpha$,H28$\alpha$,(H35$\beta$)\\
34&AGAL043.889$-$00.786&19:14:26.7&$+$09:22:31&(H25$\alpha$),H28$\alpha$,H30$\alpha$,(H35$\beta$)\\
35&AGAL045.071$+$00.132&19:13:22.0&$+$10:50:53&H30$\alpha$\\
36&AGAL045.121$+$00.131&19:13:27.8&$+$10:53:35&H25$\alpha$,H28$\alpha$,H35$\beta$\\
37&AGAL045.454$+$00.061&19:14:20.9&$+$11:09:13&H25$\alpha$,H28$\alpha$,(H35$\beta$?)\\
38&AGAL045.474$+$00.134&19:14:07.4&$+$11:12:25&(H30$\alpha$)\\
39&AGAL048.991$-$00.299&19:22:26.1&$+$14:06:37&(H30$\alpha$)\\
40&AGAL049.214$-$00.342&19:23:01.1&$+$14:17:10&H30$\alpha$\\
41&AGAL049.369$-$00.301&19:23:10.3&$+$14:26:27&H30$\alpha$\\
42&AGAL049.489$-$00.389&19:23:43.6&$+$14:30:32&(H25$\alpha^{c}$?),H26$\alpha$,H28$\alpha$,(H30$\alpha^{c}$?),(H35$\beta^{c}$?)\\
43&AGAL285.264$-$00.049&10:31:30.2&$-$58:02:16&H26$\alpha$\\
44&AGAL289.881$-$00.797&11:01:00.6&$-$60:50:22&H26$\alpha$\\
45&AGAL291.272$-$00.714&11:11:51.3&$-$61:18:39&H25$\alpha$,H26$\alpha$,H28$\alpha$,H29$\alpha$,H30$\alpha$,H35$\beta$\\
46&AGAL298.224$-$00.339&12:10:01.0&$-$62:49:53&H25$\alpha$,H26$\alpha$,H28$\alpha$,H29$\alpha$,H30$\alpha$,H35$\beta$ \\
47&AGAL298.859$-$00.437&12:15:24.3&$-$63:01:24&H25$\alpha$,H26$\alpha$,H28$\alpha$,H29$\alpha$,H30$\alpha$,H35$\beta$ \\
48&AGAL301.116$+$00.959&12:36:01.9&$-$61:51:28&(H25$\alpha$),H27$\alpha$,H28$\alpha$,(H35$\beta$)\\
49&AGAL301.136$-$00.226&12:35:35.0&$-$63:02:31&H25$\alpha$,H28$\alpha$,(H35$\beta$)\\
50&AGAL305.196$+$00.034&13:11:13.7&$-$62:45:04&H25$\alpha$,H28$\alpha$,H30$\alpha$,H35$\beta$\\
51&AGAL305.271$-$00.009&13:11:55.5&$-$62:47:16&H25$\alpha$,H27$\alpha$,H28$\alpha$,(H35$\beta$?)\\
52&AGAL305.272$+$00.296&13:11:43.3&$-$62:29:00&(H25$\alpha$),H28$\alpha$,H30$\alpha$,(H35$\beta$)\\
53&AGAL305.357$+$00.202&13:12:31.2&$-$62:34:10&H25$\alpha$,H28$\alpha$,H30$\alpha$,(H35$\beta$?)\\
54&AGAL305.361$+$00.186&13:12:33.5&$-$62:35:03&(H25$\alpha$),(H28$\alpha^{c}$?),(H35$\beta$)\\
55&AGAL305.367$+$00.212&13:12:35.6&$-$62:33:33&(H25$\alpha$),H27$\alpha$,H28$\alpha$,(H35$\beta$)\\
56&AGAL311.899$+$00.084&14:07:35.3&$-$61:27:20&(H25$\alpha$),H27$\alpha$,H28$\alpha$,(H35$\beta$)\\
57&AGAL312.108$+$00.309&14:08:41.7&$-$61:10:45&(H25$\alpha$?),H28$\alpha$,H30$\alpha$,H35$\beta$ \\
58&AGAL316.786$-$00.037&14:45:11.3&$-$59:48:49&(H25$\alpha$),H28$\alpha$,(H35$\beta$)\\
59&AGAL316.799$-$00.056&14:45:20.5&$-$59:49:31&H25$\alpha$,H28$\alpha$,H30$\alpha$,H35$\beta$\\
60&AGAL316.811$-$00.059&14:45:27.8&$-$59:49:22&(H25$\alpha$),H28$\alpha$,H30$\alpha$,H35$\beta$\\
61&AGAL318.914$-$00.164&15:00:34.7&$-$58:58:10&H25$\alpha$,H28$\alpha$,H30$\alpha$,H35$\beta$\\
62&AGAL320.319$-$00.176&15:10:01.4&$-$58:17:29&(H25$\alpha$),H27$\alpha$,H28$\alpha$,(H35$\beta$)\\
63&AGAL321.719$+$01.176&15:13:48.4&$-$56:24:44&(H25$\alpha$),H27$\alpha$,H28$\alpha$,(H35$\beta$)\\
64&AGAL322.158$+$00.636&15:18:37.3&$-$56:38:11&(H25$\alpha$),(H27$\alpha$),H28$\alpha$,(H35$\beta$)\\
65&AGAL322.164$+$00.622&15:18:39.6&$-$56:38:59&H25$\alpha$,H28$\alpha$,H30$\alpha$,H35$\beta$\\
66&AGAL323.459$-$00.079&15:29:19.6&$-$56:31:28&(H25$\alpha$),H27$\alpha$,H28$\alpha$,(H35$\beta$)\\
67&AGAL324.201$+$00.121&15:32:53.2&$-$55:56:10&(H25$\alpha$),H28$\alpha$,H30$\alpha$,H35$\beta$\\
68&AGAL326.446$+$00.907&15:42:16.5&$-$53:58:27&(H25$\alpha$),H28$\alpha$,H30$\alpha$,(H35$\beta$)\\
69&AGAL326.657$+$00.594&15:44:42.8&$-$54:05:43&H25$\alpha$,H28$\alpha$,H30$\alpha$,H35$\beta$\\
70&AGAL327.293$-$00.579&15:53:08.7&$-$54:37:08&(H25$\alpha^{c}$?),(H28$\alpha^{c}$?),(H35$\beta^{c}$?)\\
71&AGAL327.301$-$00.552&15:53:02.6&$-$54:35:42&H25$\alpha$,H28$\alpha$,H30$\alpha$,H35$\beta$\\
72&AGAL328.308$+$00.431&15:54:08.4&$-$53:11:40&(H25$\alpha$),H28$\alpha$,H30$\alpha$,H35$\beta$\\
73&AGAL328.566$-$00.534&15:59:37.4&$-$53:45:58&(H25$\alpha$),H27$\alpha$,H28$\alpha$,(H35$\beta$?)\\
74&AGAL328.809$+$00.632&15:55:49.6&$-$52:42:60&H25$\alpha$,H26$\alpha$,H28$\alpha$,H30$\alpha$,(H35$\beta$)\\
75&AGAL330.294$-$00.394&16:07:38.8&$-$52:31:06&(H25$\alpha$),H28$\alpha$,(H35$\beta$)\\
76&AGAL330.879$-$00.367&16:10:20.3&$-$52:06:15&H25$\alpha$,H26$\alpha$,(H28$\alpha^{c}$?),(H29$\alpha^{c}$?),(H30$\alpha^{c}$?),(H35$\beta$)\\
77&AGAL330.954$-$00.182&16:09:52.9&$-$51:55:01&H25$\alpha$,H26$\alpha$,H28$\alpha$,H35$\beta$\\
78&AGAL331.521$-$00.081&16:12:06.3&$-$51:27:13&(H25$\alpha$?),H28$\alpha$,(H35$\beta$)\\
79&AGAL331.546$-$00.067&16:12:10.6&$-$51:25:45&(H25$\alpha$),H28$\alpha$,(H35$\beta$)\\
80&AGAL332.156$-$00.449&16:16:40.0&$-$51:17:06&H25$\alpha$,H28$\alpha$,H35$\beta$\\
81&AGAL332.647$-$00.609&16:19:37.6&$-$51:03:20&H25$\alpha$,H28$\alpha$,(H35$\beta$)\\
82&AGAL332.826$-$00.549&16:20:11.4&$-$50:53:19&H25$\alpha$,H26$\alpha$,H28$\alpha$,H35$\beta$\\
83&AGAL333.018$-$00.449&16:20:36.4&$-$50:40:56&(H25$\alpha$),H28$\alpha$,(H35$\beta$)\\
84&AGAL333.134$-$00.431&16:21:02.6&$-$50:35:13&H25$\alpha$,(H26$\alpha$?),H27$\alpha$,H28$\alpha$,H35$\beta$\\
85&AGAL333.284$-$00.387&16:21:31.5&$-$50:27:02&H25$\alpha$,H26$\alpha$,H28$\alpha$,H35$\beta$\\
86&AGAL333.308$-$00.366&16:21:32.2&$-$50:25:09&H25$\alpha$,H28$\alpha$,H35$\beta$\\
87&AGAL333.604$-$00.212&16:22:09.9&$-$50:06:07&H25$\alpha$,H26$\alpha$,H28$\alpha$,H35$\beta$\\
88&AGAL337.121$-$00.174&16:36:42.9&$-$47:31:34&(H25$\alpha$),H28$\alpha$,(H35$\beta$?)\\
89&AGAL337.922$-$00.456&16:41:06.1&$-$47:07:01&(H29$\alpha^{c}$?)\\
90&AGAL338.074$+$00.011&16:39:38.7&$-$46:41:33&(H25$\alpha$),H28$\alpha$,(H35$\beta$)\\
91&AGAL338.332$+$00.131&16:40:07.1&$-$46:25:10&(H25$\alpha$),H28$\alpha$,(H35$\beta$)\\
92&AGAL338.434$+$00.012&16:41:02.0&$-$46:25:19&H29$\alpha$\\
93&AGAL340.784$-$01.016&16:54:19.7&$-$45:17:22&H29$\alpha$\\
94&AGAL344.424$+$00.046&17:02:09.0&$-$41:46:59&(H25$\alpha$?),H28$\alpha$,(H35$\beta$?)\\
95&AGAL345.408$-$00.952&17:09:35.2&$-$41:35:45&H25$\alpha$,H28$\alpha$, (H35$\beta$?)\\
96&AGAL345.488$+$00.314&17:04:27.7&$-$40:46:21&H25$\alpha$,(H26$\alpha$?),H28$\alpha$,H35$\beta$\\
97&AGAL345.649$+$00.009&17:06:16.5&$-$40:49:39&H25$\alpha$,H28$\alpha$,H35$\beta$\\
98&AGAL348.698$-$01.027&17:19:59.3&$-$38:57:56&(H25$\alpha$),(H28$\alpha$),(H35$\beta$)\\
99&AGAL351.161$+$00.697&17:19:56.2&$-$35:57:45&(H26$\alpha$)\\
100&AGAL351.244$+$00.669&17:20:18.9&$-$35:54:38&H25$\alpha$,H26$\alpha$,H28$\alpha$,H35$\beta$\\
101&AGAL351.416$+$00.646&17:20:52.3&$-$35:46:50&(H25$\alpha^{c}$?),(H26$\alpha^{c}$?),(H28$\alpha^{c}$?),(H35$\beta^{c}$?)\\
102&AGAL351.581$-$00.352&17:25:25.2&$-$36:12:45&H26$\alpha$\\
103&AGAL353.189$+$00.899&17:24:44.9&$-$34:10:36&H25$\alpha$,H28$\alpha$,H35$\beta$\\
104&AGAL353.409$-$00.361&17:30:26.8&$-$34:41:51&H25$\alpha$,(H26$\alpha$),H28$\alpha$,H35$\beta$\\
\end{longtable}

\vskip 5cm

\twocolumn
\section{Potential ALMA candidates}\label{appendix:alma_candidate}

\begin{figure*}
\includegraphics[height= 0.3\textwidth]{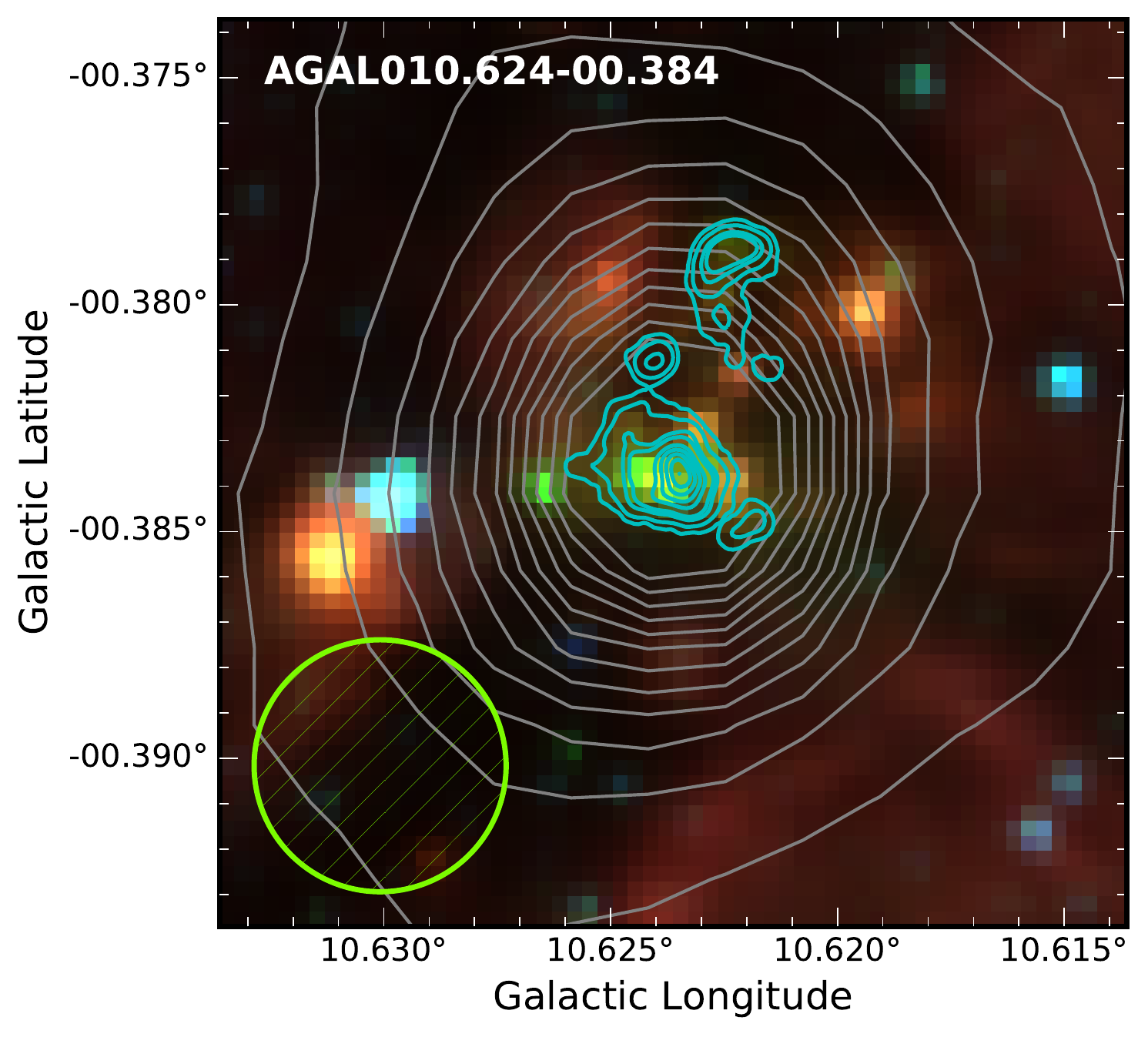}
\includegraphics[height= 0.3\textwidth]{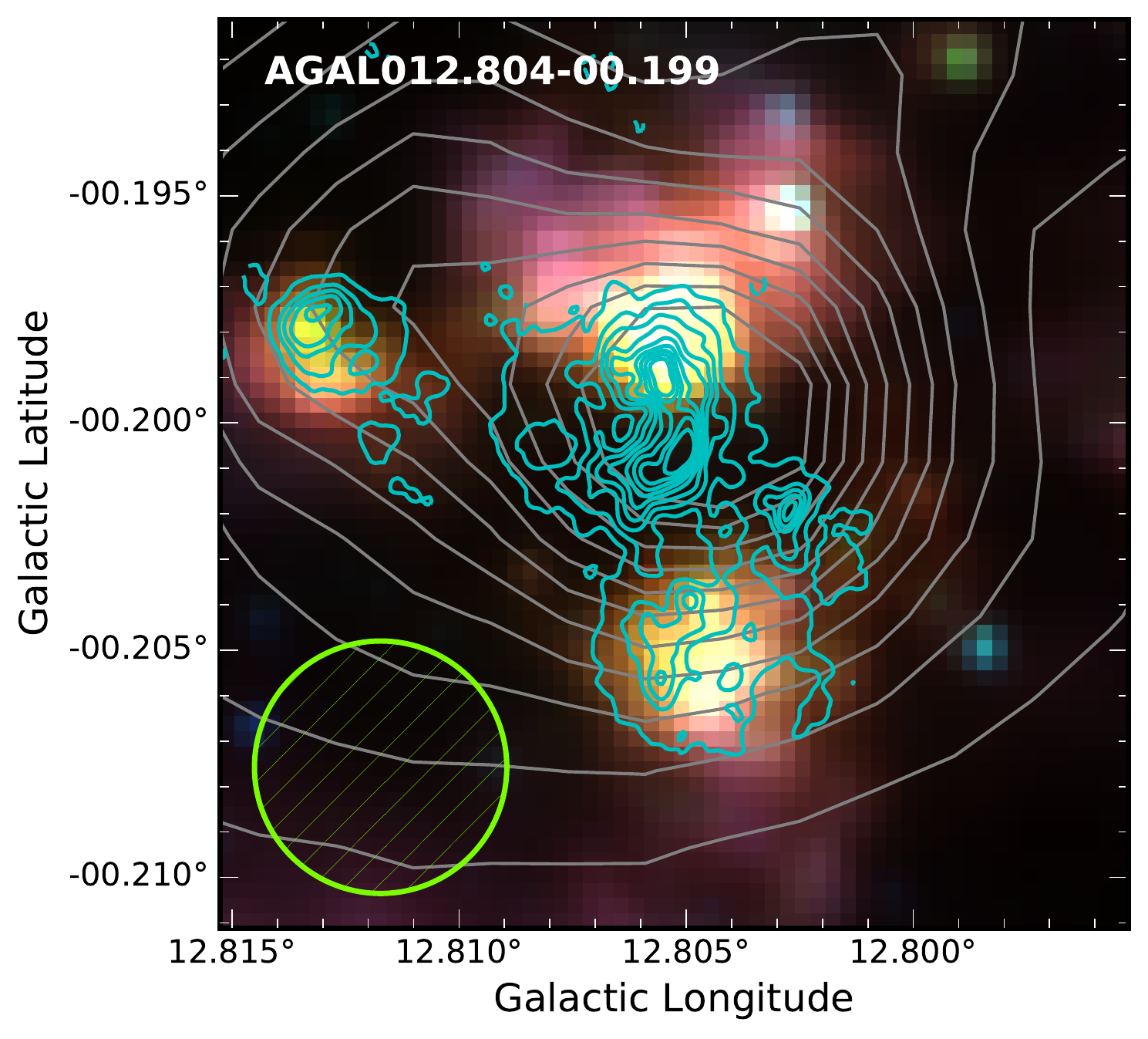}
\includegraphics[height= 0.3\textwidth]{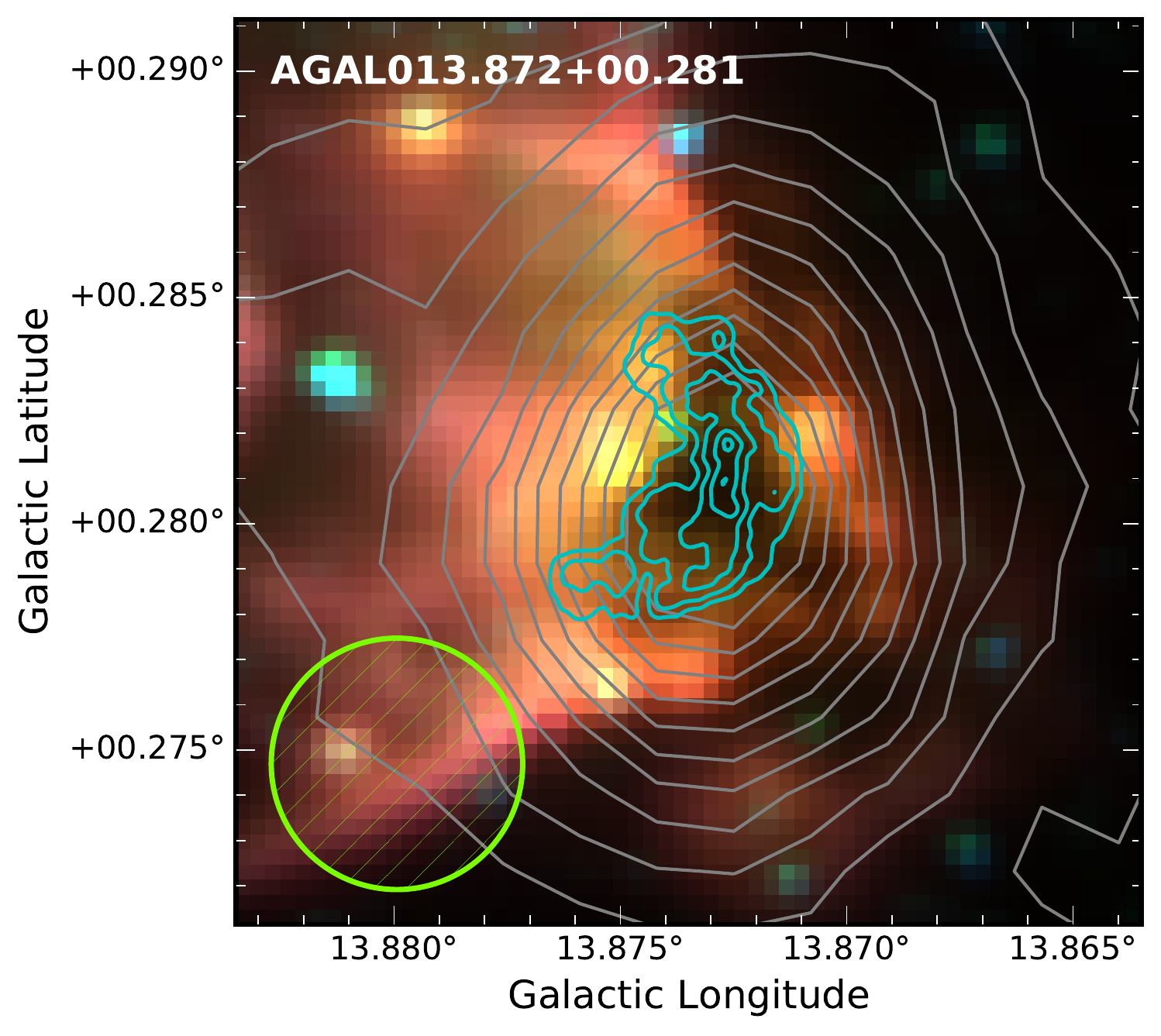}
\includegraphics[height= 0.3\textwidth]{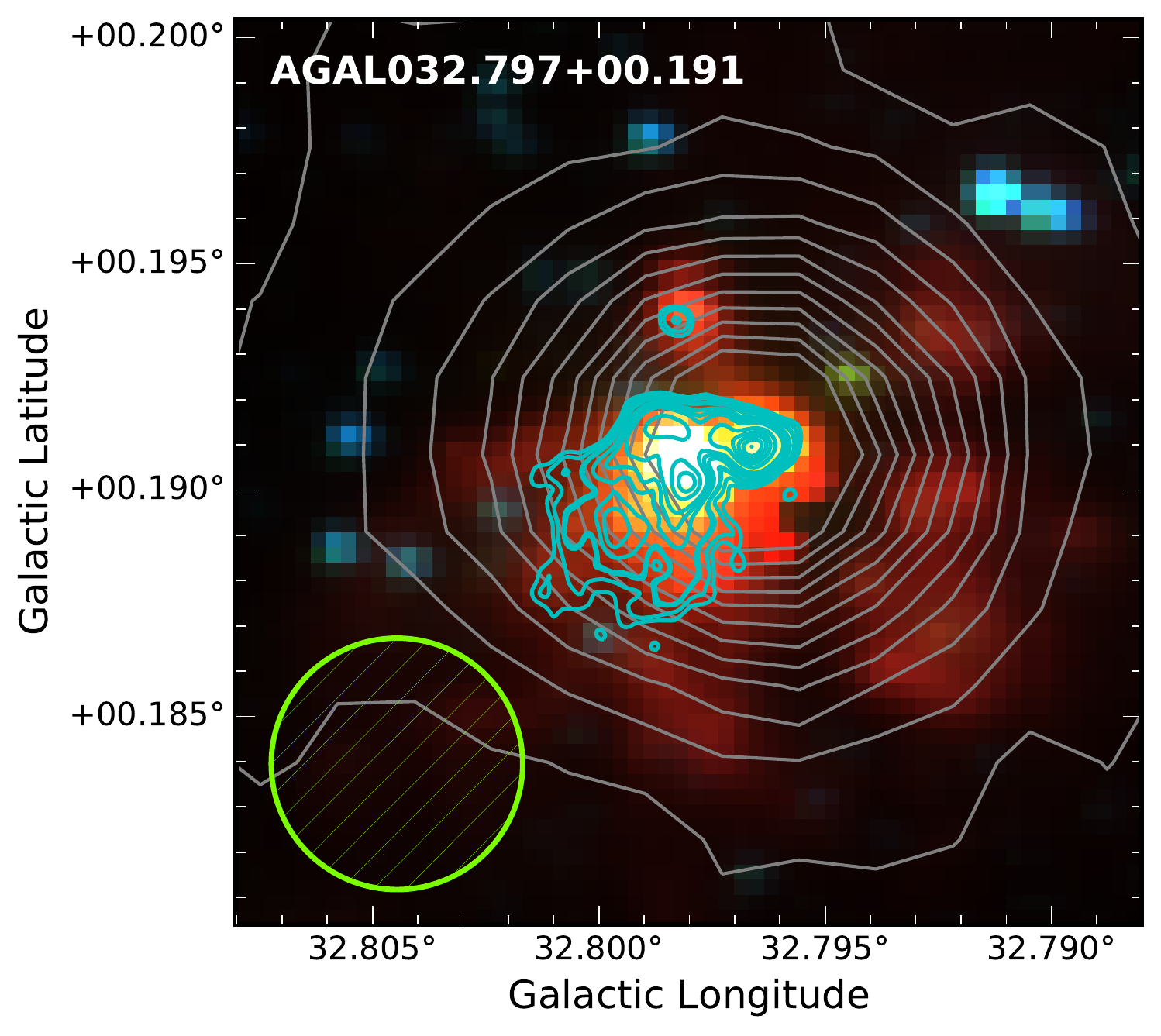}
\includegraphics[height= 0.3\textwidth]{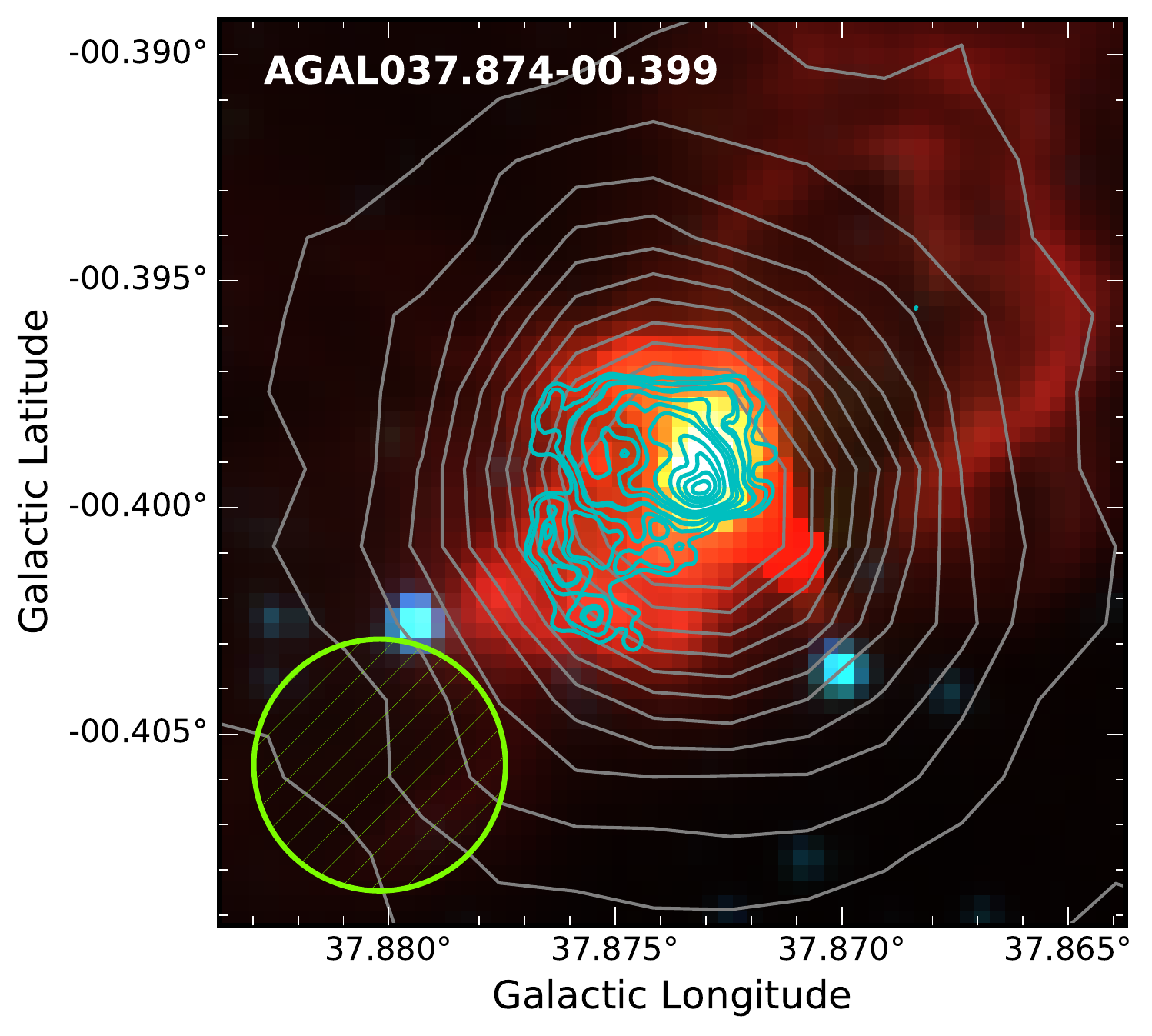}
\includegraphics[height= 0.3\textwidth]{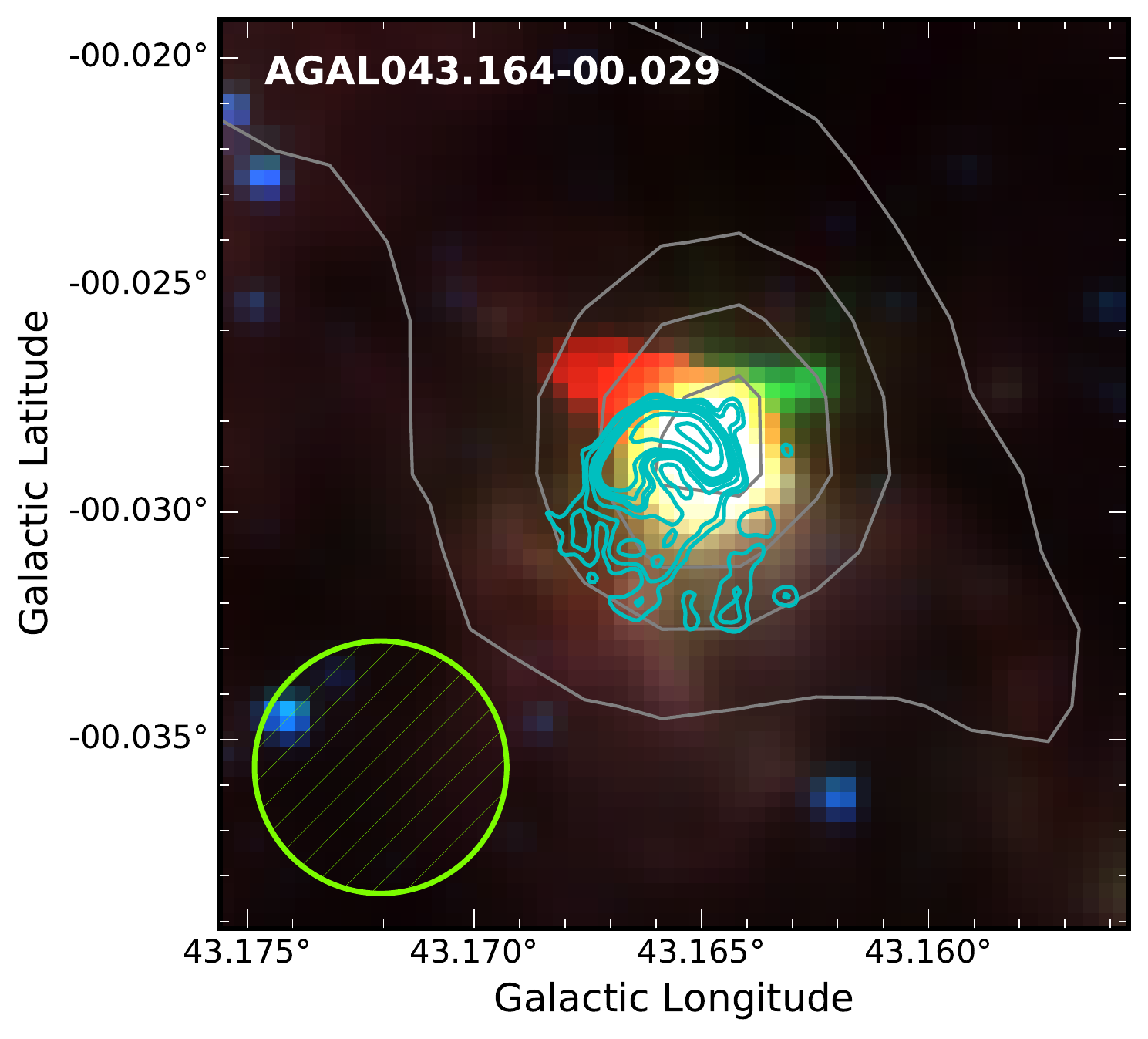}
\includegraphics[height= 0.3\textwidth]{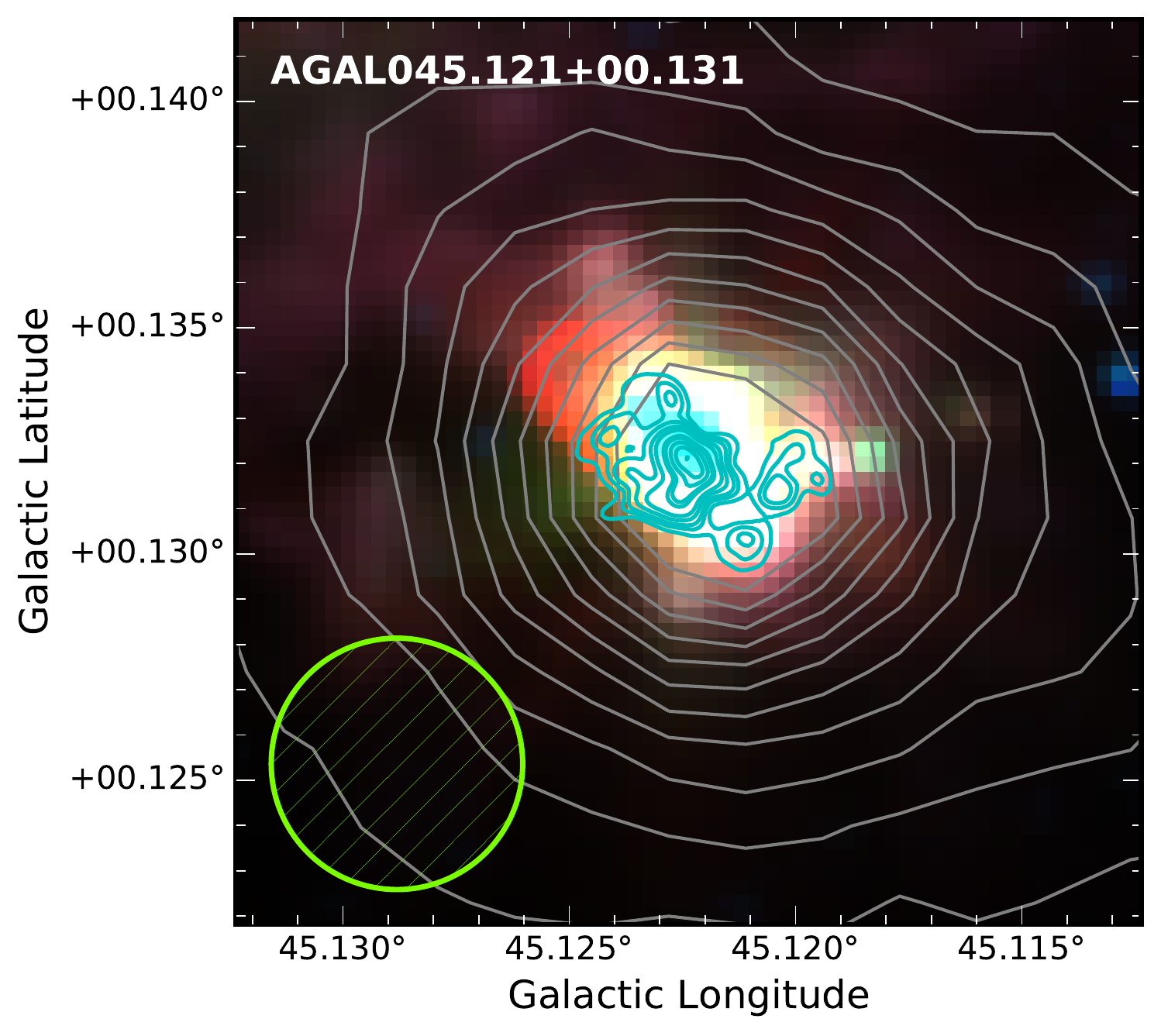}
\includegraphics[height= 0.3\textwidth]{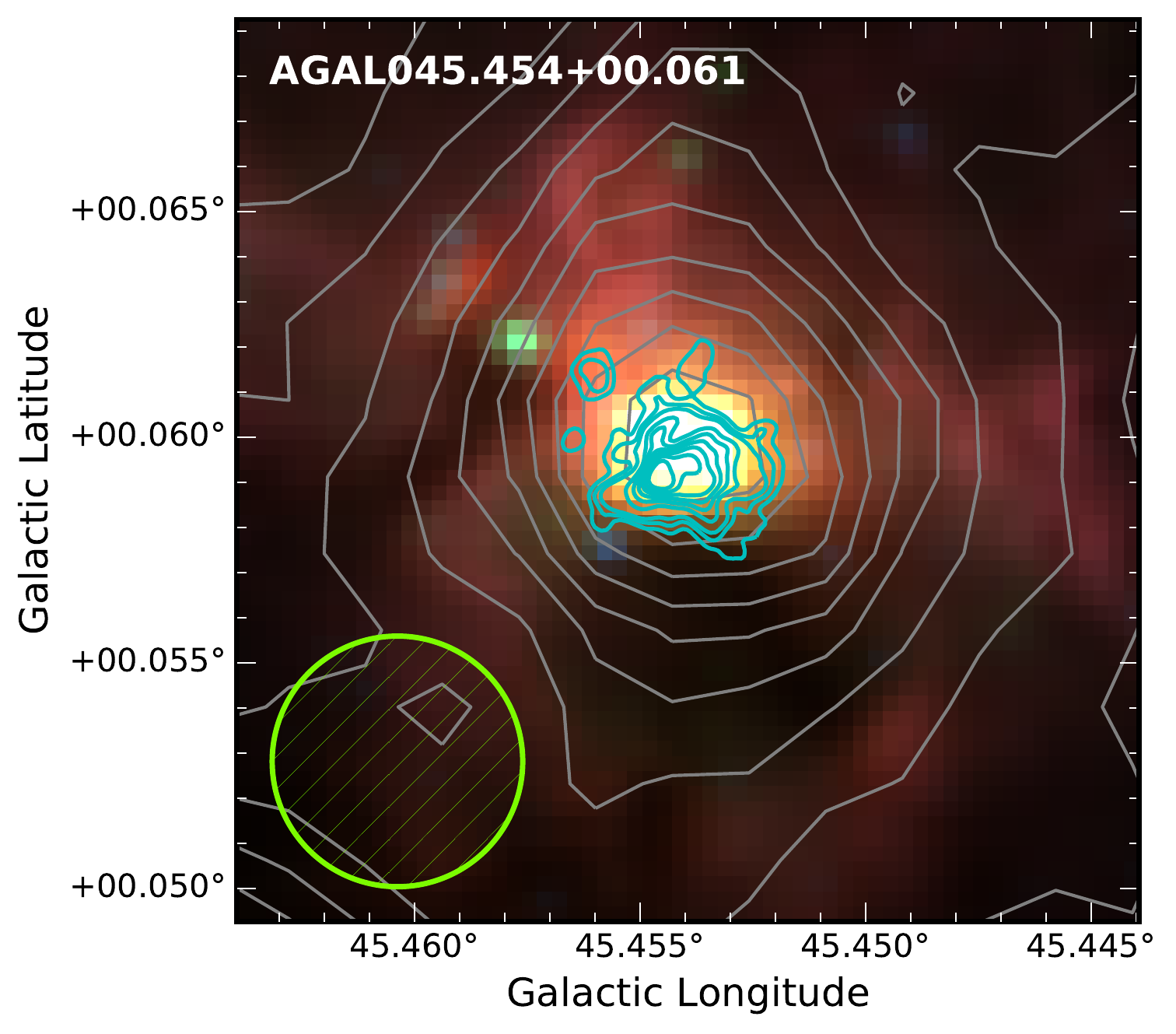}
\includegraphics[height= 0.3\textwidth]{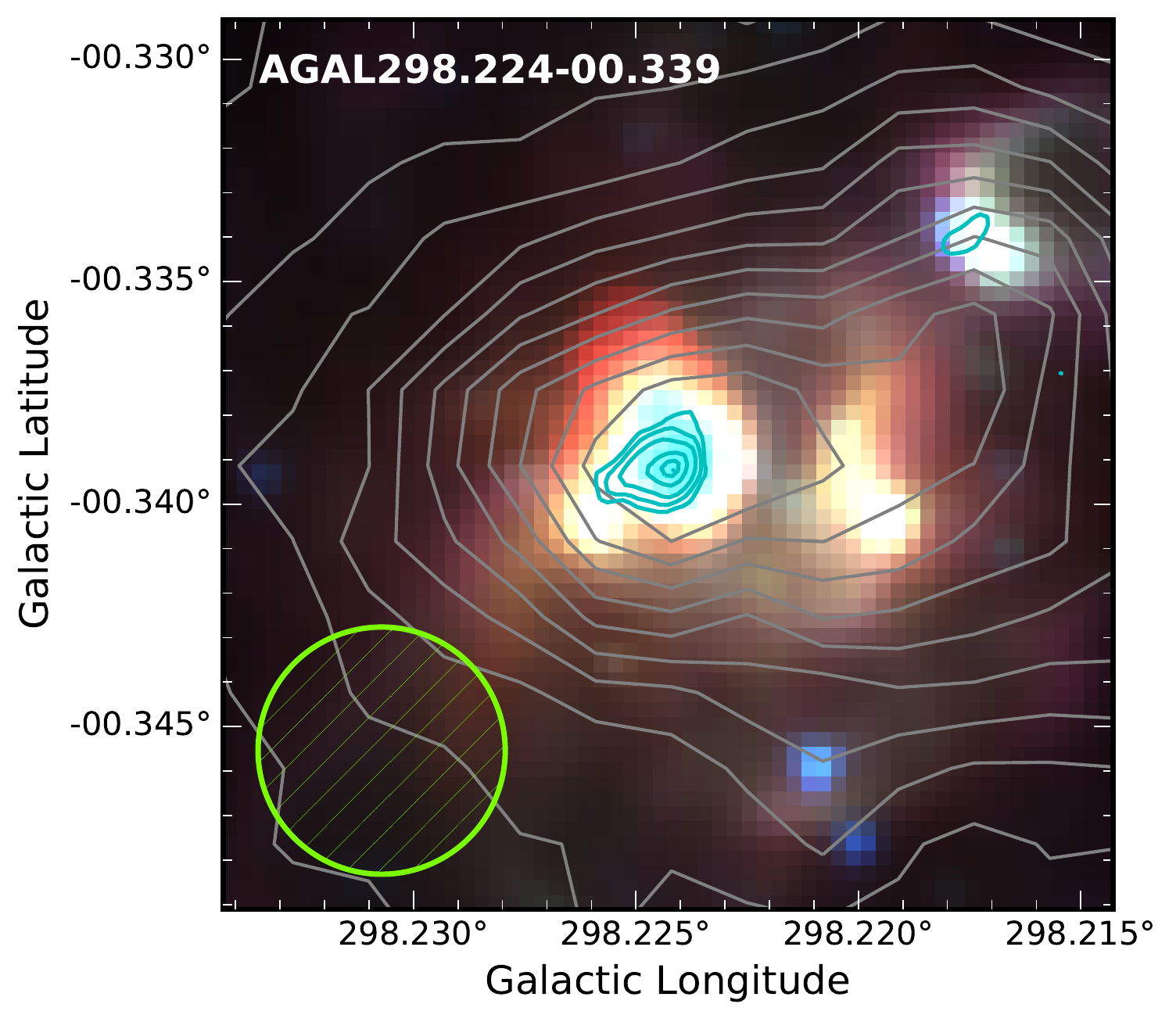}
\includegraphics[height= 0.3\textwidth]{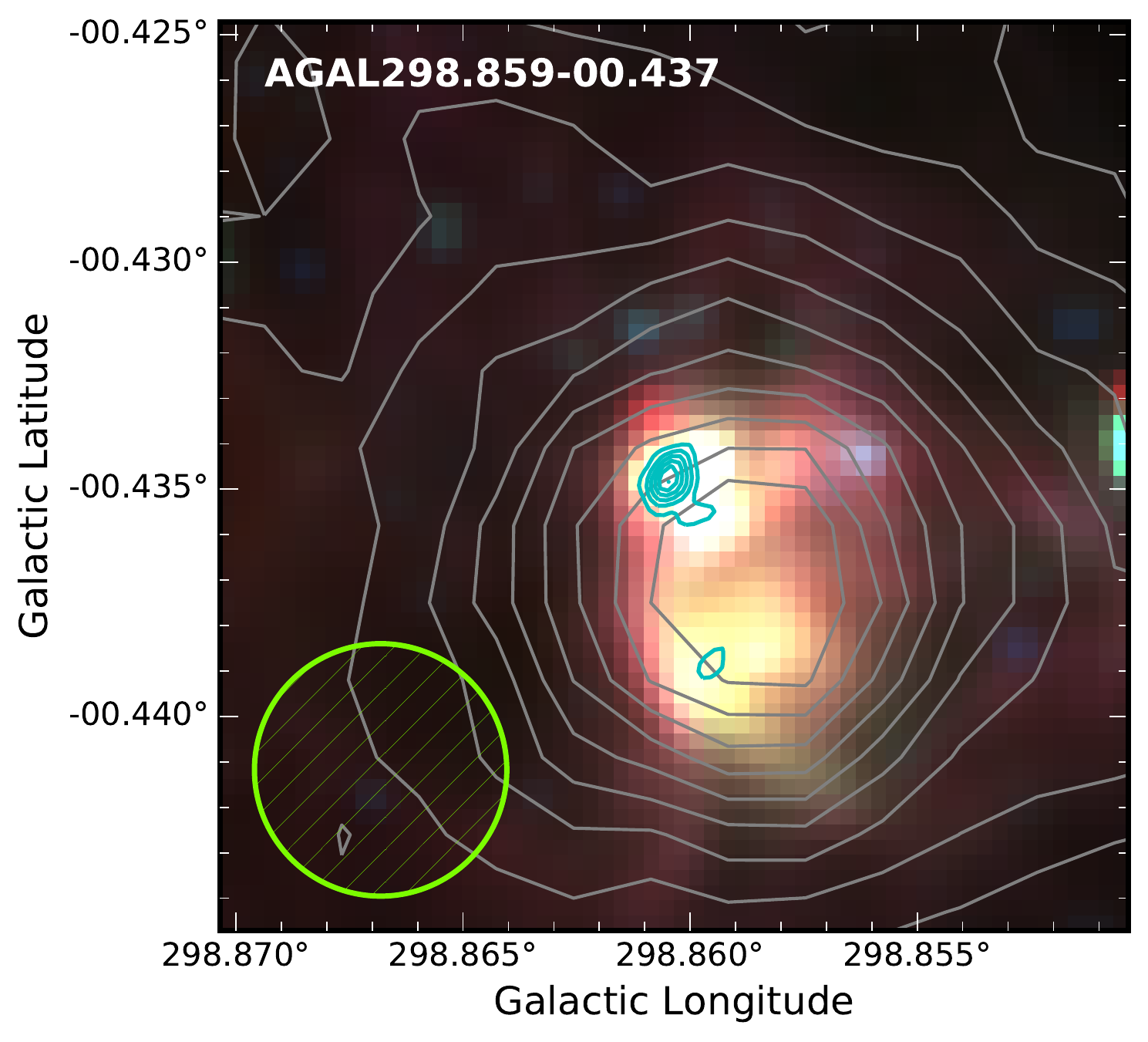}
\includegraphics[height= 0.3\textwidth]{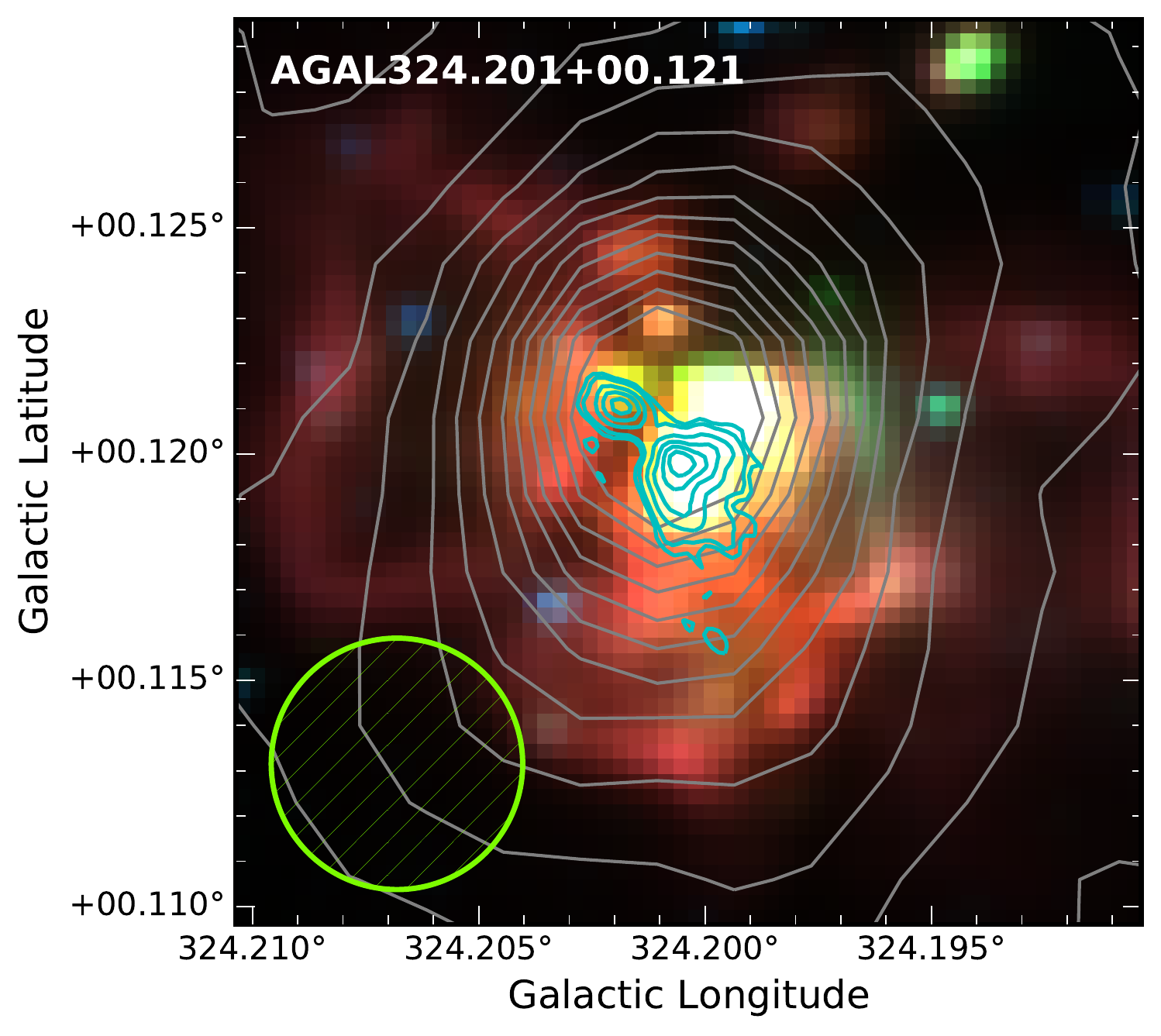}
\includegraphics[height= 0.3\textwidth]{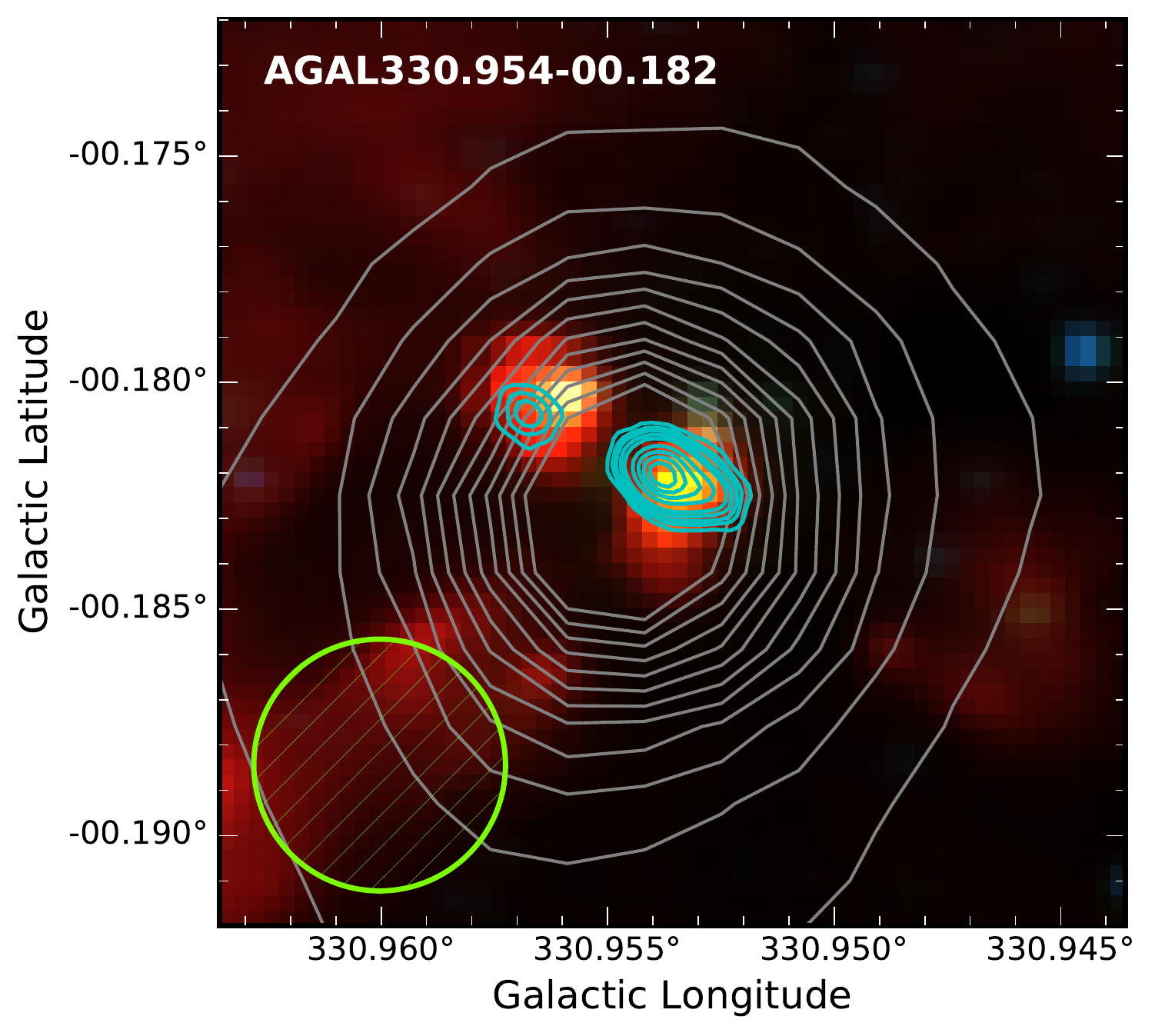}
\caption{\label{fig:mid-ir_alma} GLIMPSE IRAC three-color composite image (blue: 3.6\,\mum, green: 4.6\,\mum, and red: 8\,\mum). Gray contour represents 870\,\mum\ dust continuum emission from the ATLASGAL survey. Cyan and blue contours show 5\,GHz radio continuum emission from CORNISH and RMS surveys. The arbitrary beam of 20$''$ is indicated with a bright green hatched circle in the left-bottom corner.}
\end{figure*}
\begin{figure*}
\ContinuedFloat
\includegraphics[height= 0.3\textwidth]{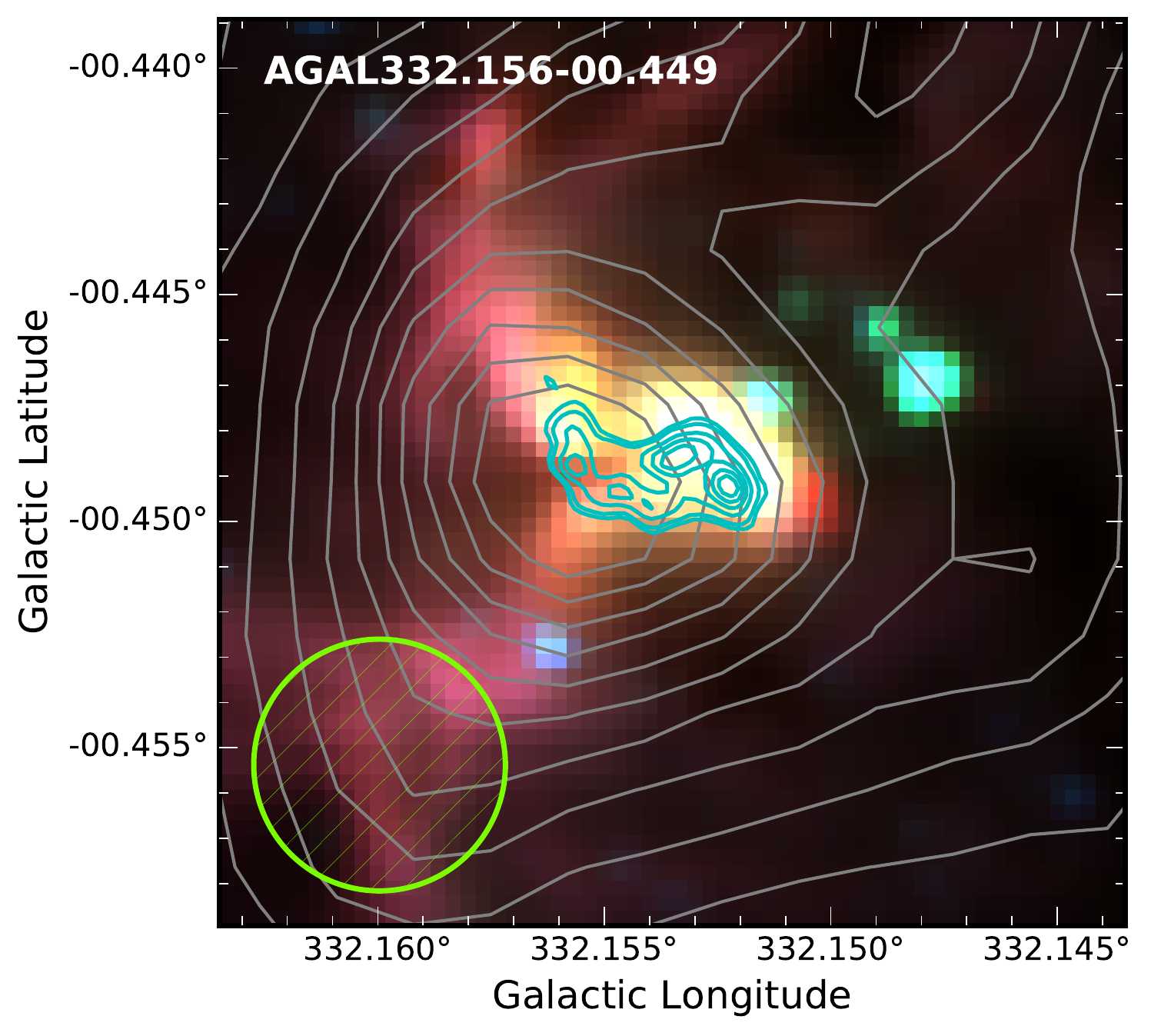}
\includegraphics[height= 0.3\textwidth]{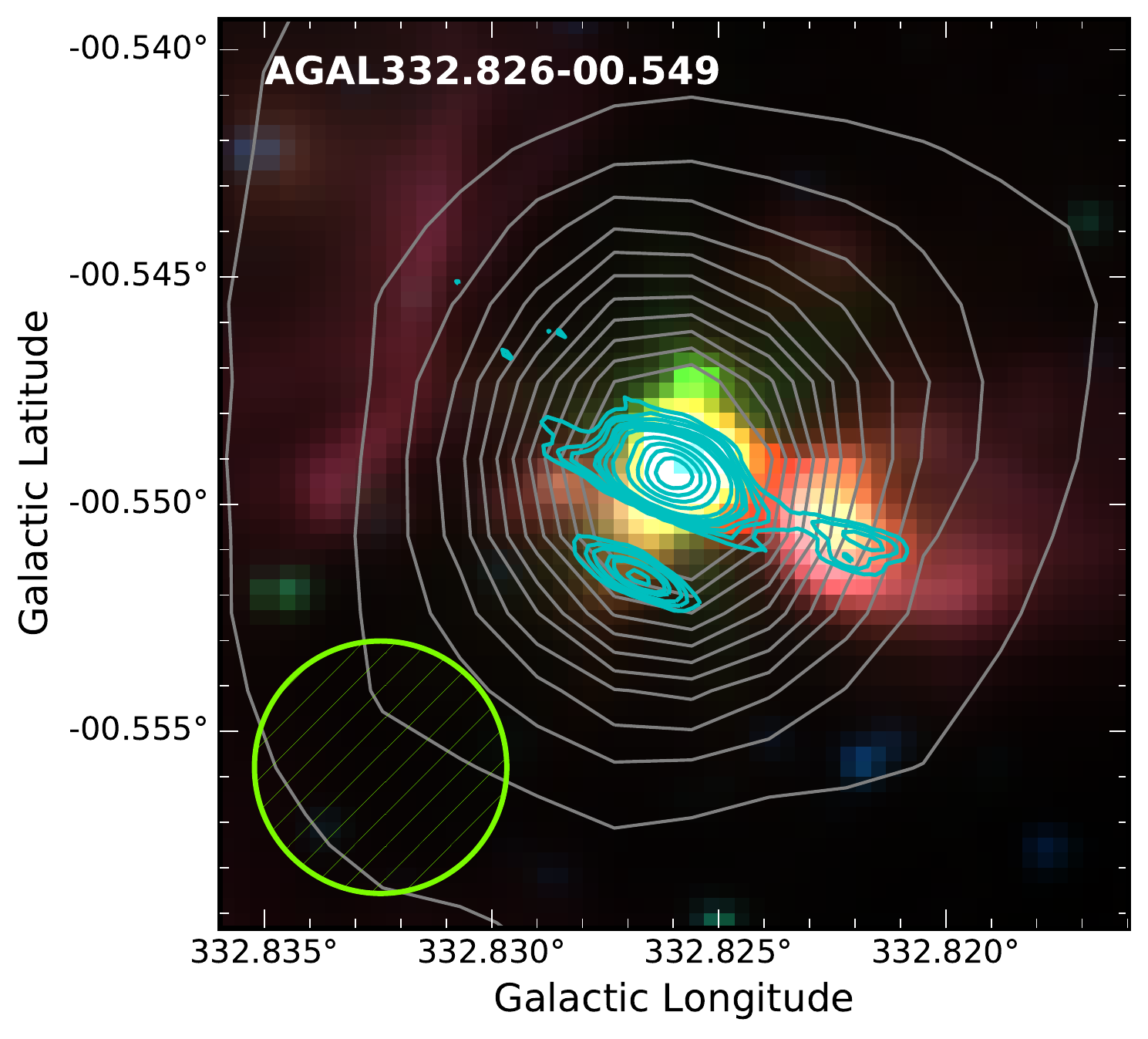}
\includegraphics[height= 0.3\textwidth]{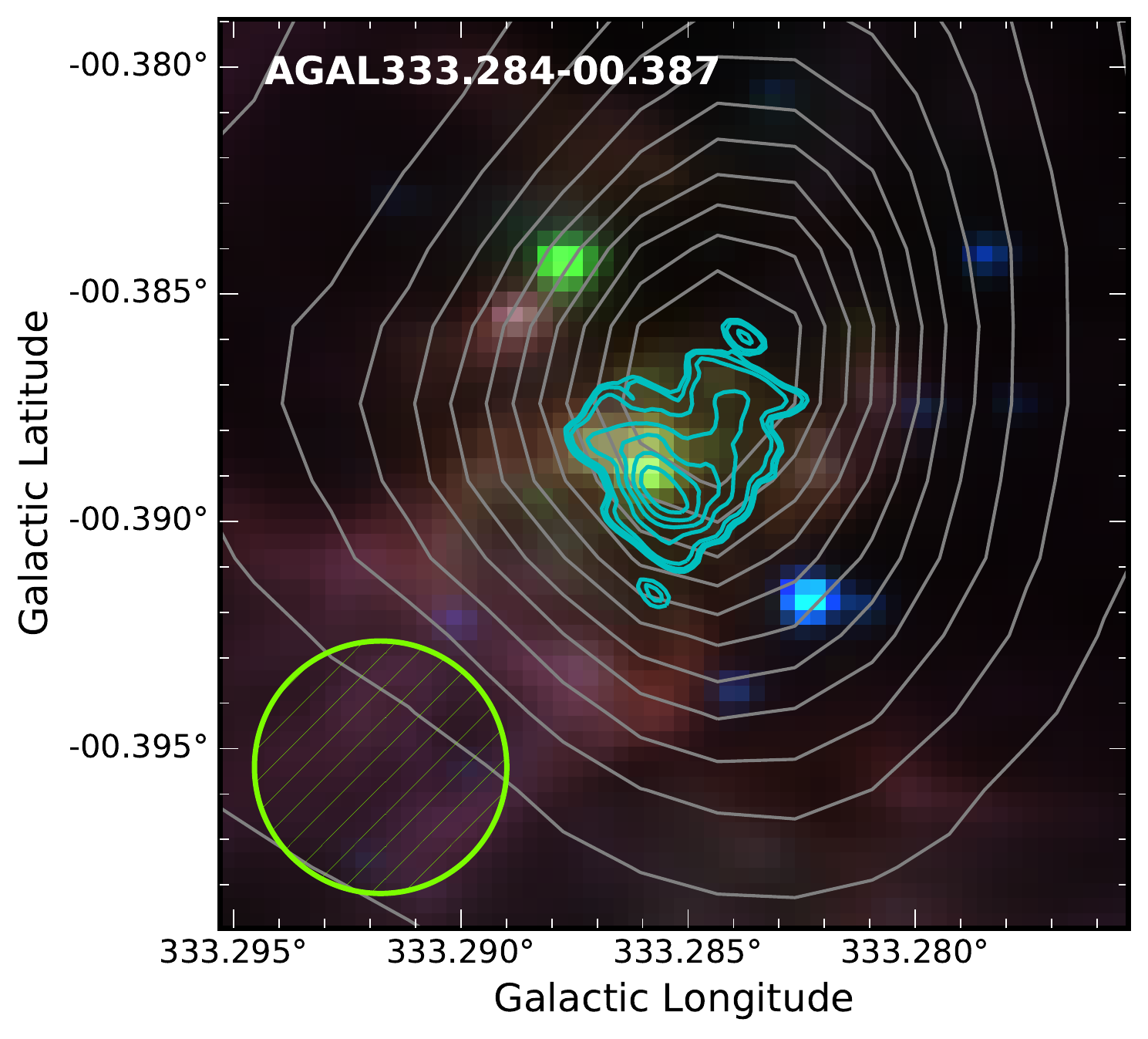}
\includegraphics[height= 0.3\textwidth]{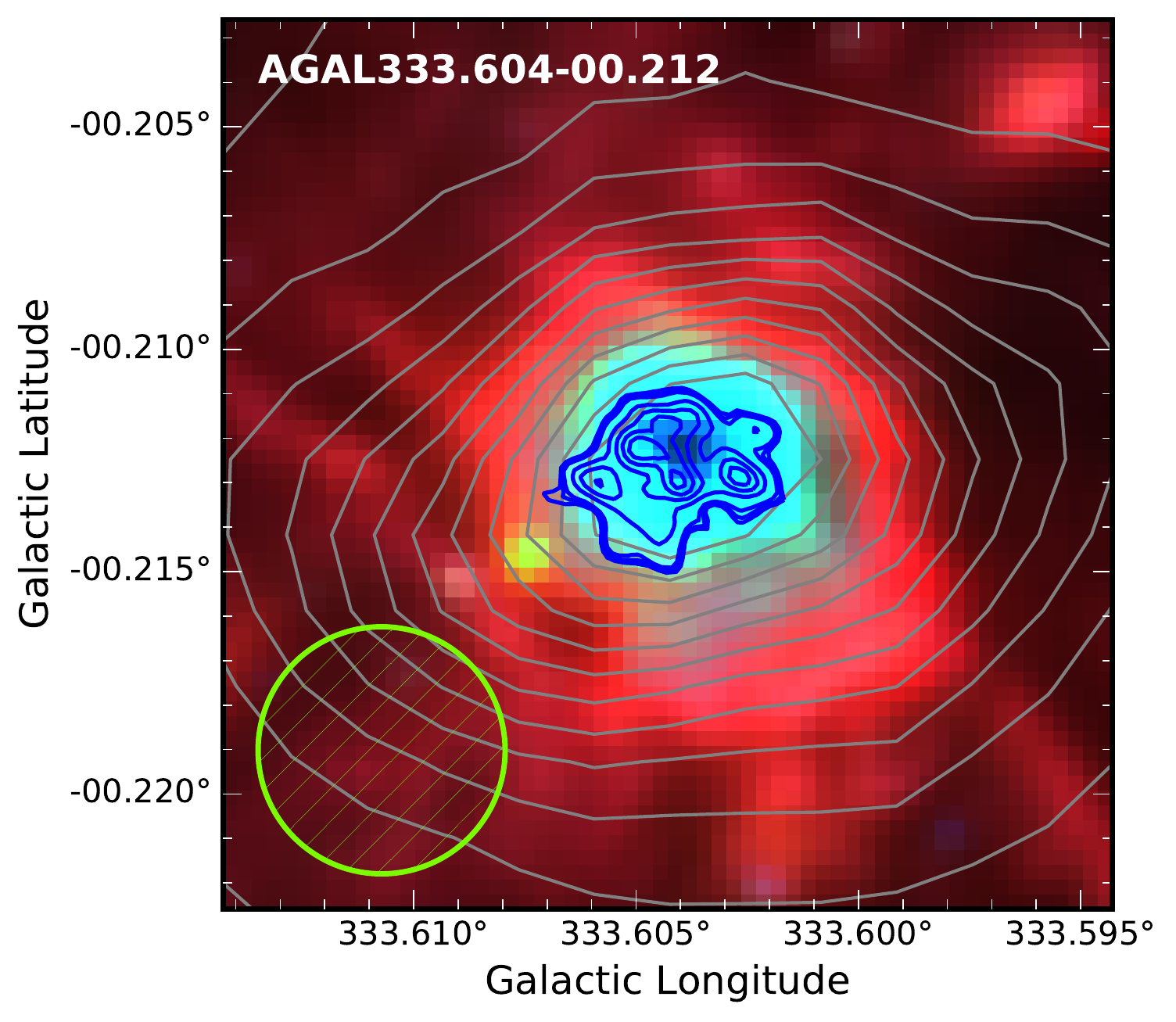}
\includegraphics[height= 0.3\textwidth]{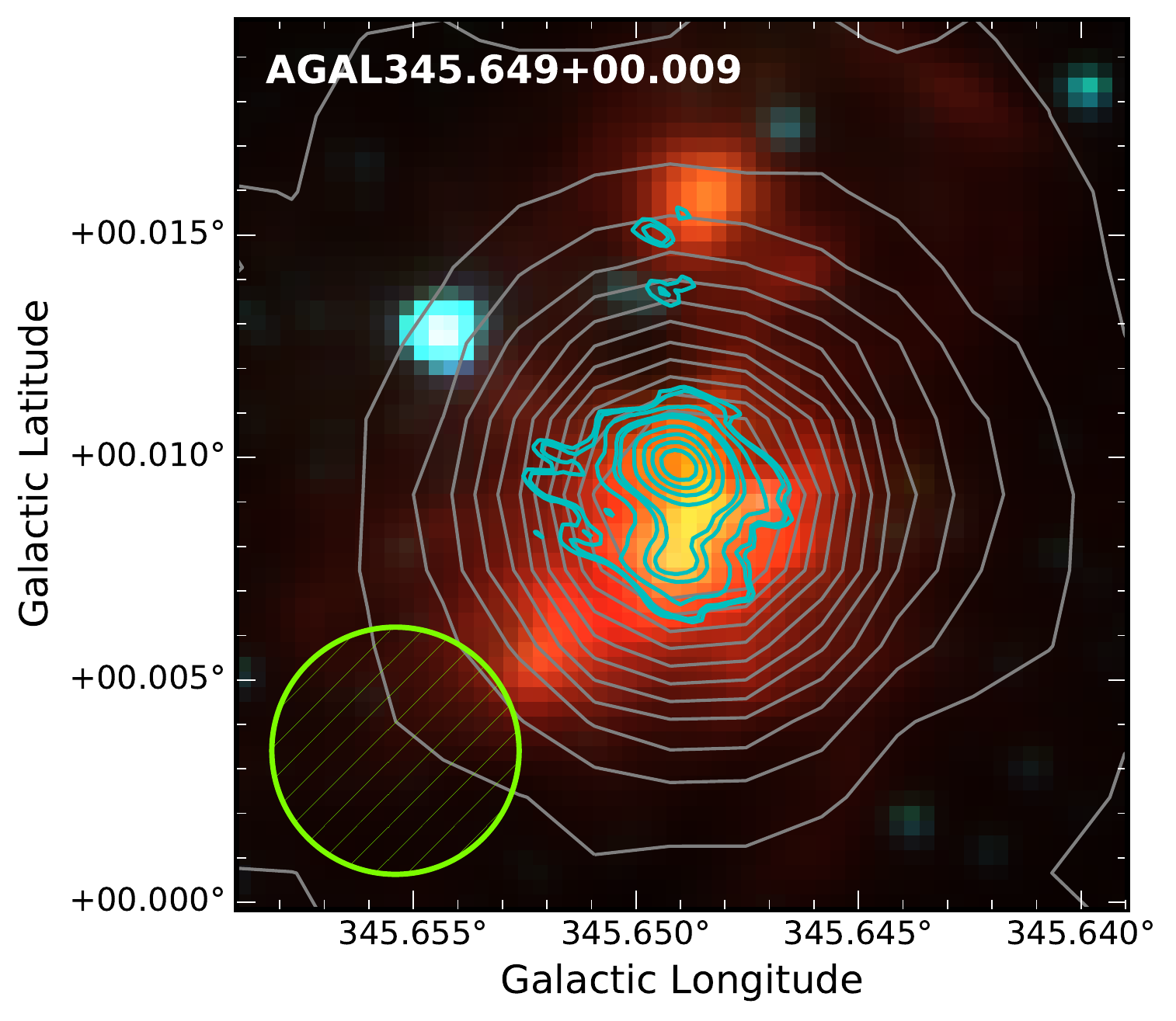}
\caption{Continued.}
\end{figure*}
\end{appendix}

\end{document}